\documentclass[apj]{emulateapj}
\newcommand{\rion}[2]{{\ensuremath{\mbox{\rm #1$\,${\small\uppercase\expandafter{\romannumeral#2\relax}}}}}}
\newcommand{\HII}{\rion{H}{2}}

\newcommand{\M} {\ensuremath{\mbox{\rm \texttt{MAPPINGS}}}}
\newcommand{\MII}{\ensuremath{\mbox{\rm \texttt{MAPPINGS\,II}}}}
\newcommand{\MIII}{\ensuremath{\mbox{\rm \texttt{MAPPINGS\,III}}}}
\newcommand{\MIV}{\ensuremath{\mbox{\rm \texttt{MAPPINGS\,IV}}}}
\newcommand{\MV}{\ensuremath{\mbox{\rm \texttt{MAPPINGS\,V}}}}
\received{}
\revised{}
\accepted{}
\shorttitle{Self-consistent pre-ionisation in radiative shocks}
\shortauthors{Sutherland \& Dopita}

\begin{document}

\title{Effects of Pre-ionisation in Radiative Shocks I: Self-Consistent Models}

%
%
%
%
\author{Ralph S. Sutherland \altaffilmark{1}, Michael A. Dopita \altaffilmark{1}}
\altaffiltext{1}{Research School of Astronomy and Astrophysics, Australian National University, Cotter Road, Weston Creek, ACT 2611, Australia}
%
%
%

\begin{abstract}
In this paper we treat the pre-ionisation problem in shocks over the velocity range $10 < v_{\rm s} < 1500$\,km/s in a self-consistent manner. We identify four distinct classes of solution controlled by the value of the shock precursor parameter, $\Psi = {\cal Q}/v_s$, where ${\cal Q}$ is the ionization parameter of the UV photons escaping upstream. This parameter determines both the temperature and the degree of ionisation of the gas entering the shock.  In increasing velocity the shock solution regimes are cold neutral precursors ($v_s \lesssim 40$\,km/s), warm neutral precursors ($40 \lesssim v_s \lesssim 75$\,km/s),  warm partly-ionized precursors ($75 \lesssim v_s \lesssim 120$\,km/s), and fast shocks in which the pre-shock gas is in photoionisation equilibrium, and is fully ionized. The main effect of a magnetic field is to push these velocity ranges to higher values, and to limit the post-shock compression. In order to facilitate comparison with observations of shocks, we provide a number of convenient scaling relationships for parameters such as post-shock temperature, compression factors, cooling lengths, and H$\beta$ and X-ray luminosity.
\end{abstract}


\section{Introduction}\label{sec:intro}
The physical structure and predicted emission line spectrum of radiative atomic J-shocks is now well understood following the pioneering work of \citet{Cox72} and subsequent studies by \citet{Dopita76,Dopita77,Dopita78}, \citet{Raymond79} and \citet{Shull79}. Whilst the coupled ionisation-cooling balance equations are readily solved in the cooling and recombination zones of the shock, the optical/IR emission line spectrum also depends on the ionisation state, temperature and magnetic pressure of the gas entering the shock. For fast shocks \citep{Dopita96,Allen08}, the ionisation and temperature of this gas is entirely determined by the ionising UV photons produced in the cooling zone of the shock, which run ahead to pre-ionize the incoming medium. For shocks with velocities $v_s \gtrsim 150$km/s, the velocity of the ionisation front is appreciably greater than the shock velocity, and an equilibrium \HII\ region develops in the precursor medium. Thus, the ionisation state and temperature of the gas entering the shock is given simply by an equilibrium photoionisation computation using the computed upstream EUV photon field.

At lower velocities, ionising photons are still produced in the cooling zone of the shock, but the upstream photon field is weak, and the photoionisation front is trapped near the shock front. This case was first treated in a semi-analytic fashion by \citet{Shull79}. In this calculation, the thermal balance of the pre-shock gas was not treated, since heavy element cooling was not included. The radiative transfer of hydrogen- and helium-ionising photons was treated using a mean photoionisation cross-section approximation. For the heavy elements, a similar formalism as used for hydrogen and helium was applied to compute the complete ionisation state of the gas entering the shock.

Since the precursor partly-ionized zone in slow shocks is relatively thin, the timescale for incoming atoms to be advected through this zone can be considered short compared to the recombination timescale. In this case, to a good approximation, we can simply balance the upstream EUV photon flux (cm$^{-2}$\,s$^{-1}$) coming out of the shock, $F_{\mathrm up}$, with the number of ionized atoms advected into the shock (cm$^{-2}$\,s$^{-1}$). Thus, considering only hydrogen, $F_{\mathrm up} \sim xn_{\rm H}v_s$, where $n_{\rm H}$ is the number density of hydrogen atoms in the pre-shock gas, $v_s$ is the shock velocity, and $x$ is the fractional ionisation of hydrogen in the gas entering the shock. The approximate validity of this  simple equation to compute the self-consistent pre-ionisation state of the gas entering the shock can be confirmed by comparison with Fig 5 of \citet{Shull79}.

In this paper we provide for the first time a fully-self consistent treatment of the pre-shock ionisation and thermal structure by iteratively solving for the fully time-dependent photoionisation, recombination, photoelectric heating and line cooling of the pre-shock (atomic) gas as it is advected into the shock. We cover a velocity range of $10 < v_{\rm s} < 1500$\,km/s which allows us to distinguish the transition to the fast-shock regime, as well as identifying various thermal and ionisation transitions which occur in slower shocks. The models presented here cover pre-shock densities up to $10^4$\,cm$^{-3}$, and we study the effects of density and pre-shock magnetic fields on cooling lengths and compression factors. In the next paper of this series,  we will apply these new models to the emission line diagnostics of Herbig-Haro shocks in the velocity range of $20 < v_{\rm s} < 150$\,km/s, and demonstrate how these models can assist in developing our understanding of the physical parameters of Herbig-Haro outflows and mass-loss from pre-main sequence stars.

This paper is structured as follows. In Section \ref{sec:code} we briefly summarise the status of the \M\ code used in these computations, in section \ref{sec:RHflow} we provide the method of solution of the Rankine-Hugoniot flow equations in terms of pressure (gas pressure, ram pressure and magnetic pressure). In  Section \ref{sec:iterates} we describe in detail our new treatment of the time-dependent shock photoionized precursors, introducing the concept of the precursor parameter in section \ref{sec:precursor_param}. Section \ref{sec:shock_types} demonstrates the existence of four distinct regimes of shock solutions of which three have partially-ionized precursor regions. We investigate in detail the effect of the magnetic field on the structure of both the shock and its precursor in Section \ref{magfield}, and in sections \ref{compress} to \ref{sec:Xray} we provide a number of useful scaling relationships that could be used to interpret the physical properties of observed interstellar shock waves.

\section{The MAPPINGS code}\label{sec:code}
The original \M\ code \citep{Binette85} was developed out of an earlier code by \citet{Dopita76,Dopita77,Dopita78} in order to provide a single framework within which to model the emission line and continuum spectra of equilibrium ionisation objects such as HII regions while at the same time being able to do the same for plasmas which are well out of collisional or photoionisation equilibrium such as the radiative shocks in supernova remnants (SNR) or Herbig-Haro shocks and their precursor zones. This is accomplished using a time-dependent matrix solution to the coupled ionisation/cooling equations, and treating equilibrium plasmas and special examples of the time-dependent case as time $t \rightarrow \infty$. The \MII\ code extended the earlier code to include many new atoms, ions, and physical processes by \citet{Sutherland93}. The \MIII\ code was further extended to include dust heating \citep{Dopita00} and its non-equilibrium heating and IR emission \citep{Groves04,Groves06,Dopita05}. The earlier version of the \MIV\ code was described by \citet{Dopita13}, and included improved treatment of temperature-dependent collision strengths, and line excitation and recombination under a $\kappa - $distribution of electrons \citep{Nicholls12}, as well as in the simple Maxwell-Boltzmann case.

\begin{figure*}
  \includegraphics[width=\textwidth]{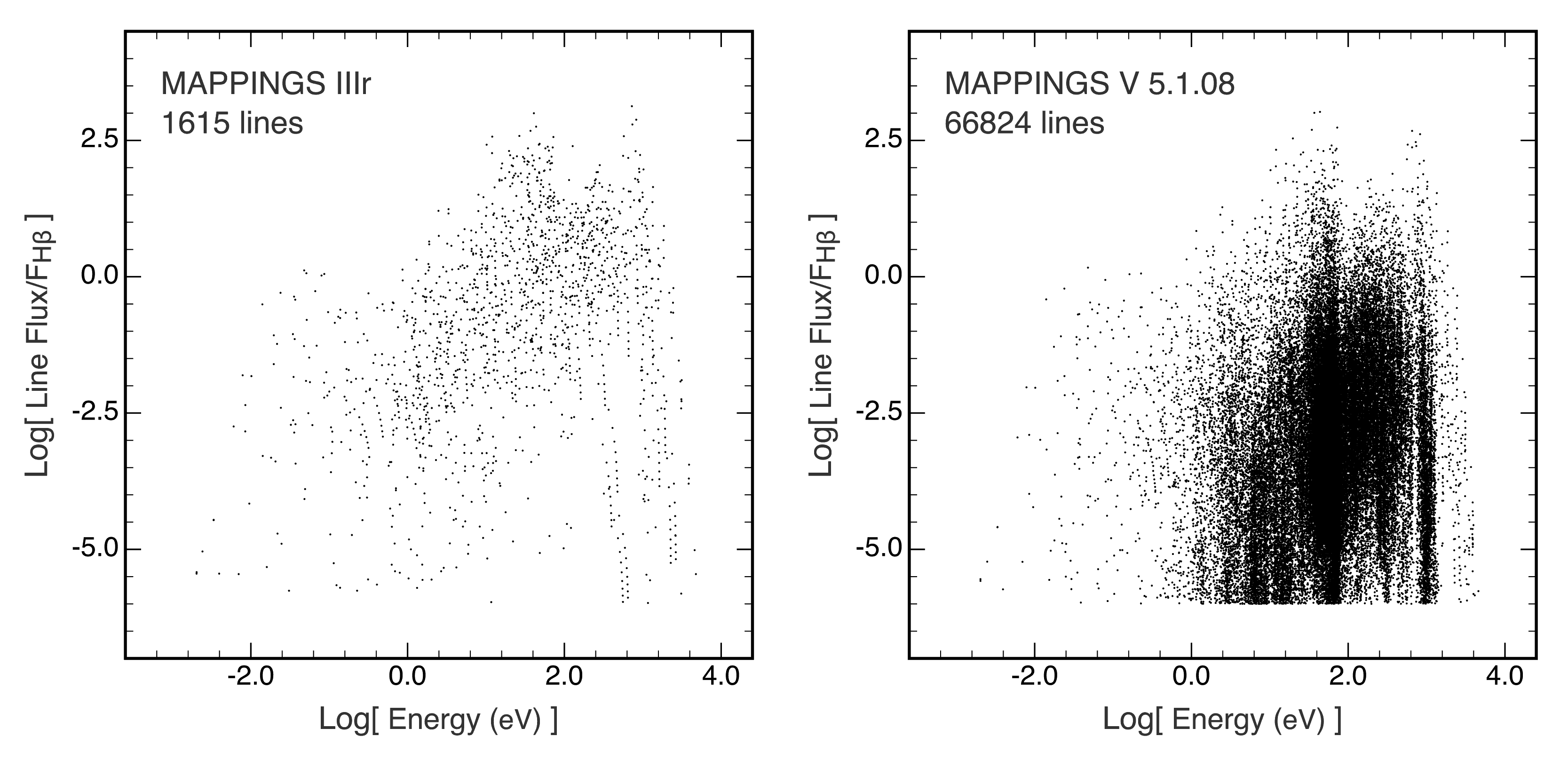}
  \caption{Emission line strengths as a function of energy computed by (left) MAPPINGS III, and (right) the latest version of MAPPINGS. These are given for a 600\,km/s shock propagating into fully ionised gas with solar composition and zero magnetic field. The ability of the new code to produce detailed X-ray spectra and to generate more reliable diagnostics is made clearly evident in this plot.} \label{fig:lines}
 \end{figure*}

This much improved \MV\ code\footnote{Available at: {\url https://miocene.anu.edu.au/mappings}} will be discussed in detail in a forthcoming paper (Sutherland \& Dopita, 2017) which provides new cooling function computations for optically thin plasmas. This is based on greatly expanded atomic data of the CHIANTI 8 database. The number of cooling and recombination lines has been expanded  from $\sim 2000$ to over $80,000$, and temperature dependent spline based collisional data has been adopted for the majority of transitions.  As en example, we present in Figure \ref{fig:lines} a plot of all emission lines with strength $ > 10^{-6}$ of H$\beta$ in a model of a 600\,km/s shock with no magnetic field and with solar composition propagating into fully ionised gas, and cooling from about $5\times10^6$\,K down to $10^4$\,K.

The new and expanded atomic dataset provides greatly improved modelling of both thermally and photo-ionized ionized plasmas.  In particular, the code is now capable of predicting detailed X-ray spectra of non-equilibrium plasmas over the full non-relativistic temperature range, which will greatly increase its utility in Cosmological simulations, in modelling cooling flows, and in generating accurate models for the X-ray emission from shocks in supernova remnants.

\section{Solving for the Shocked Flow}\label{sec:RHflow}

In the literature one can find many expressions for the Rankine--Hugoniot flow equations \citep{Rankine1870,Hugoniot1887}, both for isentropic (reversible) flows and for entropy producing shock solutions \citep{Cox72,Shull79,McKee80,Bertschinger86,Innes87,Shapiro92,Dopita96}.  Some of the many forms which have been used can also be seen in the NACA Techical Report 1135 (1953) \nocite{tr1135}.

In this work we use a variation on previous solutions by choosing an integral form of the conservation laws (which helps to eliminate numerical roundoff errors), and focus on pressure as the principal flow variable, since this provides the principal driving force in all shock phenomena. Between any two points in the flow, $f_0$ and $f_1$, denoted by subscripts, the integral conservation laws for a radiative ideal gas with a magnetic field can be written in terms of the usual primitive variables $[P,\rho,v,B]$.  However in this work we combine the mass density, $\rho$, the velocity $v$ and magnetic field strength ($B$, the perpendicular component to the velocity vector)  into pressure terms $[P, P_{\rm ram}, P_B]$ to simplify the expressions and make the relative contributions of the driving internal, dynamic and magnetic terms more apparent.  Here the variables have their usual meanings, unless noted otherwise, and generally the velocities are taken in the shock front frame of reference.   We use Gaussian magnetic c.g.s units throughout (i.e. magnetic $\mu_0 = 4\pi$).

 First, mass conservation and continuity means,
\begin{equation}
\rho_0 v_0  =  \rho_1 v_1 \; , \label{eq:massconserve}\\
\end{equation}
and so,
\begin{equation}
x = \rho_1/\rho_0 = v_0/v_1 \; ,
\end{equation}
where $x$ is the compression factor in terms of density.

 The transverse component of the magnetic field, $B$, (taken to be frozen to the matter in the flow, ideal MHD) is constrained by,
\begin{equation}
B_0/\rho_0 =  B_1/\rho_1 \; , \label{eq:frozen}
\end{equation}
and so,
\begin{equation}
B_1/B_0 = v_0/v_1 = x \;.
\end{equation}

We then define three pressures:
\begin{eqnarray}
\mbox{Gas:}\; & P & = nkT \equiv \frac{\rho k T}{\mu m} \equiv (\gamma-1)E \; , \\
\mbox{Ram:}\;  & P_{\rm ram} & = \rho v^2 \; ,\\
\mbox{Magnetic:}\; &  P_{B} & =  B^{2}/8\pi \; ,
\end{eqnarray}
where : $k$ is the Boltzmann constant, $T$ is the temperature, $n$ is the total particle number density, $\mu$ is the mean molecular weight per particle, $m$ is the atomic mass unit, and E is the internal specific energy. Both $n$ and $\mu$ are computed self-consistently from the ionization state of the plasma, and include both ions and electrons.

The conservation of momentum flux, can expressed in terms of the pressure balance:
\begin{equation}
P_0 + P_{\rm ram,0} + P_{B,0}  =  P_1 + P_{\rm ram,1} +  P_{B,1} \,. \label{eq:momentum}
\end{equation}

Finally, If the polytropic index $\gamma$ is combined into a more compact factor, $g = \gamma/(\gamma-1)$, then the specific energy equation can be written:
\begin{eqnarray}
  & & \frac{1}{\rho_0}\left[ gP_0 + \frac{1}{2} P_{\rm ram,0} + 2P_{B,0} \, - \,\left\langle  L \right\rangle \Delta t \right]   \nonumber \\
  & = &  \frac{1}{\rho_1}\left[ gP_1 + \frac{1}{2} P_{\rm ram,1} + 2P_{B,1} \right] \; . \label{eq:energy}
\end{eqnarray}
where the averaged radiative losses,  $\left\langle L \right\rangle$, are included as emissivity by volume (erg cm$^{-3}$ s), over a time interval $\Delta t$ (taken as the flow time between the points), the radiated energy lost between $f_0$ and $f_1$ is $\left\langle L \right\rangle \Delta t$ (erg cm$^{-3}$).  In practice $\left\langle L \right\rangle$ for a given step is solved iteratively, taking the average of the cooling before and after the step, allowing the ionisation state to evolve.  Since the cooling is coupled to the ionisation state and temperatures (\emph{via} the gas pressures and density here) a series of iterations for the mean cooling and step evolution is made until the results converge.  This is effectively a form of operator splitting for the cooling.  In this work the polytropic index $\gamma$ is taken to be constant, equivalent to assuming the gas remains as an ideal monatomic gas throughout.

\subsection{Solutions}

We derive a solution to the compression factor, $x = \rho_1/\rho_0$, between any two points in the flow, allowing for both the steady flow and shock jump solutions in the same solver, following \citet{Cox72}.  Since the radiative losses may not be zero, the flows are neither reversible, nor strictly isentropic. The discontinuous shock solution is assumed to occur over $\Delta t = 0$, and is thus radiatively lossless.

 The compression ratio $x$ between any two flow points, $f_0$ and $f_1$, in terms of the flow variables at $f_0$, is the largest positive root for $x = \rho_1/\rho_0$, in the cubic equation:
 \begin{eqnarray}
  & x^3&\left[\frac{}{}P_{B,0}\left( 2-g \right)\frac{}{}\right]  \nonumber \\
    - & x^2&\left[\frac{}{}gP_0 + \frac{1}{2}P_{\rm ram,0} + 2P_{B,0} -\left\langle  L \right\rangle \Delta t \frac{}{}\right]  \nonumber \\
    + & x&\left[\frac{}{}gP_0 + gP_{\rm ram,0} + g P_{B,0}\frac{}{}\right] \nonumber \\
    + & &\left[\frac{}{}P_{\rm ram,0} \left(1/2 - g\right)\frac{}{}\right] = 0 \;. \label{eq:cubicshock}
 \end{eqnarray}
 This applies to both supersonic shocked, compressible flows, (${\cal{M}} > 1$) and to steady approximately isentropic flows (${\cal{M}} < 1$).

Equation (\ref{eq:cubicshock}) above can be solved quite efficiently using a standard Newton iteration scheme. Once $x$ is known, the rest of the flow variables at $f_1$ can be determined from $[\rho_0, v_0, P_0, B_0; P_{B,0}, P_{\rm ram,0}]$, by simple substitution into equations (\ref{eq:massconserve})--(\ref{eq:frozen}) above to get $\rho_1$, $v_1$, and $B_1$, thence $P_{B,1}$ and $P_{\rm ram,1}$, and finally $P_1$ from equation (\ref{eq:momentum}).

The integral form of the momentum and energy equations makes it convenient to check that the solution is conservative to numerical roundoff precision at every step in the calculation applying equation \ref{eq:energy} both before and after the step.

\subsubsection{Quadratic Shock--Only Solution}

In this case we can describe the physics use the usual characteristic velocities,
$$\mbox{Sound speed:}\; \; c_s^2 = \gamma P/\rho\;,$$
$$\mbox{Alfv\'en speed,}\; c_a^2 = 2 P_B/\rho\;,$$
and  Mach numbers:
$$\mbox{Sonic Mach number,}\; {\cal M} = v/c_s\;,$$
$$\mbox{ Alfv\'en Mach number,}\; {\cal M}_A = v/c_a\;.$$

A shock jump occurs when $\gamma P/P_{\rm ram} = 1/{\cal M}^2 < 1$, and  $2P_B/P_{\rm ram} = 1/{\cal M}_A^2 < 1$, the larger root from equation (\ref{eq:cubicshock}) results in an increase in entropy (as $P/\rho^\gamma$), while the smaller root at $x = 1$ is trivially isentropic and steady, that is no change, since $\Delta t = 0$ and radiative losses are zero across the jump.  In the this case the steady roots can be further eliminated by polynomial division by $(x-1)$, to give a quadratic expression for the shock jump solutions with no losses.  This quadratic cannot be used for the steady flow solutions, but can be solved algebraically and serves as a check for the iterative solution to the cubic equation over the shock jump.
\begin{eqnarray}
       & x^2&\left[ 2P_{\rm ram,0}(2-\gamma) \right]  \nonumber \\
     + &   x&\left[ (\gamma-1)P_{\rm ram,0} + 2\gamma(P_{B,0}+P) \right] \nonumber \\
     - &    &\left[ (\gamma+1)P_{\rm ram,0} \right] = 0 \;, \label{eq:shockmhd}
 \end{eqnarray}
or in terms of ${\cal M}^2 = \gamma P_{\rm ram}/P$ and $\alpha = P_B/P$, the ratio of magnetic to gas pressure;
\begin{eqnarray}
      & x^2&\left[ 2\alpha(2-\gamma) \right]   \nonumber \\
    + &   x&\left[ \gamma(\gamma-1){\cal M}_0^2 + 2\gamma(\alpha+1)\right] \nonumber \\
    - &    &\left[ \gamma(\gamma+1){\cal M}_0^2\right] = 0 \;. \label{eq:shockmhd2}
 \end{eqnarray}

In the post shock gas the flow is subsonic, ${\cal M }< 1$ and ${\cal M}_A < 1$, and the cubic gives two roots, the larger being $\sim1$, steady and isentropic.  If the losses are non-zero then the root is shifted in the neighbourhood of 1.0.

\section{Time-Dependent Shock Precursors} \label{sec:iterates}

The shock jump solution which gives the post-shock temperature, $T_s$, as a function of the shock velocity, $v_s$, depends in part on the mean molecular weight of the particles entering the shock, $\mu$. A substantial fraction of the kinetic energy flux into the shock is converted to internal energy, and hence to the gas pressure.  The Rankine-Hugoniot relation for temperatures in terms of the shock Mach number is,
\begin{equation}
\frac{T_1}{T_0} = \frac{[(\gamma-1){\cal M}_0^2 + 2][2\gamma {\cal M}_0^2 - (\gamma-1)]} {( \gamma + 1 ) 2 {\cal M}_0^2} \, . \label{eq:TM}
\end{equation}
see NACA Technical Report 1135 eq 95 (1953) \nocite{tr1135}.

In the limit of ${\cal M}_0 >>1$, this yields an expression for the post shock temperature, eg. \cite{McKee80}, \cite{adu2003},
\begin{equation}
T_s = \frac{2(\gamma-1)}{(\gamma+1)^2}\,  \left[\frac{\mu m_u v_s^2}{k}\right] \, ,\label{Te_PS}
\end{equation}
where $m_u$ is the atomic mass unit ($1.66053904\times10^{-24}\,$\,g; CODATA 2014), $\gamma$ is the adiabatic index in an ideal gas, and  $k$ the Boltzmann constant.

The mean particle mass, $\mu m_u$, is determined by the ionisation balance of the plasma in the pre-shock region.  As the ionisation state changes, the mix of light electrons and heavier ions changes, so that in non-molecular, $\gamma = 5/3$, solar composition plasmas, $\mu$ can vary from $\sim1.2$ in neutral gas, to $\sim0.6$ in the fully ionized case.  Thus, for a given shock velocity in the local interstellar medium (ISM), the post shock temperature may vary by a factor of two depending on the ionisation state of the pre--shock, or shock precursor region.  The subsequent ionisation and collisional excitation and line emission in the cooling post--shock region depends strongly on the temperature of the plasma, so it is clearly very important to determine a self-consistent value for the pre-shock ionisation in order to correctly compute the emission line spectrum of the shock.

In addition, for fast shocks, the line emission from the photoionized shock precursor itself can become an important fraction of the total spectrum emitted by a radiative shock \citep{Dopita95,Allen08}. To compute the effect of the shock precursor emission, we need to solve for the time-dependent photoionisation, heating, and radiative transfer of the ionizing photons escaping upstream from the shock in the plasma which is being advected into the shock.

\subsection{Preheating}
Even a shock with a velocity which is  modest by astrophysical standards ($\sim 50$\,km/s) can heat plasma to many thousands of degrees.  This shocked gas emits line and continuum radiation as it cools. A fraction of the radiation consists of photons in the hard UV or X-ray bands which can then photoionize nearby gas.  Much of this ionisation occurs in the pre-shock gas which is being advected into the shock. In the frame of the shock front (following from  equation (\ref{eq:shockmhd2}), and ignoring magnetic field terms), this pre-shock material is also being compressed to the post--shock density on passage through the shock:
\begin{eqnarray}
\rho_s & = &\rho_0 \frac{(\gamma+1){\cal M}^2}{(\gamma - 1){\cal M}^2 + 2}\, , \label{eq:cmpf}\\
\rho_s   & \sim &4 \rho_0 \; \mbox{\rm in the strong shock limit , $\gamma = 5/3, \: {\cal M}>>1$ }\, \nonumber ,
\end{eqnarray}
where $\cal M$ is the Mach number in the pre--shock flow. The feedback of the UV radiation field on the surrounding plasma results in the first iterative problem when computing the resulting shock structure, since the upstream field is capable of both pre--heating and pre--ionizing  the precursor material. This changes the sound speed,  $c_s$, and reduces both the Mach number, ${\cal M}$, and the shock compression factor (see equation  \ref{eq:cmpf}). With these altered shock parameters, the computed total emission in the shock -- which would otherwise be approximately proportional to density squared  -- can drop. This in turn reduces the computed pre--heating.  Thus, it is necessary to solve for the self-consistent pre--ionisation and heating in an iterative manner.

In the models presented here, an initially cold pre--shock region is used to compute a hot dense shock, which in turn is used to pre--heat the  precursor.  A subsequent (somewhat lower density and cooler) shock is computed with the new pre--shock sound speed, and a new pre--heating is evaluated.  If the precursor is estimated to be too hot, the Mach number drops and the shock emission is reduced via a lower compression ratio leading to lower post--shock densities.  Although the non--linear nature of this shock--precursor heating coupling could potentially lead to some instability, in the range of shock studied so far, from $10 - 2000$\,km/s, this preheating cycle generally converges to $<<1$~\% temperature variations in between 5 and 30 iterations.

\subsection{Pre-ionisation}

In addition to simply heating the pre-shock gas, the EUV shock emission can also photoionize the  pre--shock gas. If more electrons are present, $\mu$ will drop from the low-velocity limit of $\sim1.2$ to $\sim0.6$ when the pre-shock plasma becomes fully ionized.  The onset of appreciable pre-ionisation occurs at shock velocities above $\sim 65$\,km/s, and shock and precursor structure iterations are again needed to arrive at a steady-flow solution.  This ionisation limit is more or less independent of the pre-shock temperature, being primarily a function of the increasing energy of the upstream photons.  Below $65$\,km/s, the pre-heating which can heat the plasma to $\sim10^{4}$\,K produces only a small degree of collisional ionisation.

In practice, the pre--ionisation and pre--heating are closely coupled, so in \MV\ both are computed within a single iterative scheme in which we empirically iterate the shock--precursor system with numerical integration techniques until a series of parameters, including Mach number, compression factor, hydrogen and helium ionisation fractions become steady within a chosen limit between iterations.

In figure \ref{fig:iterations} the upper left panel shows the ionisation fractions of hydrogen, with a curve for each iteration. The right hand edge of the panel shows the ionisation state reaches a steady value after relatively few iterations while the outer regions continue to evolve.

\subsection{Radiative Transfer}
We consider a parcel of neutral gas in a \emph{proto--state}; temperature $T_{\rm pro}$, ionisation state $\chi_{\rm pro}$, number density $n_{\rm pro}$, etc.  Starting at a distance $x_{\rm pro}$, and moving at the shock velocity $v_s$. With $N$ steps in space  $dx = x_{\rm pro}/N$, time steps are simply, $dt = dx/v_s$. Calculations begin at $t=0, x=x_{\rm pro}$ and integrate to $t=x_{\rm pro}/v_s, x = 0$.

\begin{figure*}
  \includegraphics[width=\textwidth]{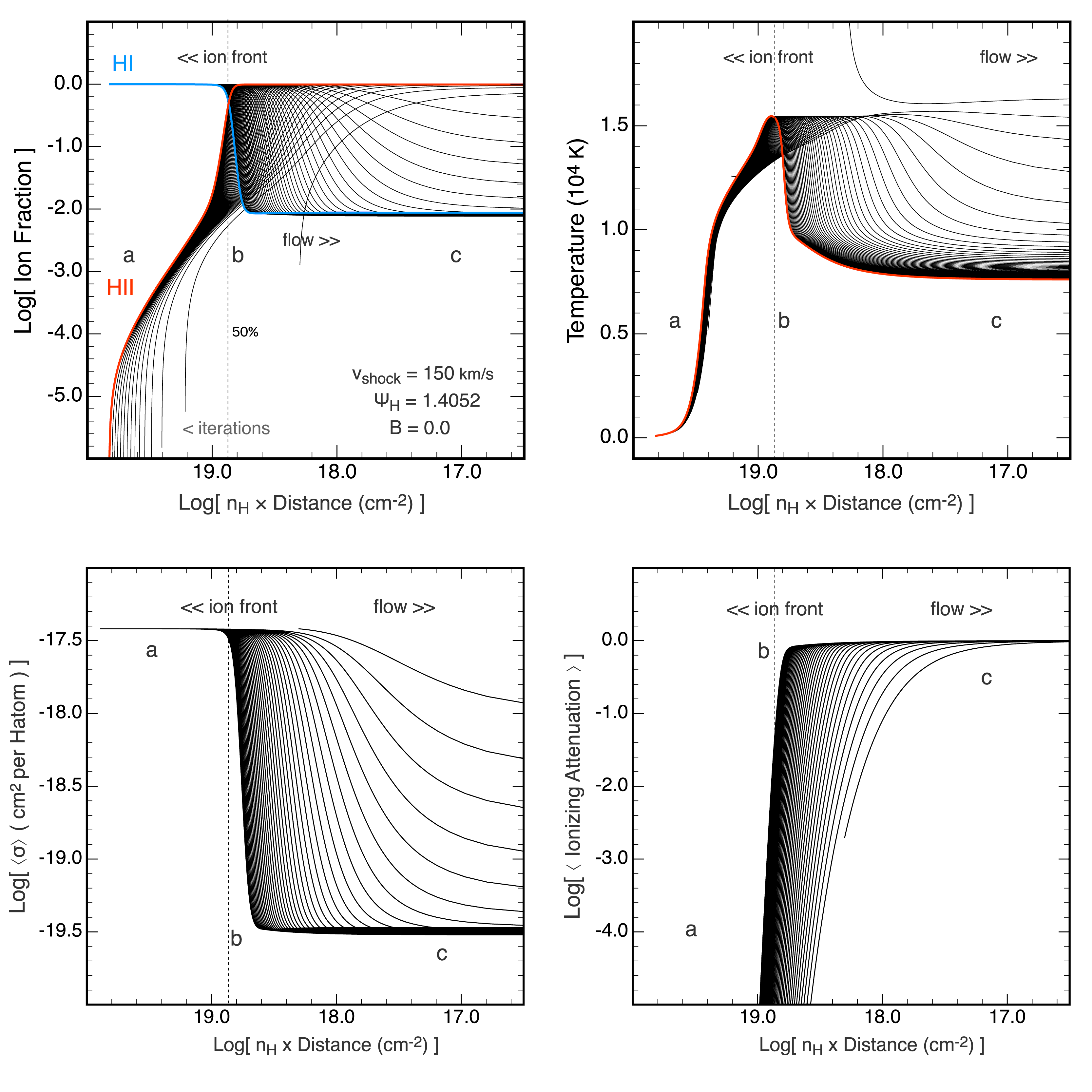}
  \caption{The multi-zone iterative approach to the solution of the pre-shock structure for a 150\,km/s shock. Upper Left, the ionisation state, Upper Right, the temperature. The colored curves are the final solution, while the other curves represent successive iterations. Lower Left the mean absorption cross section per H atom, averaged and weighted by the local radiation field. Lower Right the mean attenuation of the ionising field through the precursor.
For these faster shocks, where the precursor is thick, many iterations are needed to obtain the complete ionisation structure, but relatively few are needed to determine the pre-ionisation conditions of the gas entering the shock. ($\Psi$ is the precursor parameter, see section 5 for its full definition.)} \label{fig:iterations}
 \end{figure*}

As a parcel of gas approaches the shock ( from left to right in Figure \ref{fig:iterations}, seen in the frame of the shock front at $x=0$ ), it sees the shock  ionizing photon field though the intervening parcels of gas that will precede it, but which have not yet been computed.  So, to compute the local ionisation, we need to determine the upstream shock radiation field, $\phi$ absorbed by the intervening gas not yet traversed.  To calculate this we need to know the ionisation state of the plasma throughout to compute the local absorption cross sections.  The ionisation state gives the abundances of each species and the photoelectric cross sections sum to obtain a total cross section and hence opacity at each point.  In the lower left panel, the mean cross section, averaged over $i$ species and weighted by the radiation field, $\phi(\nu)$, notionally:
$$
\langle \sigma \rangle = \sum_i \frac{\int \sigma_i(\nu) \phi(\nu) d\nu}{\int \phi(\nu) d\nu} \,
$$
 is shown evaluated numerically locally. Internally the \MV\ code does not use this mean quantity, but instead integrates at 7977 frequency/energy bins from $10^{-6}$ to $10^{5}$ eV, of which approximately 3000 bins are between 13.598\,eV and 1000\,eV. This covers the vast majority of ionising flux energy even at the highest shock velocities.  In these shock models, the precursor opacity due to hydrogen dominates, and the mean cross section drops from $\sim 3.0\times 10^{-18}$\,cm$^{-2}$ per hydrogen atom to approximately $\sim 3.0\times 10^{-20}$\,cm$^{-2}$/H atom at the point indicated by (b) in the panels, and is less in the ionised region (c).  Summing the ion populations into ion columns for all the plasma between each point and the shock front allows the total cross section and opacity per ion as a function of frequency to be computed. The shock field is thus attenuated at each point, which varies in each iteration as the columns change. The lower right panel shows the averaged attenuation from the shock front to point of the ionising field (that is everywhere the cross sections are non--zero, predominantly above the ionising threshold of hydrogen at 13.598\,eV.

In the past, \citet{Shull79} used the idealised radiative transfer method described in Section \ref{sec:intro}, while  \citet{Binette85} approximated the entire precursor as a single zone with a mean field.  Initially neutral, the pre-shock gas was exposed to the upstream radiation field, and the ionisation was solved with single-slab time-dependent ionisation and temperature evolution across the zone.  The photons absorbed in this single slab were matched to the upstream ionizing photon field. The shock structure was iterated with the computed incoming pre-ionisation until convergence was attained. This approach worked fairly well within its limitations, but the shock precursor structure remained highly approximate. In general, using the mean field will underestimate the true field at the shock front, and potentially underestimate the ionisation.  In addition the single zone method depends on a good estimate of the zone depth, or $x_{\rm pro}$.  If the single zone is too large the mean field is too low and the time step is too long, and if too small  then the field may be too high and the time step too short.  Fortunately these two effects cancel to some degree, but the method still remains dependent on the initial guess at $x_{\rm pro}$.

\subsubsection{Multi-Zone Iterative Scheme}

With the greater computer resources now available, we can now take a multi-zone approach to the precursor, solving for $x_{\rm pro}$ fully self--consistently, and integrating the field (over $N \sim 100$ steps) to achieve a stable solution for the ionisation, temperature and line emission properties throughout the precursor region.  This approach addresses the key shortcoming of using a single mean field on the final time dependent ionisation ($\mu$) and temperature $T$, and hence $c_s$ and $\cal M$.

In the case where the precursor is small, and the plasma does not have time to become fully ionised before entering the shock we aim to capture the entire precursor structure, since the ionisation state never reaches equilibrium and the state at the shock front depends on the entire structure and the time integral through it. In this case the time-dependent multi-zone structure can be summed to recover the (faint) precursor spectrum.

In the case (described later) where the precursor becomes large and the ionisation reaches equilibrium before it enters the shock, we need only to capture enough of the precursor to achieve the equilibrium state entering the shock to get a steady shock model.  In this case the time-dependent multi-zone structure may only be a partial representation of the true precursor.  It is important to note that in this case, once equilibrium is reached, information about the the proto--ionisation conditions are lost to the shock, however the ionisation state at the shock front is well defined.

The method used here is essentially a relaxation method. The complete plasma state: temperature, density and ionisation state of all species, at each point, plus the integrated column densities from that point to the shock front (to derive opacities using photoelectric cross sections for each species) of all species, are allocated to memory. Initially all the zones are filled with the proto--conditions, generally a neutral and cold state, and an initial shock is run, which because of the large $\mu$ gives an over--dense hot initial solution.

Next, assuming the entire precursor is in the neutral proto--state, a precursor starting point position, $x_{\rm pro}$ is chosen, corresponding to an optical depth between 5 and 10 on the basis of the sum of all photoelectric and dust opacities for all the species present.  A time-dependent ionisation calculation \citep{Binette85,Sutherland93} is then performed in which the field is re-evaluated in each step to be consistent with the changing column densities towards the shock as the parcel approached the front.  The resulting state in each zone is saved and after the first {bf integration of the ionisation state and the computation of the resulting column densities of the ions}, the array of species column densities is recomputed using the new array of states.  The opacities drop near the front since the ionisation increases there.  A second iteration, starting again at the  proto--state, at $x_{\rm pro}$ is then run using the revised column densities and the radiation field as modified by the time--dependent ionisation solution.

In subsequent iterations, if the mean opacities drop due to ionisation of hydrogen in particular, the outer $x_{\rm pro}$ is adjusted outwards and new zones added until its ionisation remains close to the proto--state between successive iterations, and within each zone the column density, temperature and ionisation structure remains unchanged between iterations.  In this way a stable solution of the temperature and ionisation structure of the entire precursor is obtained.

\subsection{Equilibrium Pre-ionisation}

When shocks are fast enough, the ionising photon production rate in the cooling zone can become sufficiently large as to push an ionisation front into the upstream gas at a velocity greater than the shock velocity. To deal with this case, the multi-zone model could be extended outwards indefinitely, limited only by memory and cpu constraints. However, in those cases where a detailed knowledge of the emission from the precursor is not important, all we really need to know is the temperature and ionisation state of the plasma entering the shock.  \MV\ can optionally attempt to compute the entire precursor in this case, but more commonly the precursor model is only extended until the ionisation and temperature of its inner zones near the shock reach equilibrium. For fast enough shocks, this procedure provides pre-shock conditions very similar to the photoionisation equilibrium model described in  \citet{Dopita96} and \citet{Allen08}.

\subsection{Computed Shock Grid}
We have run a series of shock models with the Local Galactic Concordance abundance set as given by \citet{Nicholls2017}.  These abundances  are listed in Table \ref{t:abund}.  The key difference with respect  to the solar photospheric abundance \citep{Asplund2009} is a +0.05\,dex enhancement of oxygen and a +0.03\,dex enhancement in iron.

\begin{table*}[ht]
\caption{Abundances, logarithmic by number relative to $\log_{10}[H] = 0.00$}
\begin{center}
\begin{tabular}{*{8}{l}}
\hline
    H  &\,\,0.000  &  He &-1.010  &   Li &-8.720  &   Be &10.680\\
    B  &-9.190  &  C  &-3.577  &   N  &-4.208  &   O  &-3.240\\
    F  &-7.440  &  Ne &-3.910  &   Na &-5.790  &   Mg &-4.440\\
    Al &-5.570  &  Si &-4.500  &   P  &-6.590  &   S  &-4.880\\
    Cl &-6.500  &  Ar &-5.600  &   K  &-6.960  &   Ca &-5.680\\
    Sc &-8.840  &  Ti &-7.100  &   V  &-8.110  &   Cr &-6.380\\
    Mn &-6.580  &  Fe &-4.480  &   Co &-7.070  &   Ni &-5.800\\
    Cu &-7.820  &  Zn &-7.440  &  &  &  &\\
\hline
 \multicolumn{4}{l}{Metallicity : $1.0017 \times$Solar} &  \multicolumn{4}{l}{ $n_{\rm ion}/n_h$ = 1.0989} \\
 \multicolumn{3}{l}{Neutral $\mu$ : 1.2916 } &  \multicolumn{4}{l}{ Fully Ionized $\mu$ :  0.61622 }\\
  \multicolumn{2}{l}{Mass Fractions }  &\multicolumn{2}{l}{ X:0.71015}& \multicolumn{2}{l}{  Y:0.27559 }&  \multicolumn{2}{l}{ Z:1.42606E-02}\\
\hline
\end{tabular}
\end{center}
\label{t:abund}
\end{table*}%

The ram pressure, $P_{\rm ram}$ may be usefully scaled to that of a shock with velocity of 100\,km/s moving into an ISM with pre-shock density $n_{\rm H} = 1.0$. Defining a ram pressure variable, $R = (n_{\rm H}/1.0 {\rm ~cm}^{-3})\times (v_s/{\rm ~km s}^{-1})^2$, three pressure ranges were computed with $R = 10^4, 10^6$ and $10^8$, corresponding to  $n_{\rm H} = 1, 100$, and $10^4$ at 100\,km/s. This allows us to explore the effect of density collisional de-excitation on the radiation field of the shock, and the corresponding effect on pre-ionisation. Finally, to adequately explore the effect of magnetic field, five sets of models with ranging from to zero, weak and moderate, to very strong magnetic fields, were computed.

We define a magnetic to ram pressure ratio:
\begin{eqnarray}
\eta_M &= &2P_B/P_{\rm ram} \;, \\
      & = & B^2/(4\pi \rho v_s^2)  \;,\\
      & = & 1/{\cal M}_{\rm A}^2  \;,.
\end{eqnarray}
This is analogous to the gas pressure ratio  $\eta_G = \gamma P/P_{\rm ram} = 1/{\cal M}^2$. This quantity, $\eta_M$, is the inverse of the Alfv\'en Mach number squared, and indicates the relative strength of the pre-shock magnetic field for a given shock.  The benefit of using this form is that  $\eta_M$ can go to zero when the magnetic pressure considered becomes insignificant, avoiding the infinities of the more common Mach number form.

So, the shock grid consists of 15 cases, set by the parameters $\eta_M = 2P_B/P_{\rm ram}$, and ram pressure $R = n_H v_s^2$.   The grid of $\eta_M$, and $R$ are given in table \ref{t:shockgrid}, along with the associated values for:
$B_0$, the pre-shock magnetic field ($\mu$G), ${\cal M}_A$, the Alfv\`en Mach Number, $P_{\rm ram}$, ram pressure in c.g.s units, plus the pre-shock density $n_H$ (cm$^{-3}$) at a reference 100 km/s.
For each $\eta_M$, and $R$, a series of models form a velocity grid from $1.0 < \log_{10} v_s \le 3.18$ in 0.02\,dex steps.  Of these, the 95 models from $\log_{10} v_s = 1.30$ ($\sim 20$km/s) to $\log_{10} v_s = 3.18$ ($\sim 1500$km/s) are listed in the tables that follow, being the models were the post-shock temperature exceed $10^4$K.  The models with $v_s$ less than 20km/s produce very little emission, and the exact values of the shock structural properties, such as pre-shock temperature are numerically uncertain, and temperatures become limited numerically by the code to 10K, and so are not as reliable as the faster shock models.

\begin{table*}[ht]
\caption{Shock Model Grid Parameters}
\label{t:shockgrid}
\begin{center}
\begin{tabular}{*{7}{r}}
\multicolumn{1}{c}{Model} &                              &                         &                           &                                  &                                   & \multicolumn{1}{c}{{\small $n_H$}} \\
\multicolumn{1}{c}{Series} & \multicolumn{1}{c}{$\eta_M$} & \multicolumn{1}{c}{$R$} & \multicolumn{1}{c}{$B_0$} & \multicolumn{1}{c}{${\cal M}_A$} & \multicolumn{1}{c}{$P_{\rm ram}$} & \multicolumn{1}{c}{{\small @100km/s}} \\
\hline
\hline
 1 & 0.0000 & 1.00+04 & 0.00+00 & $\infty$ & 2.357-10 & 1.0+00\\
 2 & 0.0000 & 1.00+06 & 0.00+00 & $\infty$ & 2.357-08 & 1.0+02\\
 3 & 0.0000 & 1.00+08 & 0.00+00 & $\infty$ & 2.357-06 & 1.0+04\\
 4 & 0.0001 & 1.00+04 & 5.442-01 & 1.000+02 & 2.357-10 & 1.0+00\\
 5 & 0.0001 & 1.00+06 & 5.442+00 & 1.000+02 & 2.357-08 & 1.0+02\\
 6 & 0.0001 & 1.00+08 & 5.442+01 & 1.000+02 & 2.357-06 & 1.0+04\\
 7 & 0.0010 & 1.00+04 & 1.721+00 & 3.162+01 & 2.357-10 & 1.0+00\\
 8 & 0.0010 & 1.00+06 & 1.721+01 & 3.162+01 & 2.357-08 & 1.0+02\\
 9 & 0.0010 & 1.00+08 & 1.721+02 & 3.162+01 & 2.357-06 & 1.0+04\\
10 & 0.0100 & 1.00+04 & 5.442+00 & 1.000+01 & 2.357-10 & 1.0+00\\
11 & 0.0100 & 1.00+06 & 5.442+01 & 1.000+01 & 2.357-08 & 1.0+02\\
12 & 0.0100 & 1.00+08 & 5.442+02 & 1.000+01 & 2.357-06 & 1.0+04\\
13 & 0.1000 & 1.00+04 & 1.721+01 & 3.162+00 & 2.357-10 & 1.0+00\\
14 & 0.1000 & 1.00+06 & 1.721+02 & 3.162+00 & 2.357-08 & 1.0+02\\
15 & 0.1000 & 1.00+08 & 1.721+03 & 3.162+00 & 2.357-06 & 1.0+04\\
\hline
\multicolumn{7}{l}{\small Notation: $x.xxx\pm yy$ represents $x.xxx\times 10^{\pm yy}$}
\end{tabular}
\end{center}
\label{default}
\end{table*}%

In cases where the density reaches high values (for weak or no magnetic fields and for large $R$), collisional de-excitation modifies the emission, to a variable extent depending on the energy interval or line being considered, see below.  When $B$ is weak,  $\eta_M = 0.001$, the magnetic parameter contributes little to the post-shock pressure. The post shock gas still compresses efficiently and the majority of the energy density is available for re-radiation.  As $\eta_M$ increases to $\eta_M = 0.01$, moderate $B$, and $\eta_M = 0.10$ strong B, the energy available for emission is reduced and post--shock densities are likewise limited by magnetic pressure.  Thus, the models were divided into $\eta_M$ groups, $0-0.001$, $0.01$, and $0.10$, denoted Standard, Moderate and Strong magnetic cases, each containing all the $R$ values for that $\eta_M$ range as a group.

\section{The Precursor Parameter}\label{sec:precursor_param}

The dimensionless ionisation parameter ${\cal U}$ is defined as the ratio of the number density of ionising photons (above the ionisation potential of hydrogen $\sim 13.598eV$) to the hydrogen atom number density. This can be also written as ${\cal U} = {\cal Q}/c$, where ${\cal Q}$ is the ionisation parameter in terms of the ratio of ionising photons flux (cm$^{-2}$s$^{-1}$) to the hydrogen number density (cm$^3$).  To first order  ${\cal Q}$ can be thought of as the velocity of an ionisation front that would be driven into a neutral medium by an ionising radiation field. We can compare this ionisation parameter in the upstream radiation field generated by the shock to the  shock velocity $v_s$ to form the \emph{shock--precursor} parameter first introduced by \citet{Shull79}:
\begin{equation}
\Psi = {\cal Q}/v_s =  c\,{\cal U}/v_s\, .
\end{equation}

When $\Psi < 1$, the shock velocity is greater than the velocity of the precursor ionisation front.  The photons are trapped by the flux of incoming neutral particles, and there is only a finite time for ionisation and recombination between the moment when the particles enter the precursor region and when they enter the shock. This is the case where the multi-zone model is required to capture the entire flow structure. The optical depth gradients limit the maximum extent of the precursor zone. As explained in Section \ref{sec:intro}, $\Psi$ approximates to the fractional ionisation of the gas entering the shock. This approximation is remarkably good, as can be seen from Figure \ref{fig:shockparms}, below.

When $\Psi > 1$, the precursor is faster than the shock and the photons can establish an extended ionized region ahead of the shock that will grow until the ionisations are balanced.  In this case the inner regions can become very transparent and large parts of the structure are in high intensity fields, with the optical depth changing rapidly only at the outer edge of the partially ionized precursor.  Within the extensive optically-thin pre-ionized region, recombinations set the optical depth gradient, and the structure of the photoionized precursor approaches that of an equilibrium \HII\ region.

For simplicity, consider a pure hydrogen plasma.  Within a column, $\Delta r$ , the number of recombinations equals $F$, the absorbed flux of ionising in photons/cm$^2$/s:
\begin{equation}
n_H^2 \alpha(T)  \Delta r  = F\, ,
\end{equation}
 where $\alpha(T)$ is the recombination coefficient, and $n_H$ is the hydrogen number density.

Now, from its definition,  ${\cal Q} = F/n_H$, so ${\cal Q} = n_H \alpha(T)\Delta r$ and $\Delta r = {\cal Q}/[n_H \alpha(T)]$.  The hydrogen recombination timescale in the plasma is $\tau_r = 1/[n_H \alpha(T)]$,  and the time required for a particle to cross this region is  $t = \Delta r/v_s = \tau_r {\cal Q}/v_s   = \tau_r \Psi$.  These relations can be used to pre-estimate the approximate size of the multi-zone precursor structure when $\Psi > 1$, adopting the conventional estimate of $T \sim 10^4$\,K.

It is important to note that the multi--zone relaxation iterative method described here can continuously model the transition from an optical depth limited trapped precursor to an extended  precursor in which region near the shock are optically thin and approach photoionisation equilibrium.  The approximate single zone methods used earlier cannot properly model the cases where $\Psi$ approaches and exceeds 1, although they may be adequate for the case $\Psi <<1$. From the work of \citet{Shull79, Binette85, Dopita96} and \citet{Allen08} when $\Psi > 1$  ($v_s > 120-150$\,km/s), the structure of the extended precursor can be approximated as an isochoric (constant density) equilibrium photoionisation region.  While its extended nature is clear, what is less clear is whether equilibrium conditions really apply. This can only be verified easily using a full time-dependent calculation. As will be shown below, not only do the fast auto-ionising shocks with $\sim150$~km/s and above have extended ionized precursors, but time-dependent multi-zone models confirm that the plasma entering the shock front approaches photoionisation equilibrium to a high degree of accuracy. Thus, we show that not only is the equilibrium precursor treatment for these shocks valid, but indeed is in some respects preferable to the more complex time--dependent approach adopted here.

\section{Velocity Sequence, Shock Types} \label{sec:shock_types}

\begin{figure*}
  \includegraphics[width=\textwidth]{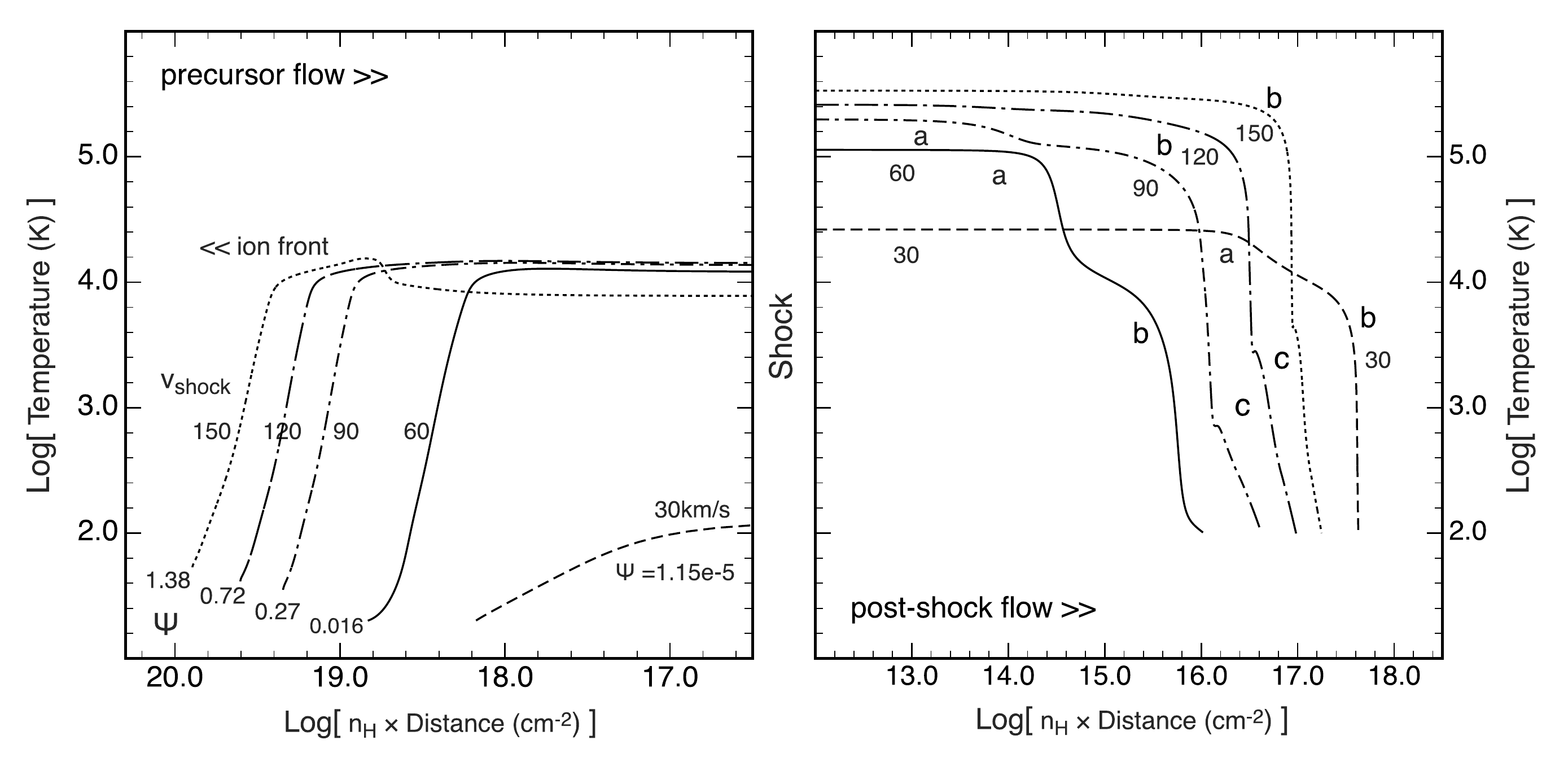}
  \caption{The pre- and post-shock parameters as a function of shock velocity. The left hand panel gives the pre-shock temperature, $T_{\mathrm{pre}}$, the precursor parameter, $\Psi$, and the fractions of ionised and neutral hydrogen ($F_{\mathrm {HII}}$ and $F_{\mathrm {HI}}$, respectively). Note how well $\Psi$ tracks $F_{\mathrm HII}$ in the regime $\Psi < 1.0$. The right hand panel gives the post-shock temperature, $T_s$, the Mach number of the shock, $\cal M$, and the mean atomic weight of the gas entering the shock, $\mu$. The deviations of the post-shock temperature with velocity from the slope = 2.0 (shown as a dashed line) are driven by changes in the molecular weight of the pre-shock plasma. The quantities with the subscript EQ show what would be obtained in the precursor if we had assumed a one-zone photoionisation equilibrium pre-ionisation, rather than solving for the full pre-shock ionisation and heating balance.
  } \label{fig:shockparms}
 \end{figure*}

\begin{figure}
  \includegraphics[width=\columnwidth]{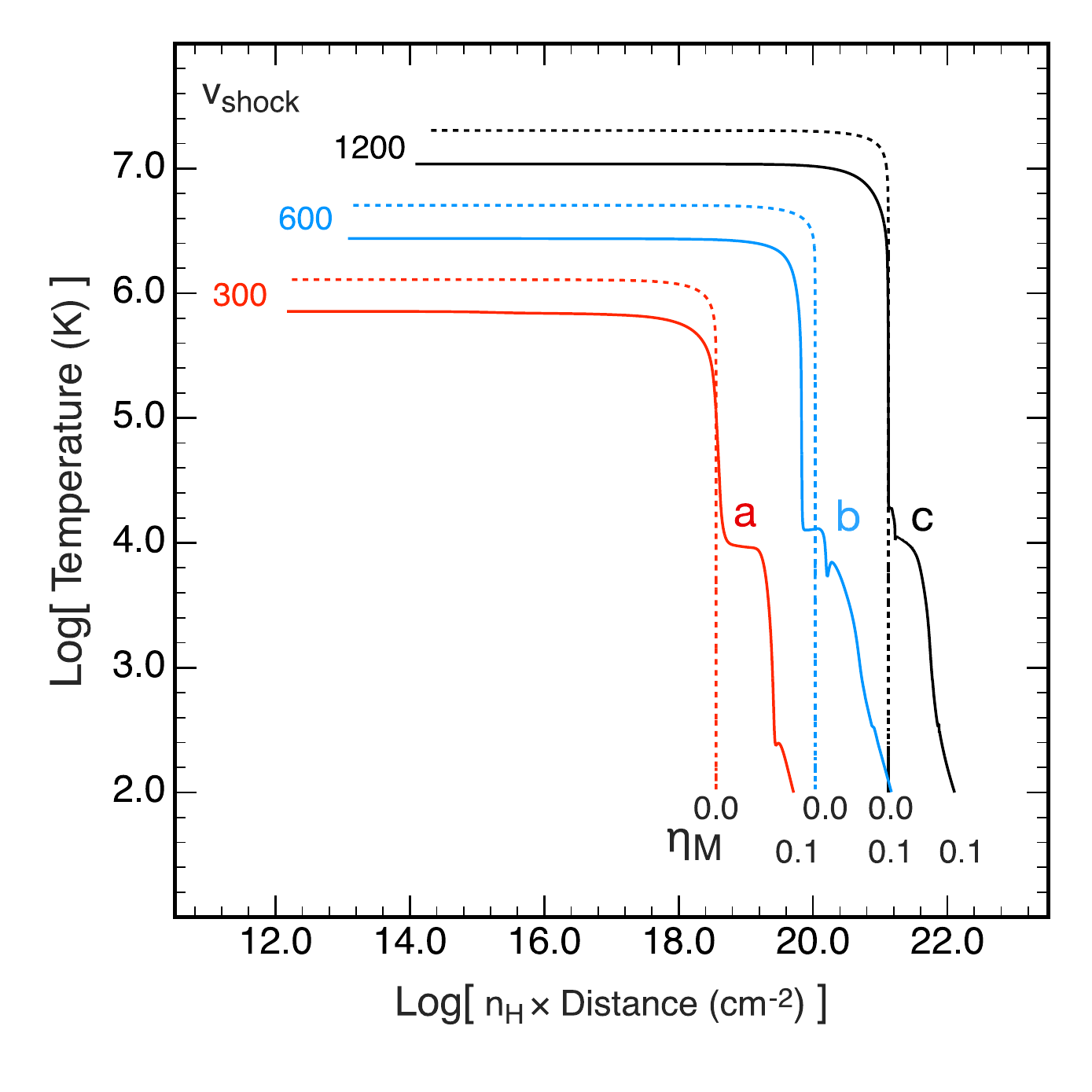}
  \caption{The temperature structure of (left) the precursor, and (right) the shock, for four characteristic shock velocities. In the 30\,km/s case, the pre-shock gas is still effectively neutral and cool, and the post-shock gas has a large cooling column, due to the scarcity of electrons. In the 60\,km/s case, the pre-shock plasma is warm but fairly un-ionised. At 90\,km/s the pre-shock gas is starting to become ionised, and by 120\,km/s the pre-ionisation is approaching unity. Finally, the 150\,km/s model is into the regime of fast shocks with equilibrium pre-ionisation. Note the lower temperature of the gas entering the shock in this case. These temperature and ionisation states are reflected in the post-shock temperature structure, where we define three shock zones; (a) where the gas is relaxing towards collisional ionisation equilibrium with strong cooling due to hydrogen. (b) the cooling zone of the shock, and (c) the recombination zone of the shock in which photoionisation effects are important.} \label{fig:shock-Te}
 \end{figure}

\begin{figure*}
  \includegraphics[width=\textwidth]{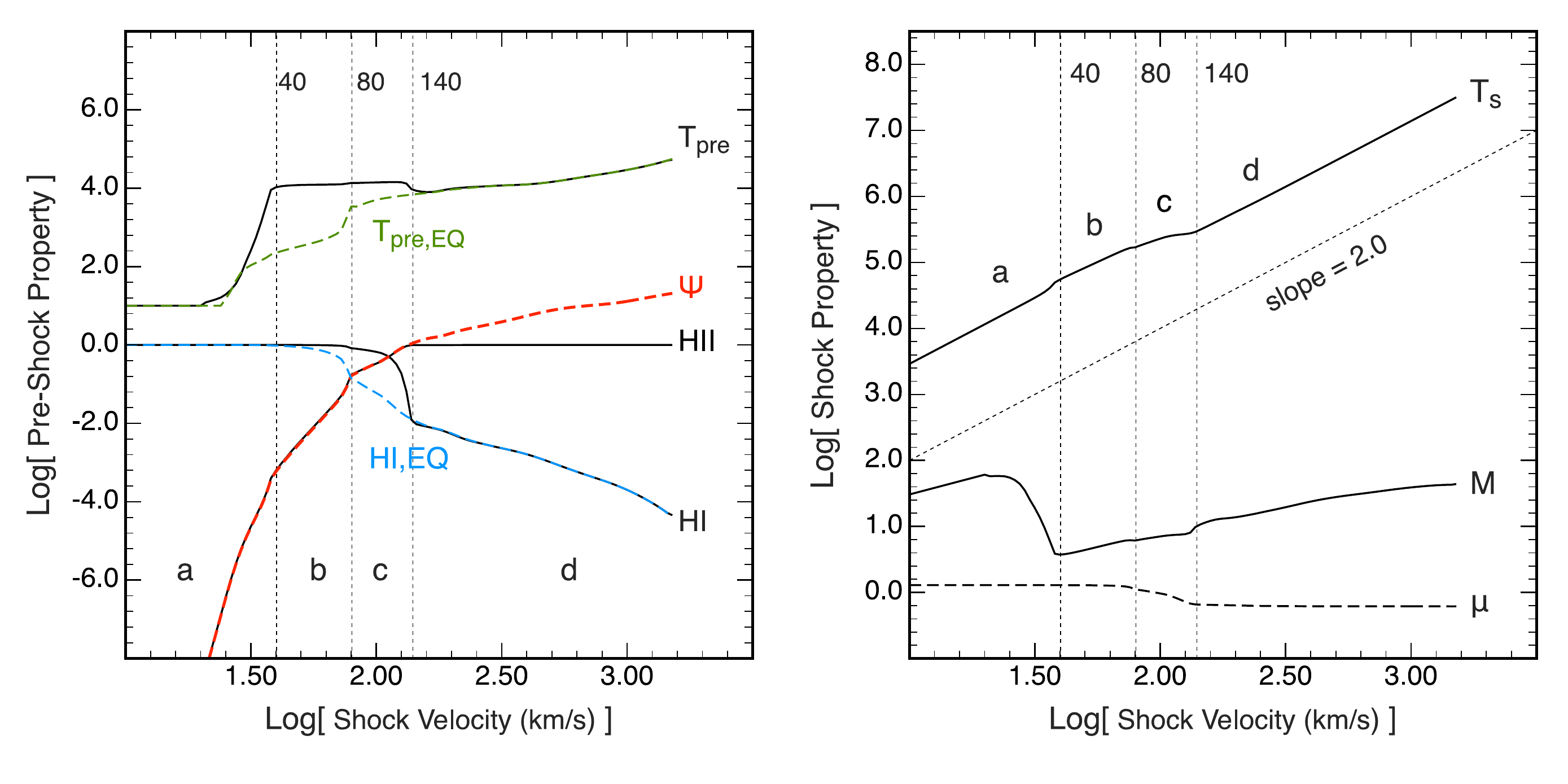}
  \caption{The temperature structure of fast $v>>100$\,km/s. As the post-shock temperature increases to over $10^7$\,K, there are further changes to the shock structures. The most notable being the development of a more complex downstream photo-heated zone (a, b and c), which can become increasingly dominated by x-ray heating above 600\,km/s, changing the single temperature structure at a, to a multiple zone structure at b and c.  The zero magnetic field models also show temperature structure in this region, but on this column density coordinate plot the extend of the region is very compressed. The vertical dashed lines at 40, 80, and 140\,km/s are marked here and in subsequent figures, to approximately divide the shock regimes discussed in the text. In later figures relating to higher energy emission, a 600\,km/s points is also marked in section \ref{sec:radiation}, which divides the high velocity fully auto-ionized shocks into two further domains; with and without significant x-ray heating.}  \label{fig:fastshock-Te}
 \end{figure*}

For shocks with $\Psi < 1$, where the ionisation front is trapped close to the shock front, we have identified three general classes of shocks.  When $\Psi > 1$ we enter into the fast shock regime, where an HII region in photoionisation equilibrium is maintained in the pre-shock region. The transition between $\Psi < 1$ and $\Psi > 1$ occurs at about 120\,km/s for shocks with low magnetic parameter, and increasing magnetic parameter drives the transition to higher velocities.  Referring to Figures \ref{fig:shockparms}, \ref{fig:shock-Te} and \ref{fig:fastshock-Te}, we now examine in more detail the structure of these classes of shock.

\subsection{Cold Neutral Precursors}
In very low velocity shocks ($v_s < 40$\,km/s), region (a) in Figure \ref{fig:shockparms}, moving into an un--ionized medium, the upstream precursor radiation field is both too weak (low ${\cal Q}$) and too soft to either ionize or heat the gas entering the shock front. The opacities in the precursor region remain nearly constant, and rapid convergence is achieved in the iterative solution of the precursor structure.

An example of the structure at 30\,km/s is shown in Figure \ref{fig:shock-Te}. Here, the ionized fraction of HI remains at $\sim1.0\times 10^{-5}$ and the electron temperature is unable to rise much above 100\,K. In passing through the shock, the electron temperature jumps from $\sim 100$\,K to 26,000\,K. Inefficient cooling, due mainly to the lack of electrons, gives rise to a relatively long overall cooling length. This is dominated initially by  fine structure cooling from neutral species plus Lyman alpha accompanied by collisionally enhanced two photon cooling (zone a in the figure). In zone b the cooling is dominated by singly ionised species until the temperature falls below 1,000\,K.

\subsection{Warm Neutral Precursors}
In shocks in the velocity range ($40 \lesssim v_s \lesssim 80$\,km/s), region (b) in Figure \ref{fig:shockparms}, the precursor radiation field contains enough high energy photons, albeit at a very low ionisation parameter, to produce some energetic electrons which can heat the gas significantly.  While the total ionisation remains low, the heating per-ionisation rises, and since this is not balanced by efficient electron collisional cooling, the electron temperature can rise over $10^4$ K. This preheating affects the pre-shock pressure and internal energy flux into the shock, altering the solutions by lowering the effective Mach number of the shock. Since in these shocks the ionisation state is low and hence the HI opacity remains high and nearly constant, the time-dependent heating and ionisation solution is well behaved, and converges rapidly (typically in less than 10 iterations).

An example of the structure of such a shock with  $v_s = 60$\,km/s is shown in Figure \ref{fig:shock-Te}.  Preheating of the precursor is strong, but without strong ionisation, the largely neutral precursor has an optical column of a few $\times10^{18}$ cm$^{-2}$. Heating of the gas is efficient and reaches more than 10,000K, reducing the Mach number of the shock.  The initial cooling from over 100,000K is very strong, and is followed by a second region of cooling by more ionized species.  Internal photoionisation plays little or no role in the structure of the cooling zone.

\subsection{Warm Partially Ionized Precursors}

When the shock velocity rises above 80\,km/s, $0 < \Psi <1.0$, region (c) in Figure \ref{fig:shockparms}, and the fractional ionisation of hydrogen in the precursor plasma closely tracks the value of $ \Psi $. The  pre-shock ionisation fraction is sufficient to modify the shock structure, and the plasma is both less ionized and appreciably hotter when entering the shock than would be the case if the gas had time to come into equilibrium with the precursor field.  This ensures that $\mu$ is somewhat higher than an equilibrium model would predict. Consequently the post-shock temperatures are higher than would be found in an equilibrium model, or in the case that the precursor is kept fully ionized by an external source of EUV photons.

\subsection{Fully ionized Precursors, Fast Shocks}
Once $\Psi$ rises above 1.0, for shock velocities above 140\,km/s,  region (d) in Figure \ref{fig:shockparms}, hydrogen is effectively fully pre-ionised. For faster shocks, additional changes in the mean molecular weight of the gas entering the shock occur as helium becomes first singly, and later doubly ionized. These changes result in small changes in the pre-shock temperature. By $\sim 200$\,km/s, the pre-shock hydrogen is also effectively fully ionized and we have entered the fast shock regime modelled by \citet{Dopita95,Allen08}, in which the pre-ionisation problem can be treated as one of equilibrium photoionisation.

While the precursors remain similar for the fast shocks, changes occur in the photoionised tails of the fast shock, which is effectively a lower ionisation parameter version of the precursor, as shown in Figure \ref{fig:fastshock-Te}.  Region c is similar to Figure \ref{fig:shock-Te}, but above 600\,km/s the temperature structure becomes more complex as soft x-ray heating becomes important to the thermal balance. Thus, 600\,km/s represents a division in the fast shocks where x-rays generated by the cooling post-shock gas come to significantly influence the ionisation state of the recombination tails and to a certain extent the precursors.

These four regimes are well-illustrated in Figure \ref{fig:shockparms} in which we plot pre-shock and the post-shock variables as a function of shock velocity (the magnetic field is set to zero). Within each shock regime, changes in the post-shock temperature, molecular weight and Mach number are driven by the pre-shock temperature and the mean atomic weight $\mu$ of the gas entering the shock.

\section{Effect of Magnetic Fields}\label{magfield}

 \begin{figure*}
  \includegraphics[width=\textwidth]{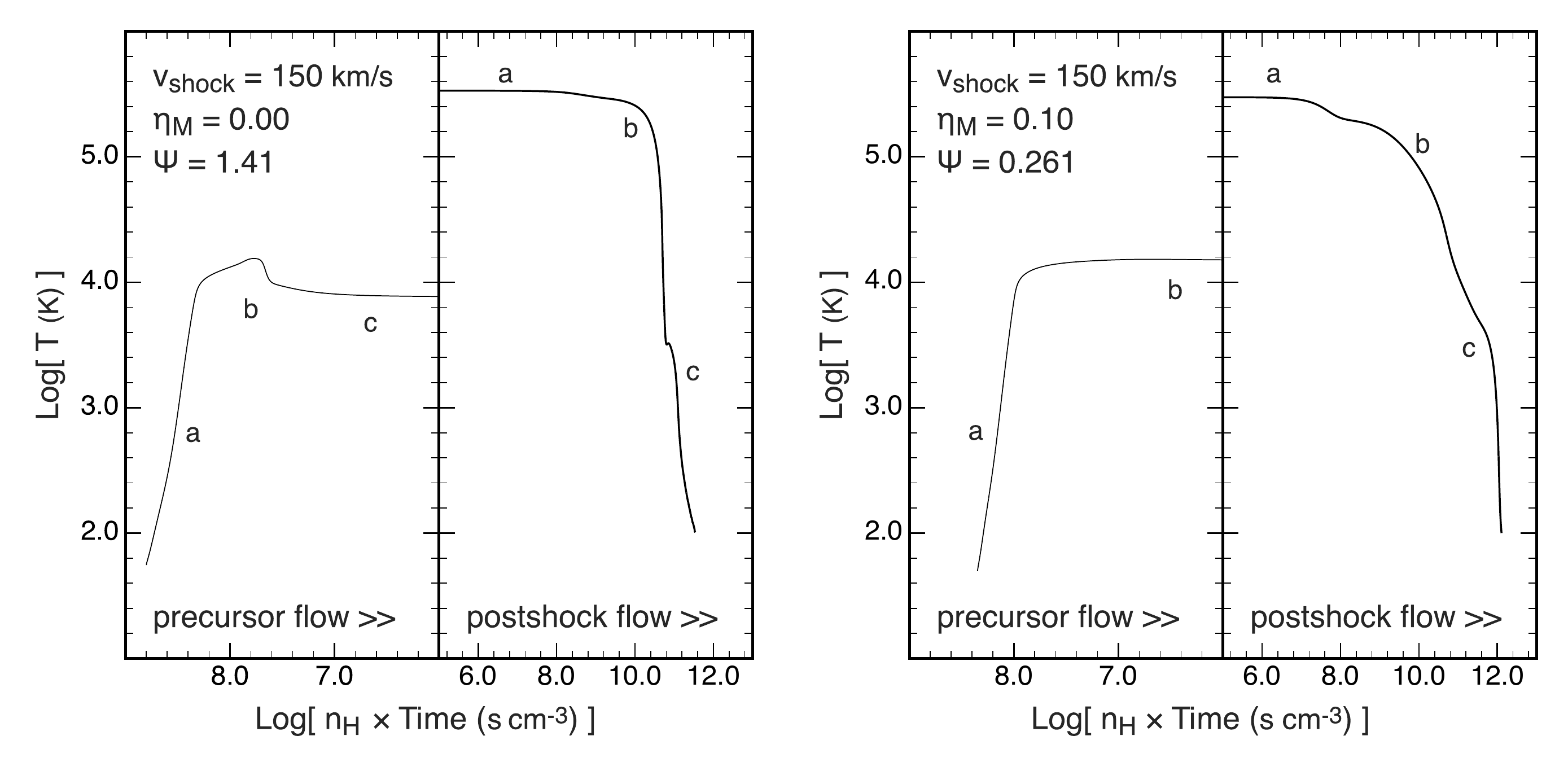}
  \caption{The effect of the magnetic field on the pre-shock (left) and post-shock (right) structure for a 150\,km/s shock. The magnetic field reduces the radiative efficiency of the shock, in this case driving $\Psi$ to well below unity. The profile now resembles the 60-90\,km/s profile in Figure \ref{fig:shockparms}, consistent with a $\Psi \approx 0.27$ value, showing the importance of the pre-ionsiation governed by $\Psi$. The effect of magnetic pressure support in the post-shock gas is to both lengthen the cooling zone of the shock (b) and to much extend the recombination zone of the shock (c).} \label{fig:mag-psi}
 \end{figure*}

The shock converts a kinetic energy flux largely into internal thermal and magnetic energy in the post shock region. The thermal component of this energy is available to be re-radiated upstream, unlike the magnetic post-shock energy. Thus, for the same total pressure, as the relative fraction of the internal energy of the shocks in magnetic pressure increases, the post shock temperature drops, and consequently higher velocities are needed to preheat and pre-ionise the precursor region.

We define a magnetic to ram pressure ratio $\eta_M = 2P_B/P_{\mathrm{ram}} = B^2/(4\pi \rho v_s^2) = 1/{\cal M}_{\mathrm A}^2$, analogous to the gas pressure ratio  $\eta_G = \gamma P/P_{\mathrm{ram}} = 1/{\cal M}^2$. This quantity, $\eta_M$, is the inverse of the Alfv\'en Mach number squared, and indicates the relative strength of the pre-shock magnetic field for a given shock.  The benefit of using this form is that  $\eta_M$ can go to zero when the magnetic pressure considered becomes insignificant, avoiding the infinities of the more common Mach number form.

In figure \ref{fig:mag-psi}, we show the shock and precursor temperature structure for the  $\eta_M = 0$ ($B= 0$) case , and for the  $\eta_M = 0.1$  case (corresponding to ${\cal M}_{\mathrm A} =\infty$ and $\sqrt 10$). Note that the recombination zone (c) is appreciably thicker in the high magnetic field case due to the magnetic field both limiting the compression factor, and reducing the cooling rate, while the decrease in the production of ionising photons in the cooling zone of the shock causes $\Psi$ in the precursor region to drop to well below below unity in the case of high magnetic field.

The structure of the precursor is profoundly modified, see left hand panel of figure \ref{fig:mag-psi}. In the case of zero magnetic field we have an initial phase of heating of the pre-shock gas (a), followed by an increase in ionisation and superheating of the plasma (b). Finally, ionisation and temperature reach equilibrium at a lower temperature with enhanced cooling in the presence of abundant electrons (c). In the high magnetic field case, the initial heating occurs, but phase (b) is only partial.  The overall structure here resembles a 60--90\,km/s shock with $\eta_M = 0$; compare, for example, with figure \ref{fig:shock-Te}.

\section{Compression Factors}\label{compress}
In this, and in the following Sections, we present some useful scaling relationships for radiative shocks covering the full velocity range liable to be encountered in real astrophysical objects. These parameterised fits enable the relatively straightforward computation of physical properties of radiative shocks which may be compared with observations.

\subsection{Compression at the Shock}

\begin{figure*}
  \includegraphics[width=\textwidth]{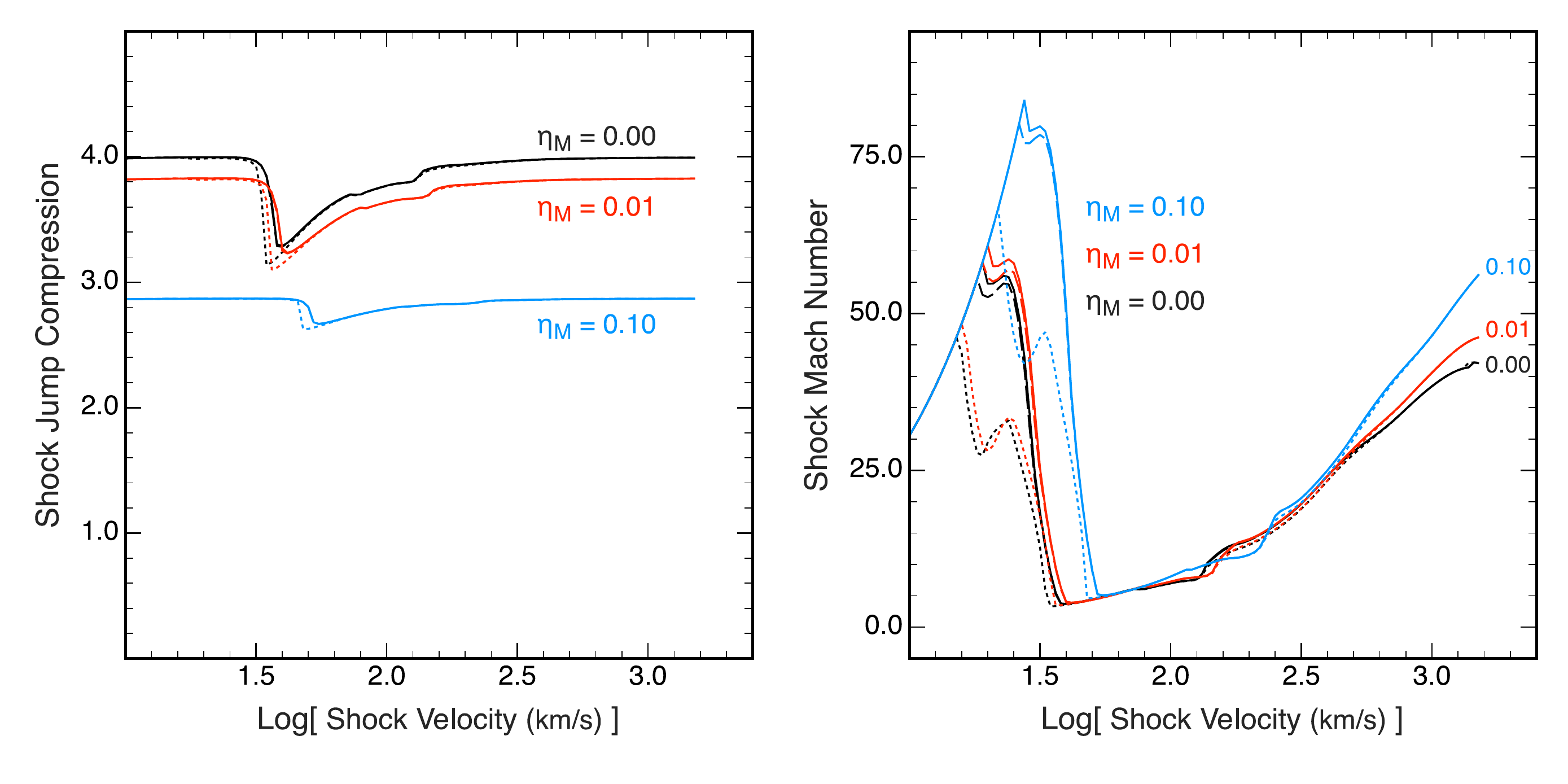}
  \caption{(Left) shock jump compression factor, and (right) the Mach number of the shock as a function of shock velocity and $\eta_M = 2P_B/P_{\mathrm{ram}}$.  Solid lines,  $n = 1.0$ and $n = 100$\,cm$^{-3}$ at 100\,km/s, dotted lines $n_{\mathrm H} = 10^4$\,cm$^{-3}$ at 100\,km/s. Black corresponds to the $\eta_{\cal M} = 0$ ($B= 0$) case , red the  $\eta_{\cal M} = 0.01$ case, and blue the  0.1 $\eta_{\cal M} = 0.01$ case. These correspond to ${\cal M}_{\mathrm A} =\infty, 10$ and $\sqrt 10$). Note the sharp drop in both Mach number and compression factor as the pre-shock gas is heated to $T_e \sim 10^4$\,K. This transition is suppressed to higher velocity in the case of high magnetic field strength. High density suppresses OI and NI cooling in the precursor, allowing pre-shock heating of the gas at lower velocity.} \label{fig:comp-mach}
 \end{figure*}

 For strong shocks in monoatomic gases without magnetic fields, the compression factor across the shock is 4.0. However, in these self-consistent shocks with pre-ionisation and heating, the compression factor across the shock may differ appreciably from this figure. This is shown in Figure \ref{fig:comp-mach}.

Taking the zero magnetic case first ($\eta_M = 0.0$). At very low velocity, the cool un-ionised gas in the precursor suffers a large Mach number shock, so the compression factor is very close to 4.0, that is $(\gamma+1)/(\gamma-1)$, for $\gamma = 5/3$. When the pre-shock gas is heated, the Mach number falls steeply, and the compression factor is reduced. Note that, at high density, the cooling in the pre-shock gas is suppressed by collisional de-excitation of, in particular OI and NI, and the transition to a heated precursor occurs at lower velocity. Above $\log v_s \sim 1.5$, both the Mach number and the shock compression factor rise monotonically back to the strong shock limit. The fluctuations in this progression are caused by changes in the mean molecular weight of the gas entering the shock as a consequence of pre-ionisation. At higher values of the magnetic field, the behaviour is qualitatively similar, but the limiting compression factors are correspondingly lower, and the effect of pre-heating in the precursor is less marked.

\subsection{Compression Through the Shock}

 \begin{figure*}
  \includegraphics[width=\textwidth]{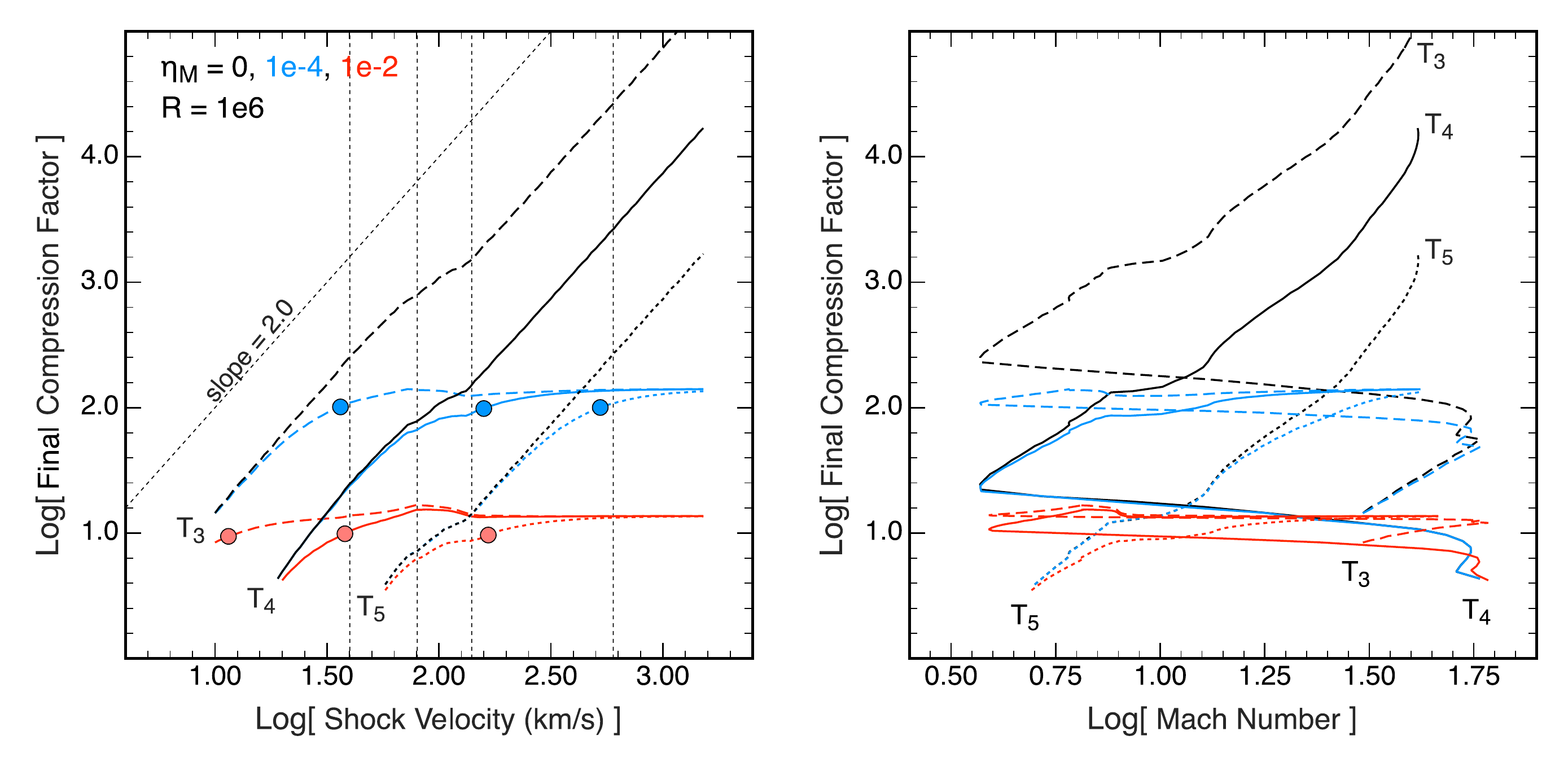}
 \caption{Final compression factors, cooling to temperatures of $10^5$\,K (T5), $10^4$\,K (T4), and $10^3$\,K (T3) as a function of shock velocity (left) and shock Mach number (right), for zero, weak  and stronger magnetic field strength; $\eta_M = 0$ (black), $\eta_M = 1e-4$ (blue) and $\eta_M = 0.01$ (red). The densities are moderate with $R = 10^6$ (preshock $n_H = 100$\,cm$^{-3}$ at 100\,km/s).
  Left Panel: Apart from small variations driven by the pre-ionisation physics, the compression for the non-magnetic case is proportional to $v_s^2$, as the post shock temperature scales as $v_s^2$ and hence the ratio to a fixed final temperature follows.  The effect of magnetic field pressure support in the cooling region of the shock is more marked. The circles mark points where the post-shock magnetic pressure equals the post-shock gas pressure, $\alpha_s = 1.0$.  At this point the magnetic support has begun to significantly reduce the compression factors, and at higher velocities the final densities are more or less constant.
 Right Panel: The compression factors are multi-valued with respect to Mach number due to pre-ionisation physics; see Figure \ref{fig:comp-mach}, and with defined final temperatures, there is no simple Mach number -- Compression relationship.  The vertical lines mark the key shock type velocities of 40, 80, 140 and 600\,km/s for comparison with earlier figures.} \label{fig:final compress}
 \end{figure*}

 \begin{figure*}
  \includegraphics[width=\textwidth]{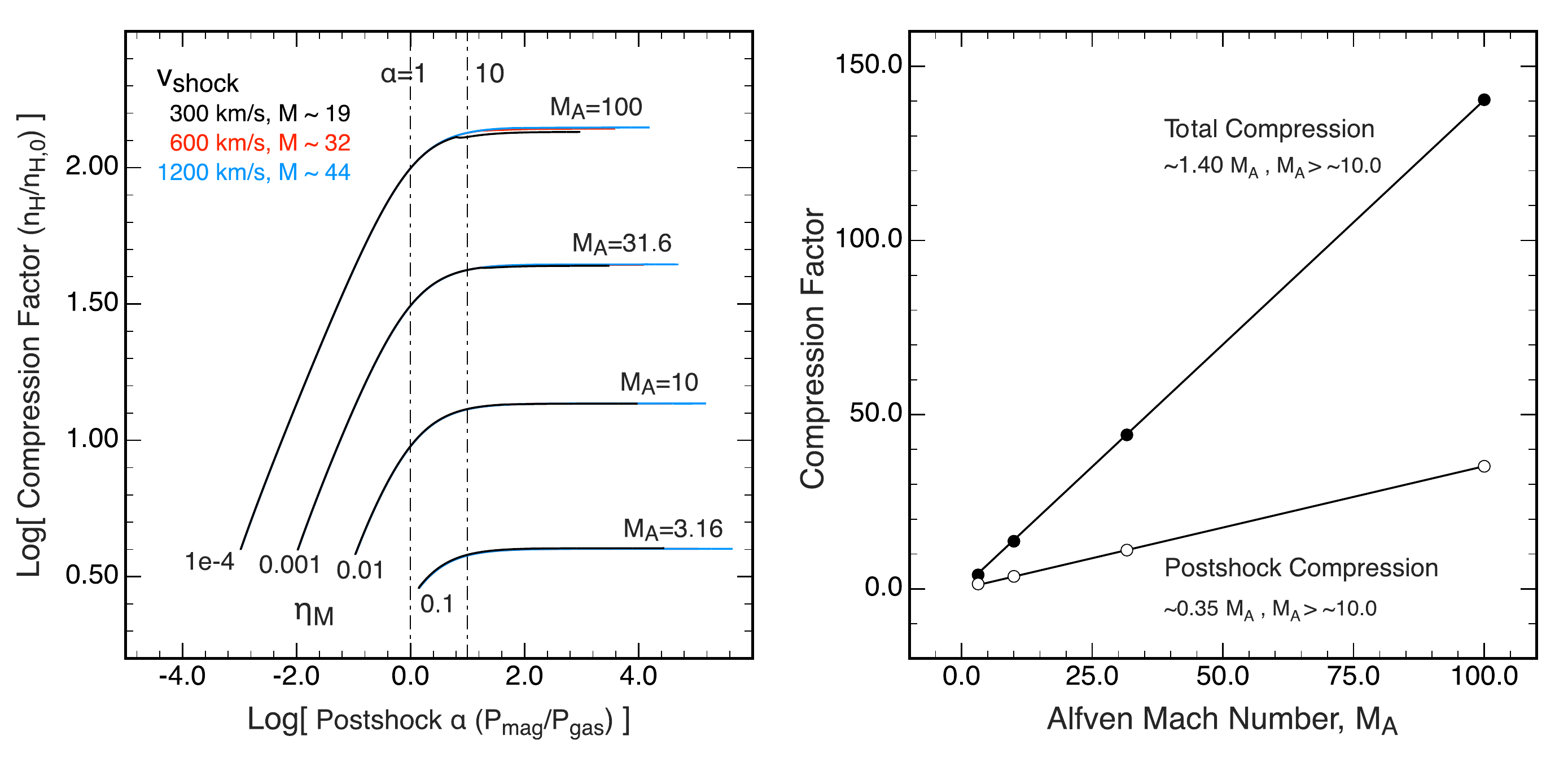}
  \caption{The left-hand panel shows compression factors as a function of $\alpha = P_B / P_{\rm gas}$ for different values of initial magnetic field strengths $\eta_M$. The corresponding Alfv\`en Mach numbers are marked. Note that these relations are effectively independent of shock velocity. On the right-hand panel are shown simple linear fits to the limiting compression factor as a function of Alfv\`en Mach number, with respect to the shock jump compression and the total compression through the shock, recovering the approximation of $\sqrt 2{\cal M}_A$ of equation \ref{eq:comp}.} \label{fig:Comp-02}
 \end{figure*}

Very frequently, the final compression factor through the shock is estimated from a strong isothermal approximation in which the post-shock gas returns to the pre-shock temperature, and the initial and final temperatures  are constant at all velocities. In this case, for an isothermal shock without magnetic field, the shock Mach number is proportional to the shock velocity, and  it is trivial to show that the final compression factor varies as ${\cal M}^2$.

A magnetic field provides additional pressure support, which strongly limits the compression factors that can be achieved. If the magnetic field is frozen in the flow (equation \ref{eq:frozen}), then ignoring the gas pressure in the post-shock gas relative to the magnetic pressure, we can equate the pre-shock ram pressure to the post-shock transverse magnetic pressure to obtain the maximum compression factor in the shock in terms of its Alfv\`en Mach number:
\begin{equation}
\frac{\rho _{1}}{\rho _{0}}=2^{1/2}{\cal M}_{\rm A}.  \label{eq:comp}
\end{equation}
Extending the strong isothermal approximation to an arbitrary ratio of magnetic pressure to gas pressure in the pre-shock gas, \citet{Draine93} obtained:
\begin{equation}
\frac{\rho _{1}}{\rho _{0}}=4\left\{ 2{\cal M}^{-2}+{\cal M}
_{\rm A}^{-2}+\left[ \left( 2{\cal M}^{-2}+{\cal M}_{\rm A}^{-2}\right) ^{2}+8{\cal M
}_{\rm A}^{-2}\right] ^{1/2}\right\} ^{-1}. \label{comp_general}
\end{equation}

In our models this isothermal assumption is not made, and both the initial and final temperature are a function of the shock velocity. In addition, we impose no arbitrary limit to cooling, but note the increasing inaccuracy due to the absence of molecules in our models (which may become important below a few thousand degrees if dust conditions are favourable)  and the increasingly poor atomic data at low temperatures. The models are terminated at $T_e =100$\,K  so that any realistic temperatures are bounded from below. We will now investigate whether any relationships between velocity or Mach number exist for these models.

We computed compression factors at specific post-shock temperatures, $10^5$\,K ($T_5$), $10^4$\,K ($T_4$), and $10^3$\,K ($T_3$).  $T_5$ is usually at a point between b and c in the post-shock structures, see Figure \ref{fig:shock-Te},  where the gas is still collisionally  excited and is cooling rapidly.  $T_4$ is found at the start of the shock tail photoionisation/recombination zone, c. The $T_3$ region is representative of the zone where only mild heating by hard photons persists and the gas is largely neutral.

Figure \ref{fig:final compress} gives these computed compression factors through the shock as a function of shock velocity and of shock Mach number for different magnetic field strengths.

In the left hand panel two things are apparent.  First, apart from small variations driven by the pre-ionisation microphysics, the compression in the non-magnetic case is proportional to $v_s^2$, the particular value being dependent on the final temperature chosen. This simply reflects the nearly isobaric nature of the cooling. Second, when magnetic fields are present, compression is limited, since as $P_{\rm gas}$ in the post shock flow vanishes, $P_B$ continues to support the flow.  The gas cools at constant density once $P_B > P_{\rm gas}$, indicated by the $\alpha_s = 1$ points in the figure.

In the right panel, the situation is more complex. Preheating of the gas in the precursor means the Mach number, which depends on the sound speed and hence temperature, is multi-valued.  Without a possible external effect such as an ambient radiation field, there is no simple relationship between (sonic) Mach number and compression factor.

Figure \ref{fig:Comp-02} shows the evolution of the compression factors as a function of $\alpha = P_B/P_{\rm gas}$ for different values of initial magnetic field strengths $\eta_M$, or equivalently, the Alfv\`en Mach numbers ${\cal M}_A = 1/\sqrt(\eta_M)$, for shocks with velocities $> 150$\,km/s where pre-ionisation effects are less {pronounced, and the post shock temperature is well above $10^5$K. In all but the strongest magnetic field $\eta_M$, the initial $\alpha \ll 1$. However, as the gas cools, the magnetic field eventually limits the compression factor.  This limit is essentially independent of shock velocity or Mach number, but is clearly a function of Alfv\`en Mach number. On the right-hand panel of Figure \ref{fig:Comp-02} are shown simple linear fits to the modelled limiting compression factor as a function of Alfv\`en Mach number, with respect to the shock jump compression and the total compression through the shock. The shock jump is fit with an relation ${\rho_1} / {\rho _0} = 0.35{\cal M}_A$, while the total compression is given by ${\rho_1} / {\rho _0} = 1.4{\cal M}_A$, as expected from equation \ref{eq:comp}.

\section{Post-shock Temperatures}\label{temp}
The post shock temperature is given by eqn \ref{Te_PS}. The general variation therefore is as the square of the shock velocity, but variations in the mean atomic weight and internal energy in the pre-shock gas can affect this relationship. The post-shock temperature as a function of shock velocity is shown in the right panel of figure \ref{fig:fastshock-Te}. It shows 4 sub regions corresponding to the regimes already discussed, and a fair approximation can be made with piecewise linear fits in log-log space,  with the post-shock temperature, $T_s$, as a function of shock velocity, $v_s$, can be fit by an equation of the form:
\begin{equation}
\log T_s = A  + B \log v_s.
\end{equation}
The fitting constants $A$ and $B$ can be fixed in velocity ranges which roughly correspond to the different shock zones defined above. In the velocity range 10 --  72.8 km/s the fitting constants are $A =  1.4049,  B = 2.0508$. In the velocity range 72.8 --  147.3 km/s the fitting constants are $A =  3.5538,  B = 0.8969$, and in the range 147.3 --  1500 km/s, $A =  1.2202$ and  $B = 1.9731$ provides a good fit.

\section{Cooling Lengths}

Corresponding to the final cooling temperatures, $T_5$, $T_4$,  $T_3$ etc, are cooling columns $n_H \Delta r$ (cm$^{-2}$) denoted as $n_H \Delta r = \lambda_n$ where $n = 3,4,5,6$ etc. For a given pre-shock density $n_H$ then a cooling length can be obtained by $\Delta r = \lambda_n/n_H$.

At a given shock velocity, provided that collisional de-excitation effects are negligible, the cooling timescale (and therefore the cooling length) of the post-shock plasma varies inversely as the density. Thus the product of the cooling length and the pre-shock density (which we will term the \emph{cooling column}, since it has dimensions of cm$^{-2}$ provides a quantity which, at the low density limit, is independent of density. This quantity is plotted as a function of shock velocity in Figure \ref{fig:coolinglength} for various magnetic field strengths and post-shock temperatures -  $10^6$\,K ($\lambda6$), $10^5$\,K ($\lambda5$), $10^4$\,K ($\lambda4$), and $10^3$\,K ($\lambda3$).  Figure \ref{fig:compstructure} shows the post-shock structure for a $\sim300$\,km/s shock, and in the middle panel, the cooling columns $\lambda_4$ and $\lambda_3$ to $T_4$ and $T_3$ are indicated.

Note that Figure \ref{fig:coolinglength} shows, provided neither the magnetic field nor pre-shock density are too high, or the post-shock temperature is not too low (all of which serve to  increase the cooling column), then for shocks with $v_s > 100$\,km/s, the cooling columns are very similar. Fitting to all the $\eta_M = 0$, and $\eta_M = 0.01$ models with  $R = 1.0\times10^4, 1.0\times 10^6$ and $1.0\times 10^8$, and for both the $\lambda4$ and $\lambda5$  cases, we can find a convenient global fit for the cooling column as a function of shock velocity with $x = \log(v_s)$, $v_s$ in km/s:
\begin{equation}
\log[\lambda(x)] \sim  \left[ P_0+x \left( P_1+x \right) \right]  / \left[  Q_0+xQ_1 \right] ,
\end{equation}
where
\begin{equation}
2 <  \log(v_s)  \lesssim 3.5 \; ,
\end{equation}
and the fitting parameters are:  $P_0 = -8.0282, P_1 = 6.6475, Q_0 = -0.263$ and $Q_1 = 0.4221$.

\begin{figure*}
  \includegraphics[width=\textwidth]{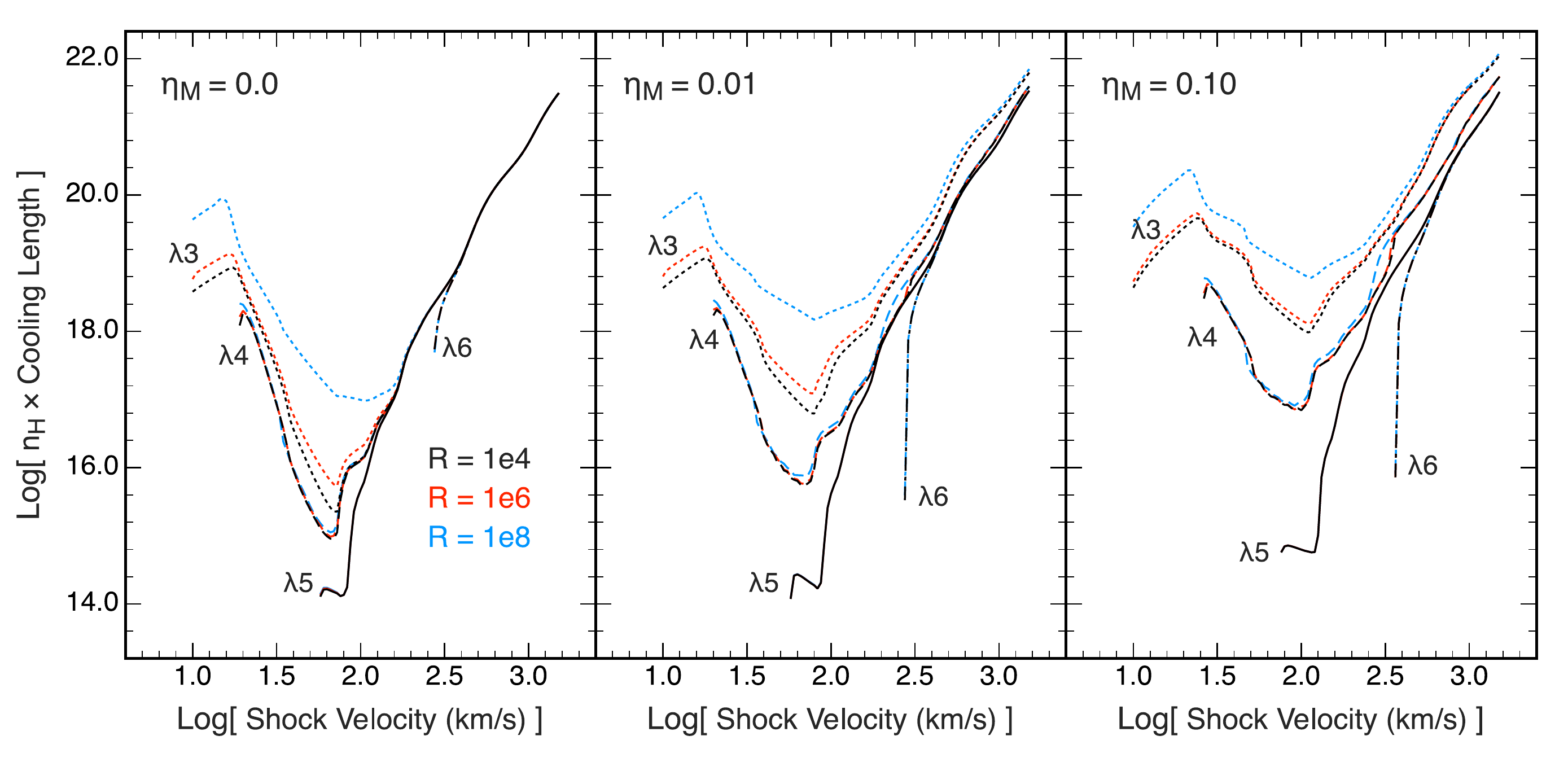}
   \caption{The cooling columns to final temperatures of $10^6$\,K ($\lambda_6$), $10^5$\,K ($\lambda_5$), $10^4$\,K ($\lambda_4$), and $10^3$\,K ($\lambda_3$) as a function of shock velocity, for different values of the magnetic field strength given by $\eta_M$. The cooling column is, to first order, independent of the pre-shock density. However, there are significant changes to $\lambda_3$ with $\eta_M$, extending $\lambda_3$ with respect to $\lambda_4$ as $\eta_M$ increases.   In addition, the high density models (blue) have greater $\lambda_3$ columns, as the effects of collisional de-excitation lower the cooling rates, and increase the cooling column for photo-heating tail regions. The cooling to $10^4$\,K, or higher, from the post-shock temperature is not as strongly effected by density or $\eta_M$ as the $\lambda_3$ values, in that each of the $\lambda_6$, $\lambda_5$ curves remain closely grouped within a density regime and across $\eta_M$ values, while $\lambda_4$ curves are mutually similar within each $\eta_M$ range, but do show some evolution between $\eta_M$ values.} \label{fig:coolinglength}
 \end{figure*}

 \begin{figure*}[ht!]
  \includegraphics[width=\textwidth]{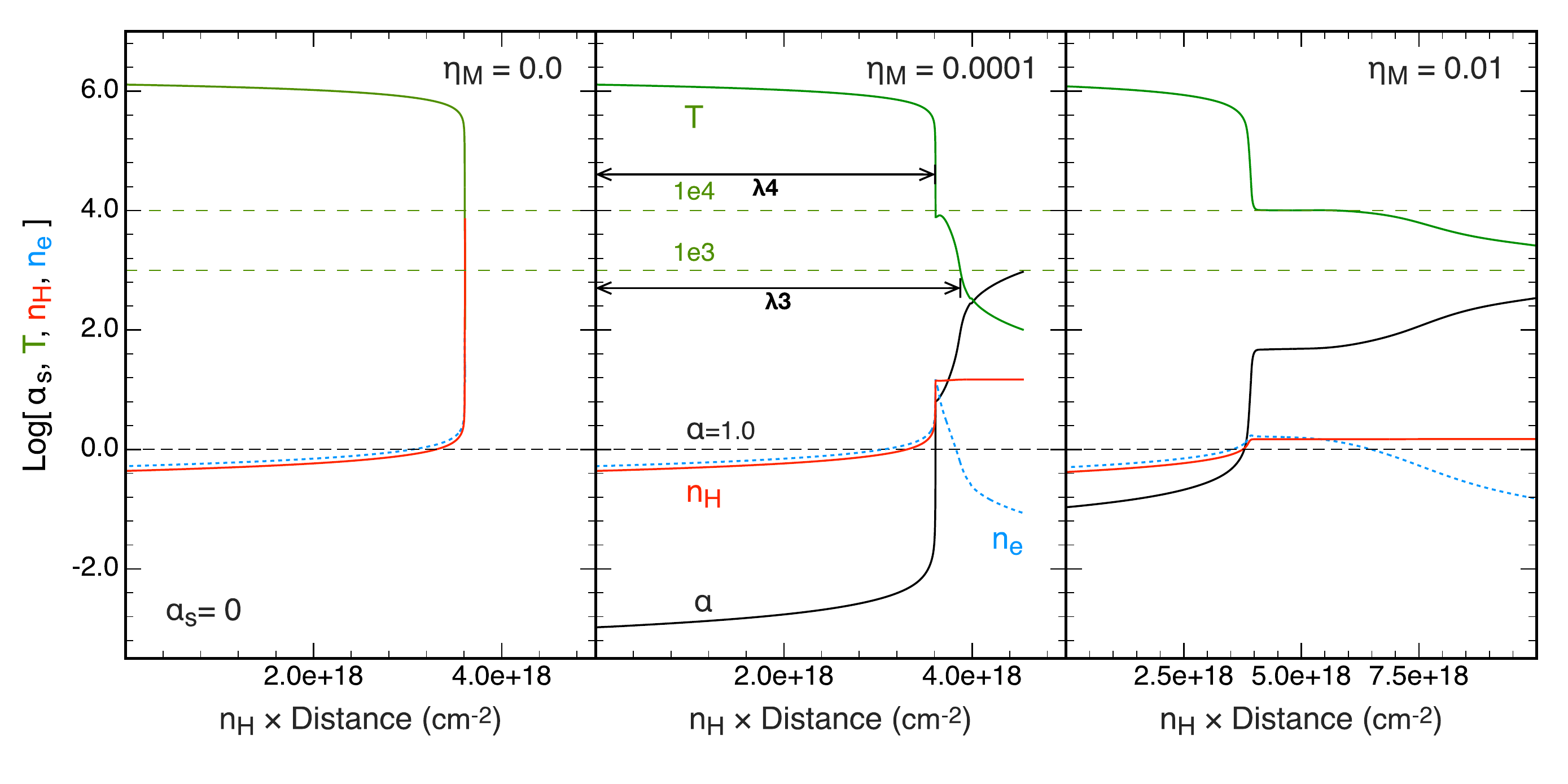}
  \caption{Post-Shock temperature and density structure of a 302\,km/s shock, for three values of $\eta_M$. Left panel, $\eta_M = 0.0$, the green line is the gas temperature, initially in excess of $10^6$\,K, then cooling to $10^2$\,K. The red line represents the hydrogen number density, blue dotted line is the corresponding electron number density, which vary proportionally to $1/T$.   Middle panel, in addition, the post-shock $\alpha_s = P_B/P_{\rm gas}$ is shown in the black curve.  Once  $\alpha_s$ exceeds 1.0, the total density rapidly stops increasing, with the low mass electrons falling away through recombination without reducing the mass density The temperature drops below $10^4$\,K before a minor rebound and the downstream photo-heated and recombination region is extended compared to the high density $\eta_M = 0$ case.  Right panel, the stronger field halts the compression at a lower density, increasing the size of the photo heated region.  The cooling columns $\lambda_4$ and $\lambda_3$ are also indicated in the middle panel, the columns where the temperature falls below $10^4$ and $10^3$\,K respectively. These are useful as the characterize the relative importance of the post-shock thermal cooling length, which is somewhat independent of $\eta_M$ ($\lambda_4$), and the photo-heated region which influences $\lambda_3$ and is a strong function of $\eta_m$. See also Figure \ref{fig:Comp-02} and the discussion of Figure \ref{fig:coolinglength} above.} \label{fig:compstructure}
 \end{figure*}

\section{Shock Structure Summary}

Full tables of the shock structural parameters, temperatures, compression factors and cooling lengths etc, are available in the online edition of this article.  Sample tables for model series 8, $\eta_M = $\,1e-3 and $R = $\,1e6 are given in Appendix \ref{app:structure} in Tables \ref{t:shckprop1} and \ref{t:shckprop2}.

\section{Scaled Radiation Quantities and Fits}
\label{sec:radiation}

\subsection{Nomenclature}
The precursor parameter defined above, is the ratio of the number of hydrogen ionising photons with $E > 13.598$~eV produced per hydrogen atom entering the shock, $\Psi = F(E>13.598eV)/(v_s\, n_H)$. This concept can be readily generalised to a \emph{photon production parameter} applicable to any chosen energy band such as for  $E > 54$~eV, in the case of HeII, or  $E > 1$~keV for the production of soft X-rays. If we define a related \emph{energy production parameter}, $\Phi =  I/(n_H \, v_s)$ (ergs/H atom), the physical quantities characterising the shock can then be readily derived from $\Psi$ and $\Phi$, using $v_s$ and $n_H$. In the following sections we adopt the following notation for both the cgs parameters of the radiation field and for the dimensionless quantities defined above:
\begin{itemize}
\item $I$~~ Intensity. This is the energy flux (ergs/cm$^2$/s), integrated over the full solid angle.
\item $I_n$~~ Specific Intensity. This is the Intensity per H atom $I/n_H$ (erg cm/s), where $n_H$ is the pre--shock hydrogen number density in these models. Over specific energy intervals, added subscripts denote the energy band referred to, for example,. $I_{\rm n, H}$ is the specific intensity for hydrogen ionising radiation.
\item $\Phi$~~ Energy Production Parameter. The energy flux per H atom flux (erg per H atom), $I_n/v_s$.
\item $F$~~  Photon Flux  (photon/cm$^2$s$^{-1}$), integrated over the full solid angle.
\item $\cal Q$~~  Ionisation Parameter; the photon flux per H atom (cm\,s$^{-1}$).
\item $\Psi$~~ Photon Production Parameter, the photon flux over a given energy band per H atom flux (photons per H atom, dimensionless).  When the integration is taken over $E > 13.598$eV it becomes the Precursor Parameter $Q/v$, defined in Section \ref{sec:precursor_param}.
\item $\langle E \rangle$~~ Mean Photon Energy, the average photon energy over a chosen energy interval, or a single value in the case of a line (erg).
\end{itemize}
These quantities can then be defined in terms of the physical parameters characterising the shock:
\begin{eqnarray}
 I   & = & n_H \, I_n\;, \nonumber \\
     & = & v_s \, n_H \, \Phi\;,\;\;\mbox{\rm (ergs/cm$^2$/s)}\; ,\\
 I_n & = & I / n_H \;,\nonumber\\
     & = & v_s \, \Phi\;,\;\;\mbox{\rm (erg cm/s)}\; , \\
   F & = &  n_H \, Q\;, \nonumber\\
     & = & v_s \, n_H \, \Psi\;,\;\;\mbox{\rm (photon/cm$^2$/s)}\; ,\\
   Q & = & F /n_H \;,\nonumber\\
     & = & v_s \, \Psi\;,\;\;\mbox{\rm (cm/s)}\; , \\
 \langle E \rangle & = & \Phi/\Psi\;,\;\;\mbox{\rm (erg)}
 \label{eq:rels}
\end{eqnarray}

Here, for the benefit of observers who would like to obtain physical parameters from commonly observed quantities, we focus on providing simple empirical fits to the following globally important key shock parameters:
\begin{itemize}
\item The ionizing flux, specifically the hydrogen ionizing flux, the integral of the radiation field above the hydrogen ionisation potential ($\sim13.598$eV).  We consider the  upstream field from the shock front that enters and drives the precursor.  This spectrum differs from the field in any other direction due to absorption within the shock structure where it is differentially absorbed and reprocessed.
\item The X--ray flux, specifically the radiation with photon energies above 1~keV.  Relatively optically thin, and observable with X--ray instruments, but also as the 2D upstream field to be consistent with the ionizing fluxes and so that relative emission between ionising and X--rays are meaningful.
\item The H$\beta\;\lambda4861.333$\AA\ flux. This is the key reference line in all spectrophotometry, often used as the divisor in spectrophotometric line studies.
\item The H$\alpha\;\lambda6562.819$\AA\ flux. This is worth giving separately, since H$\alpha$/H$\beta$ is often far from constant, and far from the HII region 'Case B' value. In addition, H$\alpha$ is often observed when H$\beta$ is either too faint, or not within the waveband of observation.
\end{itemize}
For the lines, 3D $4\pi$ steradian fluxes are considered, on the grounds that under the conditions modelled here, these lines are optically thin, and observable from all directions.

\subsection{Plane Parallel Upstream Ionising Fields}
In this section we give fits for the photon production parameter as a function of shock velocity and magnetic field for H-ionising photons with energies with $E > 13.598$~eV, $\Psi_H$, for HeII-ionising photons with energies with $E > 54.418$~eV, $\Psi_{\rm 54eV}$, and for X-rays with $E > 1$~keV, $\Psi_{\rm keV}$. These fits refer to the upstream direction only, measured at the shock front.  The 2D photon flux per unit area is presented as $\pi F_0$, where $F_0$ is the flux per steradian.  Fluxes in other directions will be modified by self absorption, and may be absent entirely in the downstream direction, having been absorbed and re--radiated at other wavelengths.

The upstream radiation fields in physical units of specific intensity, $I_n$, ionization parameter $Q$, and mean photon energy $\langle E \rangle$ are shown in Figure \ref{fig:radfields} for $R = 1e4$ and $\eta_M = 0 - 0.1$.  Important points here are the global conservation of energy, where the total intensity scales as a powerlaw with the same slope as the kinetic energy flux $\propto v^3$, and the displacements caused by the increasing magnetic fields.  This is due to the larger energy fraction carried by the magnetic fields, which are non-radiative, requiring higher velocities in general to achieve the same flux levels as weaker magnetic cases.  The ionization parameter for hydrogen $Q_H$ exceeds the shock velocity at around 140\,km/s as discussed previously. The helium ionisation parameter, for energies above 54.418\,eV, crosses the shock velocity at 320\,km/s and we might expect changes in the helium emission from the precursors in this velocity region.  At 600\,km/s the slope of the 1\,keV ionisation parameter $Q_{\rm keV}$ becomes power-law like and the mean photon energy in this band begins to increase significantly above 1\,keV.

 \begin{figure*}
  \includegraphics[width=\textwidth]{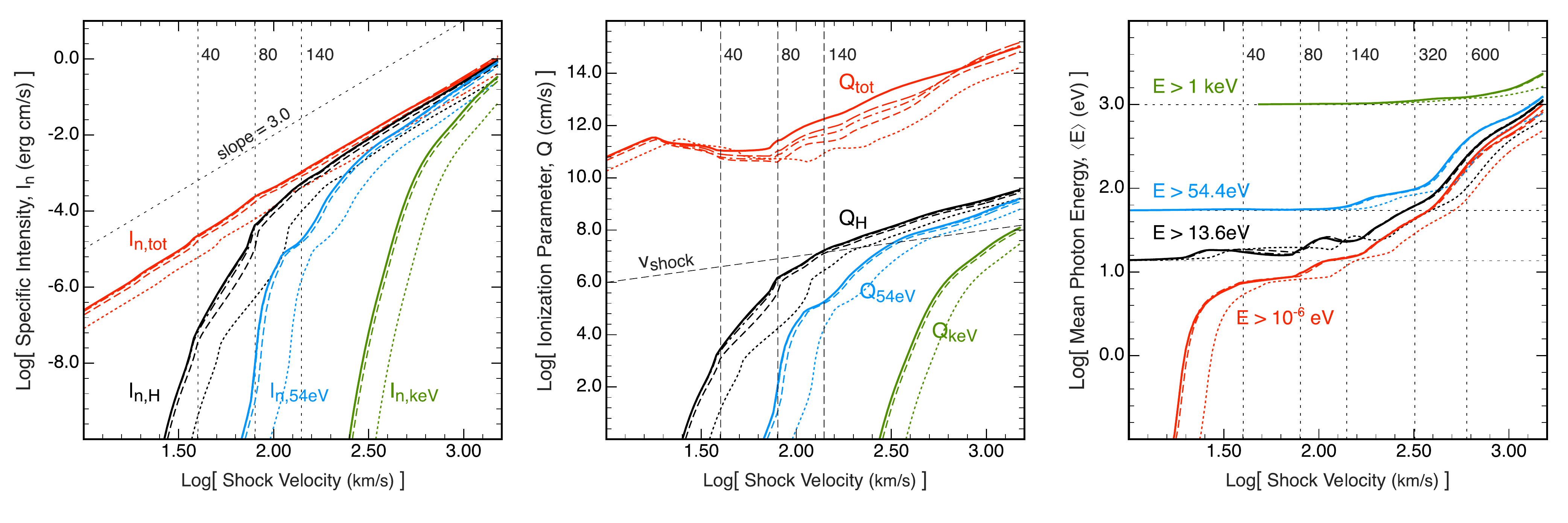}
  \caption{\emph{Left:} The velocity variation of the specific intensity, $I_n$ for the keV band (green), the HeII-ionising band (blue), the H-ionising band (black) and the specific intensity summed over all wavelengths (red). Note that the total closely tracks a slope of 3, which is proportional to the total energy flux through the shock. The short dashed lines are for the high magnetic field case, the long dashed for moderate magnetic field case, the dot-dashed lines for the weak field case, and the solid line for zero magnetic field. \emph{Centre panel} The corresponding ionisation parameters. \emph{Right:} The corresponding mean photon energies. } \label{fig:radfields}
 \end{figure*}

 \begin{figure}
  \includegraphics[width=\columnwidth]{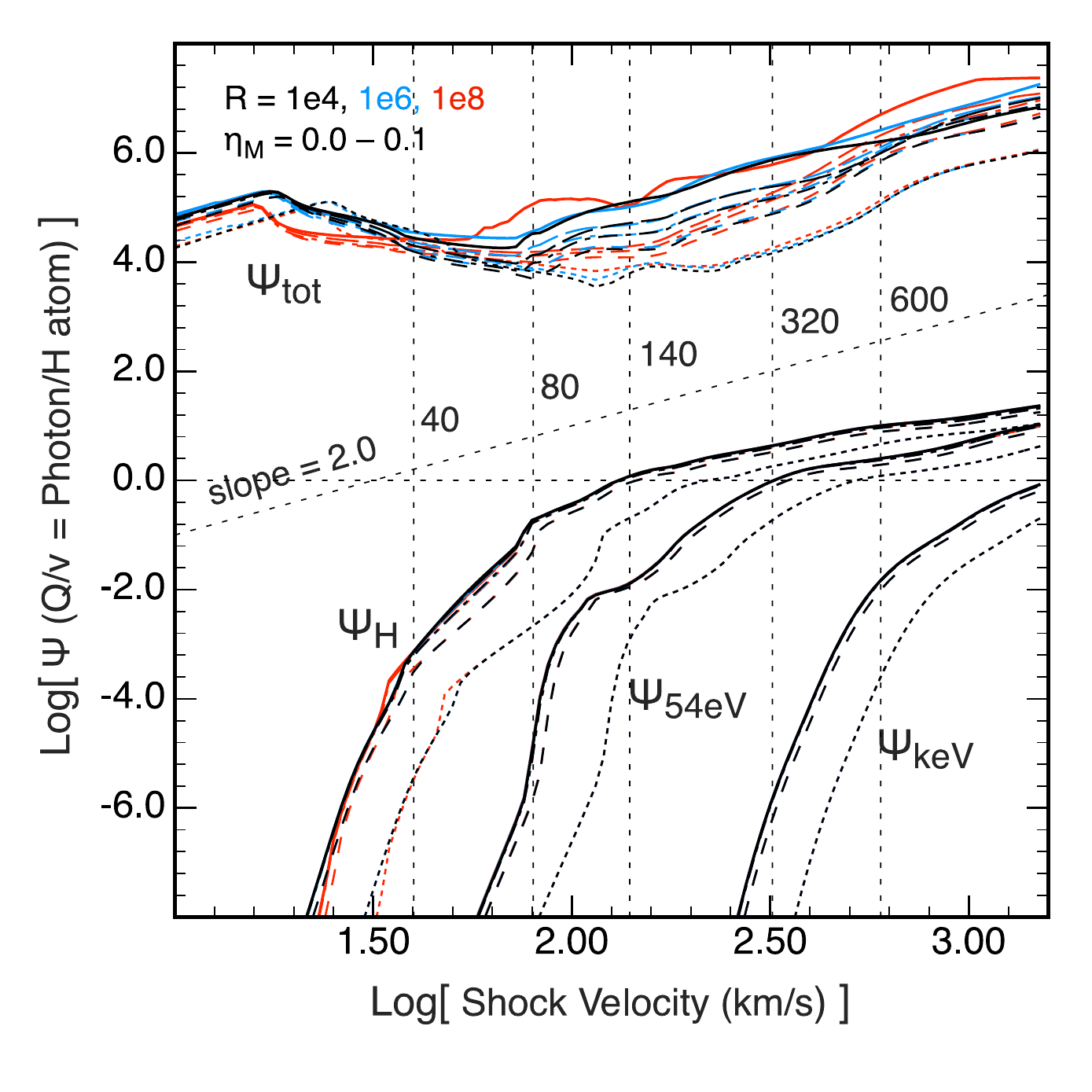}
   \caption{ The dimensionless photon production parameter $\Psi = c\,{\cal U}/v_s = {\cal Q}/v_s = F/(n_h\,v_s)$.  $\Psi_H$, $\Psi_{\rm 54eV}$, and $\Psi_{\rm keV}$ are not strongly effected by collisional de-excitation effects, and so scale nearly perfectly, especially the x-rays represented by $\Psi_{\rm keV}$. The total photon production $\Psi_{\rm tot}$ however is a complex function of density and $\eta_m$, as the total number of lines is dominated by optical and infrared lines which have lower transition probabilities and are both copious photon producers and density sensitive.   As noted earlier, $\Psi_H$ passes through 1.0 at approximately 140\,km/s, where as the $\Psi_{\rm 54eV}$, helium II ionising photons pass one at 320\,km/s and the x-rays turn over at $\Psi_{\rm keV} ~0.01$ at 600\,km/s. This panel shows all 15 model series, including the three ram-pressure or density regimes, in low-density $R = 10^4$ (black, $n_H = 1$\,@\,100\,km/s), medium density $R = 10^6$(blue, $n_H = 10^2$\,@\,100\,km/s) and high-density $R = 10^8$ (red, $n_H = 10^4$\,@\,100\,km/s)} \label{fig:psieta}
 \end{figure}

 \begin{figure}
  \includegraphics[width=\columnwidth]{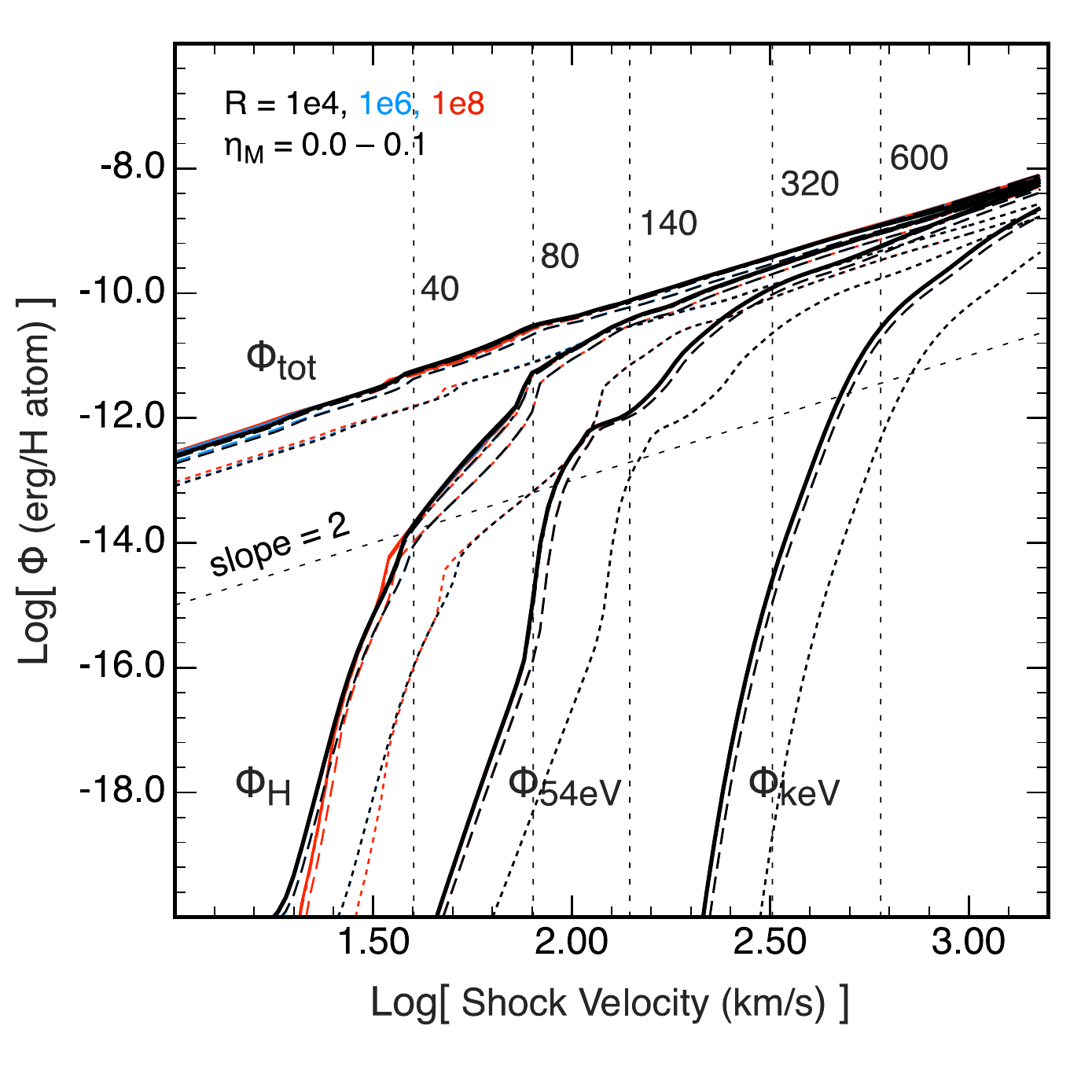}
   \caption{The energy production parameter $\Phi = I/(n_h\,v_s)$  $\Phi_H$, $\Phi_{\rm 54eV}$, and $\Phi_{\rm keV}$ are not strongly effected by collisional de-excitation effects, and so scale nearly perfectly, especially the x-rays represented by $\Phi_{\rm keV}$. Unlike photons, in Figure \ref{fig:psieta} the total energy production $\Phi_{\rm tot}$ tends towards a slope of 3.0, being constrained to be a fraction of the total energy flux into the shock. Thus energy conservation removes the variability of the integral represented by $\Phi_{\rm tot}$, and all the energy parameters appear black, as the black lines overlay the blue and red curves nearly perfectly.  The same velocity markers and reference slope are shown as for figure \ref{fig:psieta}  This panel shows all 15 model series, including the three ram-pressure or density regimes, in  low-density  $R = 10^4$ (black, $n_H = 1$\,@\,100\,km/s), medium density $R = 10^6$(blue, $n_H = 10^2$\,@\,100\,km/s) and high-density $R = 10^8$ (red, $n_H = 10^4$\,@\,100\,km/s)} \label{fig:phieta}
 \end{figure}

  Figure \ref{fig:psieta} shows the photon parameters from Figure \ref{fig:radfields} but is representative of all densities and magnetic fields in the entire grid,  in terms of the dimensionless $\Psi$ photon parameter and nearly dimensionless $\Phi$ energy parameters discussed previously.   Figure \ref{fig:psieta} shows the entire grid in black, $R = 10^4$, blue $R = 10^6$ and red, $R = 10^8$ curves, but the normalization is so consistent, especially for these ionising bands that they appear simply black, with the colored curved being overlaid, indicating that the ram pressure and density effects are minimal for these quantities, and scaling by $n_H$ will be quite accurate.  The total photon production $\Psi_{\rm tot}$ however is not well scaled by density or ram pressure, presumably due to collisional de-excitation for low energy (high photon production) transitions such as forbidden optical and infrared lines.

 Figure \ref{fig:phieta} is similar to Figure \ref{fig:psieta}, but shows $\Phi$ energy parameters.  From Figures \ref{fig:psieta}  and \ref{fig:phieta}, all the physical quantities can be derived using the relationships in Equation \ref{eq:rels}. Full tables of $\Psi$ and $\Phi$ are available in the online version of the article.  A Sample table for model series 8, $\eta_M = 1e-3$ and $R = 1e6$ is given in Table \ref{t:shckradiation}.

\subsubsection{Global Rational Polynominal Fits for $\Psi$}

We fit the photon parameters globally over as much of the velocity range as possible in an approximate manner with a non--linear rational polynomial fit. For higher velocities, when the $\Psi$ values are larger, more accurate fits are made, both for $\Psi_H >1.0$, and $\Psi_{\rm keV} > 0.01$.  We also provide simple power--law approximations for the high velocity regime where these are useful to be used in those applications when an approximate scaling relationship is all that is required.  Should a more accurate evaluation of the photon parameters be required, these should be carried out by spline interpolation of the model grid output files\footnote{available at: \url{https://miocene.anu.edu.au/mappings}}.

We fit $\Psi(v_s)$ in $\log - \log$ space, taking $x = \log_{10}v_s$, with $v_s$ measured in km/s.
For the rational polynomial fits, see \citet{1992nrc}, we take the ratio of two third order polynomials.
\begin{equation}
\log_{10}\Psi(v_s) = P(x)/Q(x)\;,
\end{equation}
 where
$$P(x) =[ P_0+x(P_1+x(P_2+xP_3)])\;,$$ $$Q(x) = [Q_0+x(Q_1+x(Q_2+xQ_3))]\;.$$
The derived coefficients are given in tables \ref{t:psih} and \ref{t:psikev} in the  Appendix \ref{app:Fits}, and are depicted in Figure \ref{fig:psifit}.

Note, as $P(x)/Q(x)$ is a ratio, we are free to scale the coefficients of $P$ and $Q$, and by convention $Q_0$ is often set to equal 1.  Here, some latitude is taken allowing $Q_0$ to vary in order to make the magnitude and sign of the other coefficients smaller and simpler.  It is trivial to divide all the coefficients by the $Q_0$ given to recover the more conventional $Q_0 = 1$ if desired.

 \begin{figure*}
 \includegraphics[width=\textwidth]{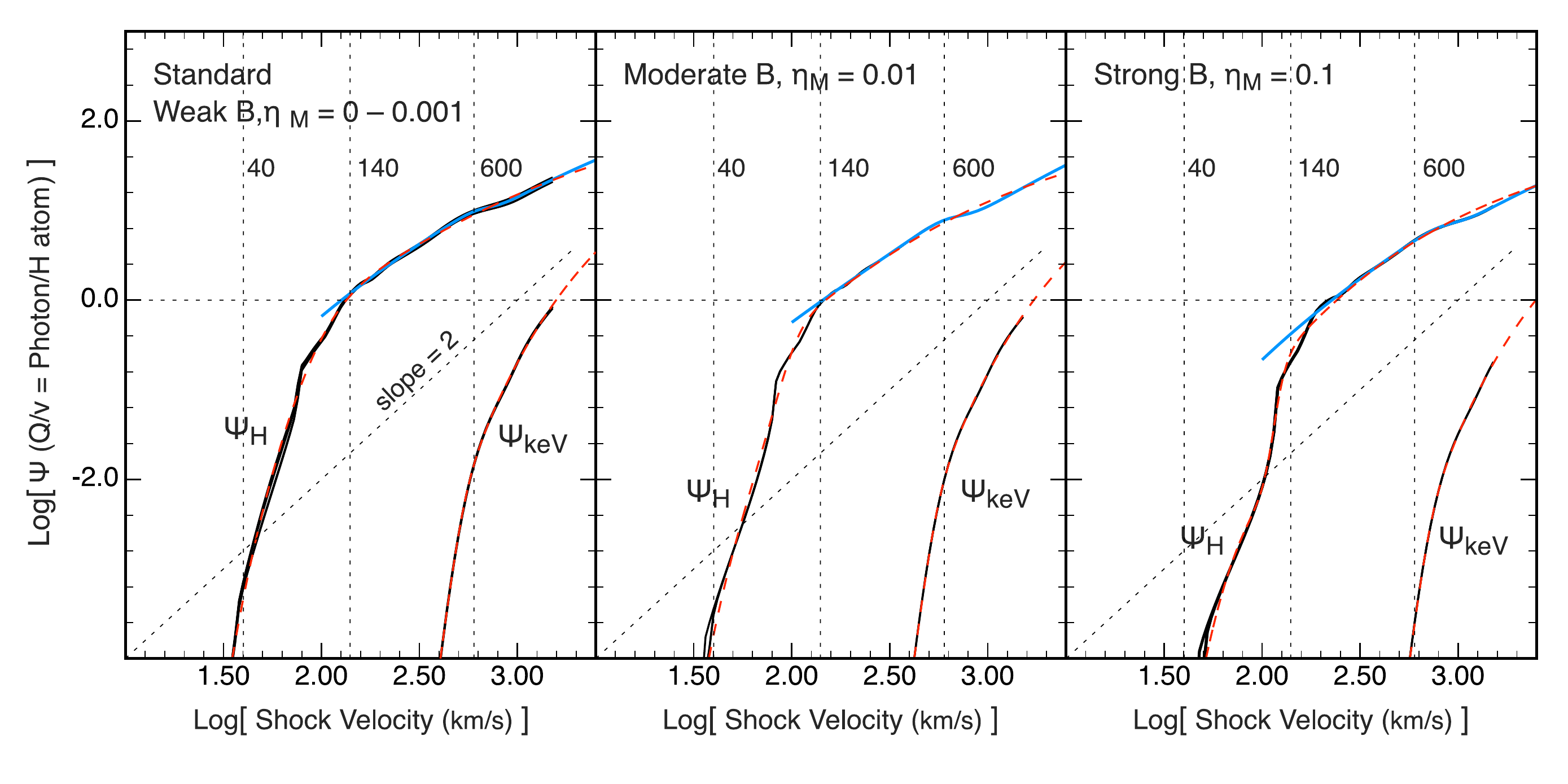}
  \caption{The variation of $\Psi_H$ and $\Psi_{\rm keV}$ as a function of shock velocity and of magnetic field strength for three cases, a standard case covering zero and relatively weak magnetic fields, a moderate field case where the magnetic pressure begins to affect the result enough to warrant a new fit, and the strong magnetic case $\eta_M = 0.1$. Global rational polynomial fits are shown in red dashed lines, and higher accuracy fits for $\Psi_H > 1$ in blue.  The global fit to the x-rays in $\Psi_{\rm keV}$ is accurate enough that an additional fit above 600\,km/s was not warranted. The sloped dotted line is a comparison power law with a slope of two, and the curves while power-law like in regions, are not well represented by this slope. Simpler piecewise power-law fits to the higher velocity ranges are given in  figure \ref{fig:psifithv}.  Key velocity markers at 40, 80, 140 and 600\,km/s are also indicated by black vertical dotted lines. } \label{fig:psifit}
 \end{figure*}

 \begin{figure}
  \includegraphics[width=\columnwidth]{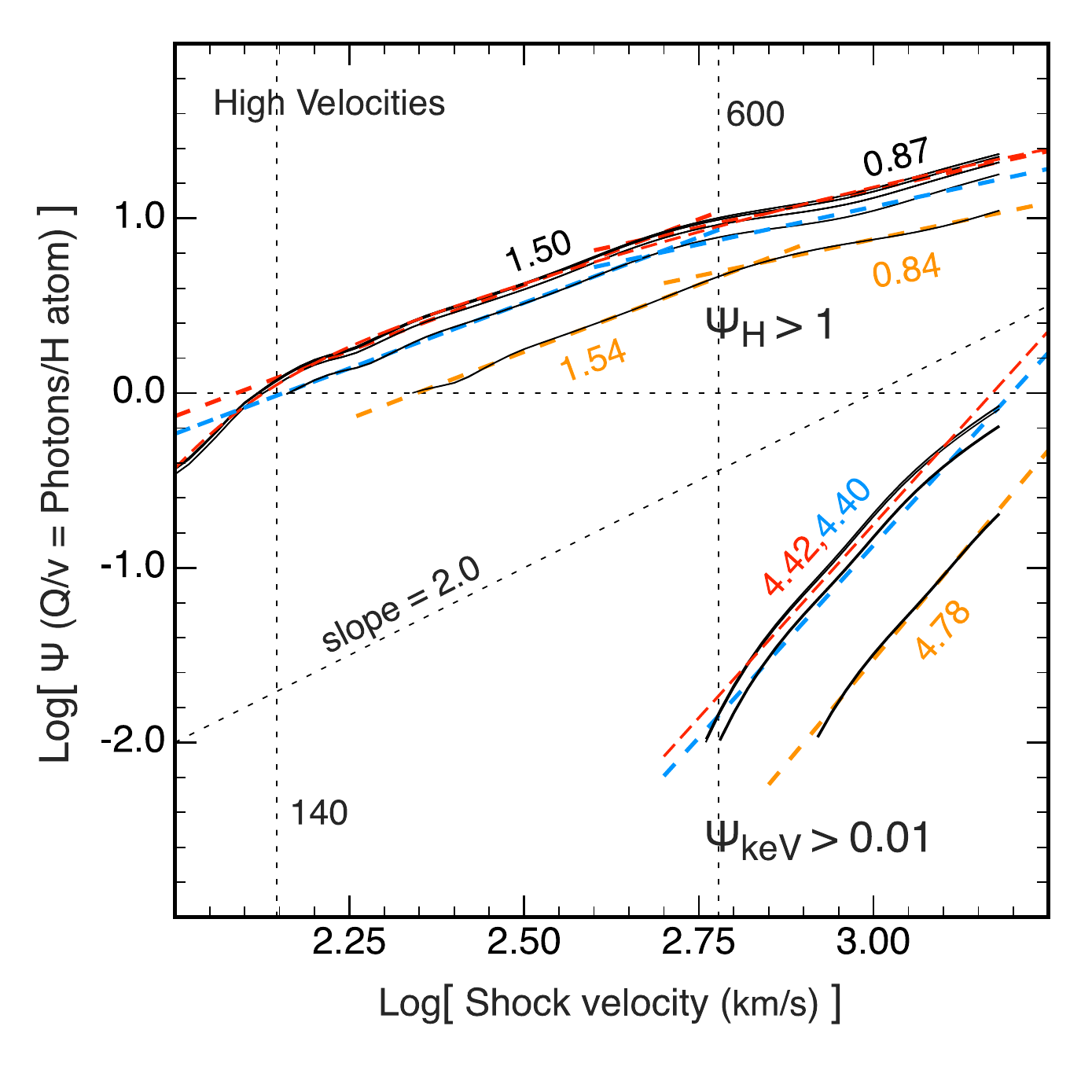}
   \caption{The variation of $\Phi_H$ and $\Psi_{\rm keV}$ as a function of shock velocity and of magnetic field strength for the three cases considered for $\Psi_H > 1.0$ and $\Psi_{\rm keV} > 0.01$. The standard model fits are shown in red, moderate models in blue and high $\eta_M$ models in orange, with slopes indicated.  Key high velocity markers at 140 and 600\,km/s are also indicated by black vertical dotted lines.} \label{fig:psifithv}
 \end{figure}

 \begin{figure*}
  \includegraphics[width=\textwidth]{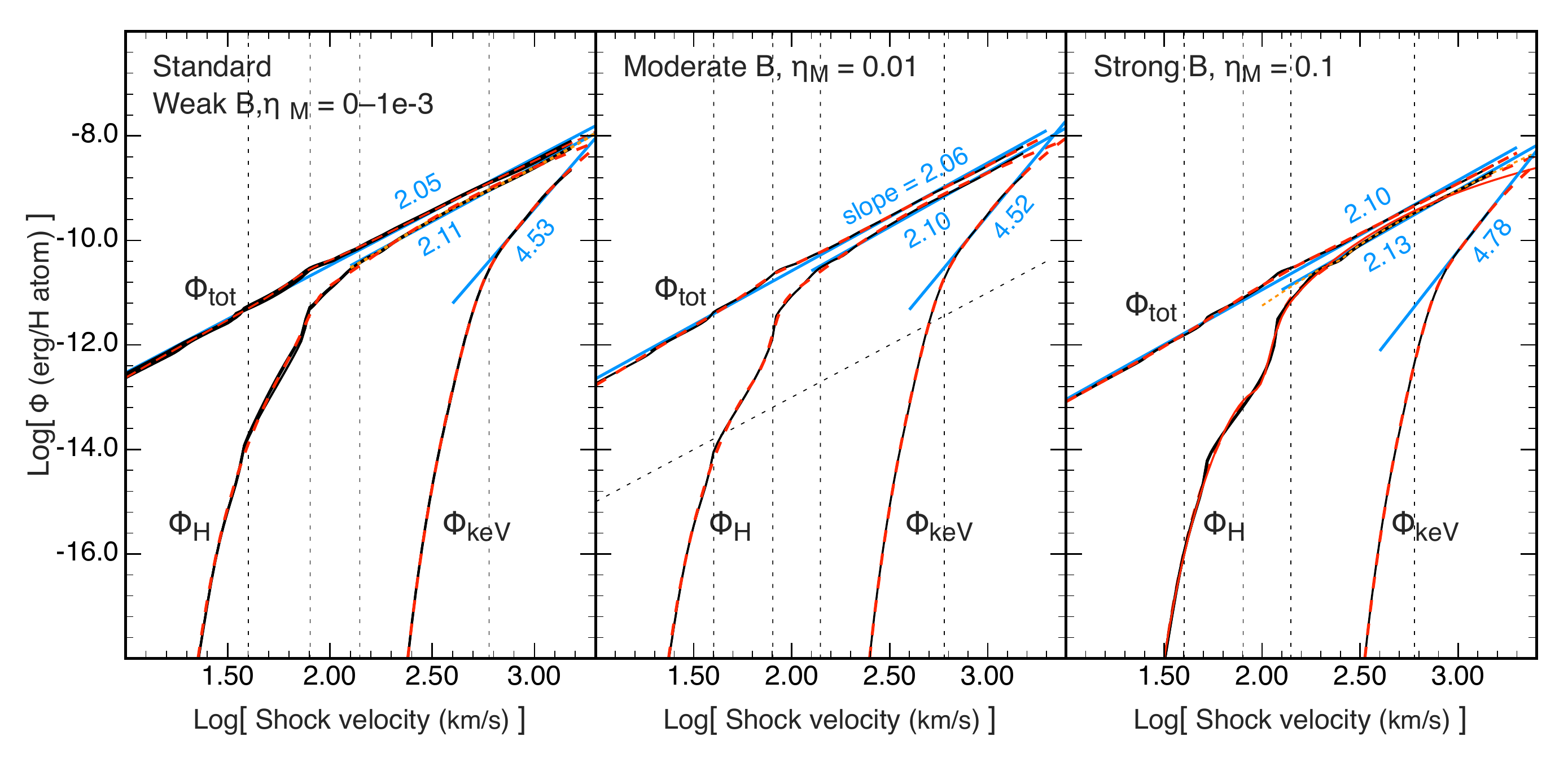}
\caption{As for figure \ref{fig:psifit}, but showing fits to the energy production parameters $\Phi_{\rm tot}$, $\Phi_H$ and $\Phi_{\rm keV}$ as a function of shock velocity and of magnetic field strength for the three cases considered. The global rational polynomial fits are shown in red dashes,  simpler power-law fits as blue lines, fitted to regions where the curves turn over and become more power-law like, with slopes indicated.  The  $\Phi_{\rm tot}$ is well fit by a power--law, unlike $\Psi_{\rm tot}$ which was not fitted above. The same velocity markers, at 40, 80,140 and 600\,km/s. as in previous figures are indicated by black dotted lines.} \label{fig:phifit}
 \end{figure*}

\subsubsection{Global Rational Polynominal Fits for $\Phi$}
\label{sec:PhiFits}
As in the case for the fitting of  $\Psi$, we fit $\Phi(v)$ in $\log - \log$ space, taking $x = \log_{10}v$, with $v_s$ in km/s.
The rational polynomial fits are taken as the ratio of a fourth-- and a third--order polynomial;
\begin{equation}
\log_{10}\Phi(v_s) = P(x)/Q(x)\;,
\end{equation}
 where
$$P(x) =[ P_0+x(P_1+x(P_2+x(P_3+ xP_4)])\;,$$ $$Q(x) = [Q_0+x(Q_1+x(Q_2+xQ_3))]\;.$$
The coefficients are given in tables \ref{t:phih} and \ref{t:phikev} in the Appendix \ref{app:Fits}, and the fits are shown in Figure \ref{fig:phifit} for both the global rational polynomial case, and for the simple power-law fits valid at high velocity, which are described below. The rational polynomial fit for $\Phi_{\rm keV}$ is accurate over the entire range $2.2 < x < 3.2$, so no specific high velocity fit is required.

\subsubsection{Approximate Power--Law Fits}
\label{sec:PLFits}
Power--law fits are made to the high velocity regimes, when $\Psi_H > 1.0$ and $\Psi_{\rm keV} > 0.01$.
These fits are made either as a broken power--law,
\begin{equation}
\log_{10} \Psi = A + B x \; , \; v_s < v_x\;,
\end{equation}
 and
\begin{equation}
\log_{10} \Psi = C + D x \; , \; v_s > v_x\;,
\end{equation} where $x = \log_{10}v_s$ with $v_s$ is measured in units of km/s. These broken power-law fits are valid for $v_s \lesssim 500-600$\,km/s. We fit simple power law for shocks faster than this:
\begin{equation}
\log_{10} \Psi = A + B x \; .
\end{equation}

Similar fits are made for the corresponding $\Phi$ quantities. The coefficients and range of validity of the fits to both $\Psi$ and $\Phi$ are given in Appendix \ref{app:Fits}, Tables \ref{t:psipowerlaw} and \ref{t:phipowerlaw}. The quality of the global rational polynomial and the power-law fits can be judged by inspection of Figures \ref{fig:psifit} and \ref{fig:phifit}, where the  global rational polynomial fits are given in red, and the power-law fits in blue.

In the standard case, relatively weak magnetic fields, and at constant density $n_H$, the hydrogen ionising fluxes, when $\Psi_H > 1.0$ ( aprox. 120\,km/s), scale as approximately $F = v_s\, n_H \, \Psi_H \propto v_s^{2.50}$, until at higher velocity still ($> 500$\,km/s) the changing photon energy distribution reduces the slope to approximately $F \propto v_s^{1.87}$ (see table \ref{t:psipowerlaw}).  The rising slope of the X-ray photon parameters implies that the turn--on for efficient X-ray shocks is rapid with increasing shock velocity. An extended grid of shock models with $\log_{10} v_s > 3.2$ could be explored in the future to investigate their properties in the regime where  $\Psi_{\rm keV} > 1.0$.

\section{3D Omnidirectional H$_\beta$ and H$_\alpha$ Fluxes}

\begin{figure*}
  \includegraphics[width=\textwidth]{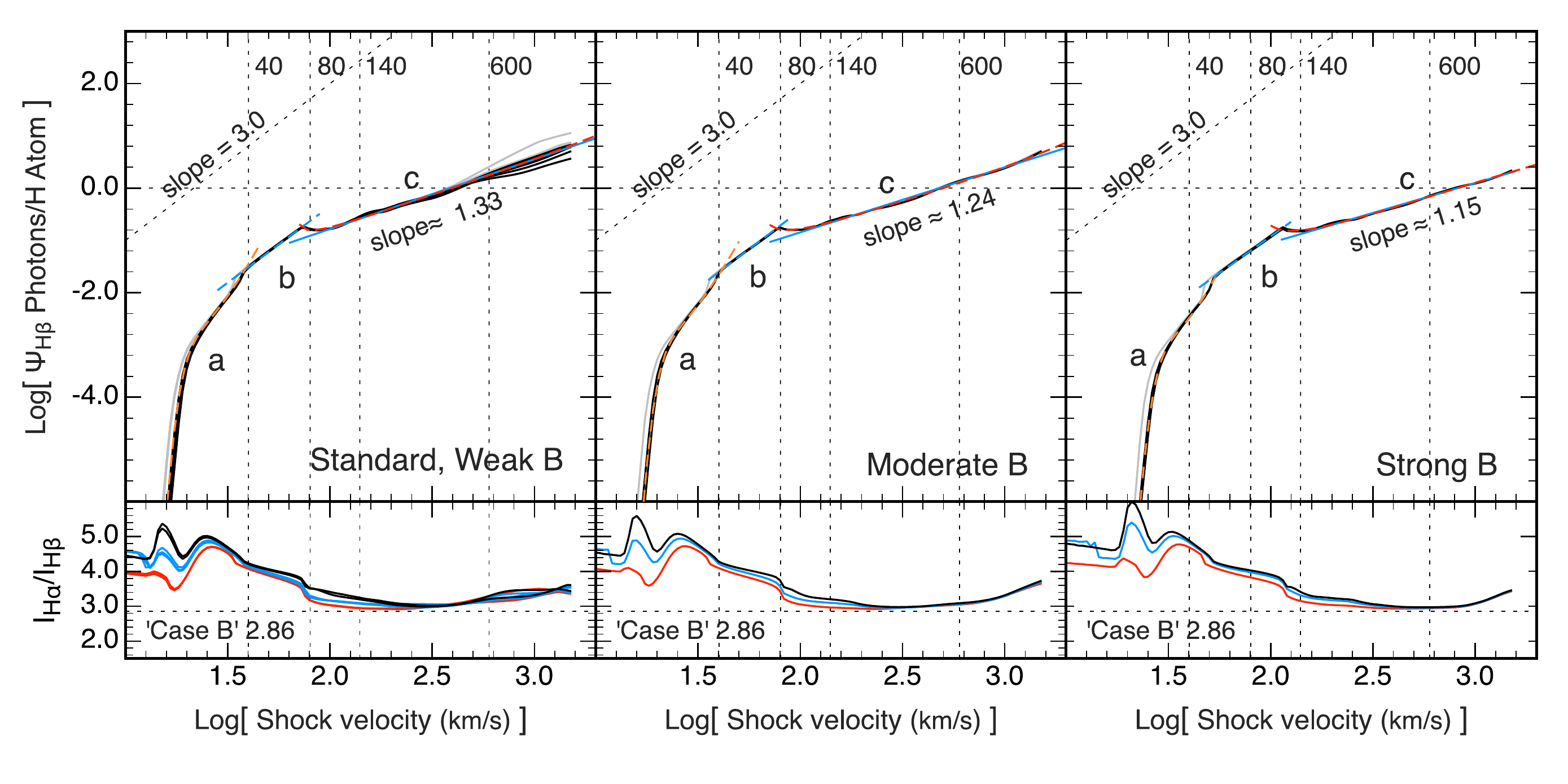}
  \caption{\emph{Upper panels:} The H$\beta$ photon production efficiency per H atom entering the shock as a function of shock velocity and magnetic field strength. There are three distinct regions marked a,b and c, described in detail in the text, which correspond to conditions prevailing in the shock precursor region. \emph{Lower panels:} The intensity ratio H$\alpha$/ H$\beta$. Note that this always exceeds the standard Case B value, and in lower velocity shocks where an appreciable fraction of the incoming H atoms are in the form of H I, the $H\alpha$/H$\beta$ ratio is strongly enhanced by collisional excitation effects in the post-shock gas.} \label{fig:Hbeta}
 \end{figure*}

The variation of the H$\beta$ photon production efficiency per H atom entering the shock as a function of shock velocity and magnetic field strength is shown in Figure \ref{fig:Hbeta}. For a given magnetic field strength, the relationships are particularly tight, and do not depend on the ram pressure. However, it is evident that there are three distinct regions with quite different slopes and curvatures. The presence of these discontinuities dictates that a piecewise fitting procedure is required.

Below $v_s \sim 40$\,km/s, where the shocks generally have cold and neutral precursors, the H$\beta$ emission in the shock is largely produced by collisional excitation of neutral hydrogen, this regime is labeled (a) in the Figure \ref{fig:Hbeta}. In this region, collisional excitation of H$\alpha$ gives rise to a very large intrinsic H$\alpha$/H$\beta$ ratio.

In the interval $40 < v_s < 80$\,km/s, the hot  but still neutral precursors, with increasing ionisation in the post-shock gas, gives rise to a very straight power--law, marked (b) in the Figure.  For $v_s > 80$\,km/s the  rapidly changing pre--ionisation has a complex effect on the post-shock temperature and hence on both hydrogen ionisation and recombination.

Above $v_s \sim 140$ \,km/s, in the region marked (c),  the shock properties settle and the hydrogen emission is a simpler function of velocity, and can be approximated by a simple power-law whose slope depends on the strength of the magnetic field.

We fit the three sections a, b and c, and select two fits in the region of the crossovers, which vary.  With the rational polynomials, the divisor $Q$ may have roots just outside the fitting regime, resulting in rapid fluctuations in value, especially below the crossover between section (b) and (c)  when the (c) fit can vary.  The tables give approximate velocity ranges for each fit section, and it is intended to only evaluate the section in these ranges, and then selecting the minimum of just the two sections in their overlap regions.

Fit (a) is a rational polynomial fit $\log_{10}\Psi_a(v) = P(x)/Q(x)\;$, fit (b) is a power--law $\log_{10} \Psi_b = A + B \log_{10} v$, and fit (c) is another rational polynomial fit . The coefficients and ranges for each component are given in tables \ref{t:psihbeta} and \ref{t:psihalpha}. Then the function is evaluated as the minimum $\Psi = \min(\Psi_1, \Psi_2)$ where  $\Psi_1$ and $\Psi_2$  are the $\Psi$ values returned by the two fits in the neighbourhood of each crossover. Finally the region (c) is also fit with a single power--law, as above, indicated in the figures as a thin blue line under the rational polynomial fit, and in the table as (c PL).

In the lower panels of figure \ref{fig:Hbeta} we also plot the modelled H$\alpha$/H$\beta$ intensity ratio for the entire grid.  In all cases, the Balmer decrement for the shock models is larger than the canonical 'Case B' H$\alpha$/H$\beta \sim 2.86$, typically seen in HII regions. This is due primarily to a significant collisional excitation contribution to the H$\alpha$ emission.  There is no obvious simple scaling relationship of the  H$\alpha$/H$\beta$ intensity ratio with shock velocity.  If accurate line ratios are required, then the full model tables should be used.

As the lines are monochromatic, a $\Phi$ function is not required as $\Phi = \langle E \rangle \Psi$ and  $ \langle E \rangle$ is simply the line energy: $E_\beta \sim 4.08616\times 10^{-12}$~ergs for H$\beta$, and $E_\alpha \sim3.02682\times 10^{-12}$~ergs for H$\alpha$.  The line ratio in photons is then $\Psi_{\rm H\alpha}/\Psi_{\rm H\beta}$ and in energy units:  $\Phi_{\rm H\alpha}/\Phi_{\rm H\beta}$ which is simply a constant times the photon ratio: $(E_\alpha/E_\beta) \Psi_{\rm H\alpha}/\Psi_{\rm H\beta}$ .

 Full tables of $\Psi_{\rm H\beta}$ and $\Psi_{\rm H\alpha}$ for all 15 grid cases at 0.02dex velocity resolution are available as extended tables in the online journal.  A Sample table for model series 8, $\eta_M =$\, 1e-3 and $R =$\,1e6 is given in Table \ref{t:shckbalmer}.

\section{The Hydrogen Ionising Relative X--ray Efficiencies}\label{sec:Xray}
\begin{figure*}
  \includegraphics[width=\textwidth]{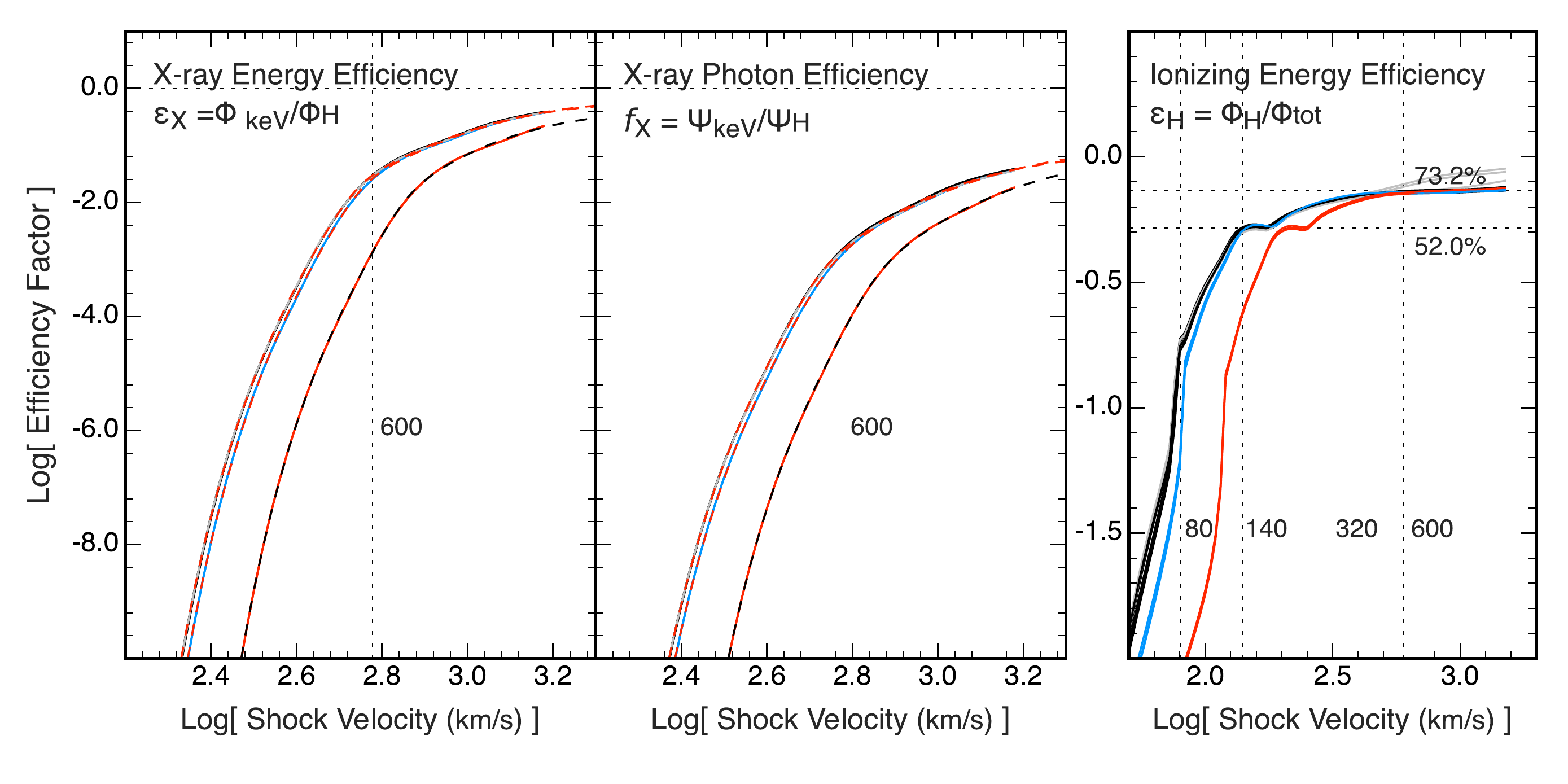}
  \caption{The X--ray and Ionising  efficiency factors.  Each panel shows 15 curves overlaid, resulting in three groups for magnetic parameter regimes, weak, moderate and strong magnetic fields, coloured black, blue and red respectively as in previous figures.  The simplicity of the plots indicate a the high degree of scaling, independent of ram pressure or density in these higher energy bands.  High accuracy global rational polynomial fits are shown in contrasting red dashes for the weak and moderate magnetic cases and black dashes for the strong magnetic cases, in the left and middle panel.  The left panel shows the ratio of the energy, $\Phi$,  production -- or equivalently the luminosity ratio -- of the x-rays compared to the hydrogen ionising band, and the middle panel shows the corresponding $\Psi$, photon ratios.  The right panel shows the  ratio of energy emission in the ionising bands compare to the total emission, without polynomial fits.} \label{fig:Epsilon-X}
 \end{figure*}

Unfortunately it will always be difficult if not impossible to observe the ionizing fluxes from fast shocks directly, the high opacity of all the material surrounding the shocks ensures that the ionization takes place in the vicinity of the shocks.  At higher energies, x-ray emission may remain optically thin, especially at energies above 1\,keV, and so a conversion factor or efficiency factor of the ratio of the ionising to x-ray fluxes may be of use when inferring ionising fluxes from x-ray observations independently of other photo-ionisation recombination optical methods.
We define  simple energy and photon ratio factors:
\begin{equation}
\varepsilon_X = \Phi_{\rm keV} / \Phi_{\rm H} \;,
\end{equation}
 and
\begin{equation}
f_X = \Psi_{\rm keV} / \Psi_{\rm H} \;.
\end{equation}

 As the curves in Figures \ref{fig:psieta} and \ref{fig:phieta} showed, once normalized to $\Psi$ and $\Phi$, the radiation parameters are essentially independent of density/ram pressure, and show only three distinct cases for $\eta_M$, that is weak $\eta_M < 0.01$, moderate, $\eta_M = 0.01$, and strong $\eta_M = 0.10$.  Figure \ref{fig:Epsilon-X} shows $\varepsilon_X$ and $f_X$ for the grid of models overlaid with global rational polynomial fits in dashed lines in the left and middle panels.  Again there is a clear break in the x-ray properties at $\sim 600$\,km/s. The curves are smooth and the fits are a very good representation of the models and extrapolate reasonably stably. The fitting parameters for the efficiencies are given in Table \ref{t:epsilon} in the Appendix.

Finally, the right hand panel of Figure \ref{fig:Epsilon-X} shows the ratio of the global ionising energy production efficiency to the total intensity, $\varepsilon_H = \Phi_H/\Phi_{\rm tot}$.  An equivalent photon ratio is not feasible, as $\Psi_{\rm tot}$ has no simple scaling (see figure \ref{fig:psieta}) and is so variable that no simple fitting is possible. This panel shows how efficient fast shocks when $\Psi_H > 1$ becomes at converting kinetic energy into ioinising radiation, and over half the radiated energy is ionising at around $\Psi_H = 1$ ($\sim 140$\, km/s), rising to over 73 percent for fast shocks over 300-600\,km/s.

\section{Conclusions}

In this paper, we have used the newly updated atomic data and microphysics incorporated into the MAPPINGS V code, and have developed a new method for computing the self-consistent pre-shock ionisation and temperature state, to map out the governing physics and various classes of radiative, plane-parallel, hypersonic, atomic astrophysical shocks.

We have identified four distinct regimes where the different precursor conditions affect the shock structures, and we have explored a grid of modelscovering a wide range of magnetic fields and ram pressures and pre-shock densities.  A key parameter, the pre cursor parameter, $\Psi = {\cal Q}/v_s$ determines much of the phenomenology of these shocks.

From this set of models, scaling relations for compression factors, temperatures and cooling lengths or column densities have been fitted using both rational polynomial fitting and simple power-laws. These provide a convenient means of estimating these key shock properties for a range of astronomical observations.  In addition, the global radiation properties in terms of photons and energy intensities have been normalised into just two key parameters, $\Psi$ and $\Phi$, which can be used to reconstruct relevant quantities such as the shock ionizing or X--ray luminosity.  We have provided convenient fitting functions and full tables of these quantities.

The modelling presented here has been focussed on the plasma state, and in particular the ionisation state at the shock front. Considering only the fast shock precursors, while these provide physical conditions which are fully accurate at the shock front, they were not fully computed to the outer boundary of the photoionisation region. In future papers, these extended precursors will be fully computed and the detailed precursor spectra, as well as the associated shock spectra will be given. In addition, a grid of shocks as a function of metallicity is currently in development, and will be the subject of a further paper in this series. These models will allow the structural and radiation parameters to extend to low metallicity contexts such as early universe star formation.

However, the modelling presented here is indeed fully applicable to the precursor structure up to 140\,km/s. These models are applied to the study of Herbig-Haro (HH) objects in the second paper in this series (Dopita \& Sutherland, 2017: in press).

\section*{Acknowledgments}
The authors wish to thank a thoughtful referee whose insights and attention to detail have improved the paper significantly. The authors also acknowledge the support of the Australian Research Council (ARC) through Discovery project DP16010363.\newline


\appendix

\section{Fitting Parameters}
\label{app:Fits}
Here we present the complete list of fitting parameters for both the rational polynomial fits and the power-law fits for the hydrogen ionising precursor parameter, $\Psi_H$, the X-Ray photon production parameter, $\Psi_{\rm keV}$, and the corresponding energy production factors, $\Phi_H$ and $\Phi_{\rm keV}$. In addition, we also gives the fits the the H$\beta$ and H$\alpha$ photon productn parameters, $\phi_{H\beta}$ and $\Phi_{H\alpha}$. See sections \ref{sec:PhiFits}, \ref{sec:PhiFits}, and \ref{sec:PLFits}

\begin{table*}[htp]
\caption{Hydrogen ionising precursor parameter, $\Psi_H = Q_H/v$}
\begin{center}
\begin{tabular}{rrrr}
\hline
\multicolumn{4}{c}{General Fit: ~ $\log_{10} v_s \gtrsim 1.3$} \\
$\Psi_H$ & Standard & Moderate $B$ & Strong $B$\\
$\eta_M$: & $0-0.001$ & $0.01$ & $0.10$\\
\hline
$P_0$ & -9.1255E+00 & -8.8896E+00 & -6.9246E+00\\
$P_1$ & 1.2480E+01 & 1.2433E+01 & 9.5080E+00\\
$P_2$ & -5.7713E+00 & -5.8249E+00 & -4.3489E+00\\
$P_3$ & 9.0301E-01 & 9.1400E-01 & 6.6234E-01\\
$Q_0$ & -1.0000E+00 & -1.0000E+00 & -1.0000E+00\\
$Q_1$ & 2.0569E+00 & 2.0990E+00 & 1.8895E+00\\
$Q_2$ & -1.2843E+00 & -1.3418E+00 & -1.1279E+00\\
$Q_3$ & 2.6035E-01 & 2.7350E-01 & 2.1687E-01\\
\hline
\multicolumn{4}{c}{High Velocity Fit; ~$\log_{10} v_s \gtrsim 2.1$} \\
$\Psi_H>1.0$ & Standard & Moderate $B$& Strong $B$\\
$\eta_M$: & $0-0.001$ & $0.01$ & $0.10$\\
\hline
$P_0$ & -9.0119E+00 & -9.4731E+00 & -1.1338E+01\\
$P_1$ & 1.0594E+01 & 1.1018E+01 & 1.2431E+01\\
$P_2$ & -4.1065E+00 & -4.2350E+00 & -4.5178E+00\\
$P_3$ & 5.2630E-01 & 5.3926E-01 & 5.4507E-01\\
$Q_0$ & 1.0000E+00 & 2.0000E+00 & 1.0000E+00\\
$Q_1$ & 2.0658E-02 & -9.8312E-01 & 1.2981E-01\\
$Q_2$ & -3.8330E-01 & -4.8952E-02 & -4.2815E-01\\
$Q_3$ & 8.9103E-02 & 5.2358E-02 & 9.1727E-02\\
\hline
\end{tabular}
\end{center}
\label{t:psih}
\end{table*}%

\begin{table*}[htp]
\caption{X-Ray photon parameter, $\Psi_{\rm keV}$, $ E > 1.0$~keV }
\begin{center}
\begin{tabular}{rrrr}
\hline
\multicolumn{4}{c}{General Fit: ~ $\log_{10} v_s \gtrsim 2.2$} \\
$\Psi_{\rm keV}$ & Standard & Moderate $B$& Strong $B$\\
$\eta_M$: & $0-0.001$ & $0.01$ & $0.10$\\
\hline
$P_0$ & -6.1864E+01 & -5.9698E+01 & -5.6792E+01\\
$P_1$ & 6.4813E+01 & 6.2138E+01 & 5.6556E+01\\
$P_2$ & -2.2599E+01 & -2.1519E+01 & -1.8731E+01\\
$P_3$ & 2.6207E+00 & 2.4777E+00 & 2.0619E+00\\
$Q_0$ & -1.0000E+01 & -1.0000E+01 & -1.0000E+01\\
$Q_1$ & 1.2354E+01 & 1.2267E+01 & 1.1665E+01\\
$Q_2$ & -5.0384E+00 & -4.9702E+00 & -4.4991E+00\\
$Q_3$ & 6.8009E-01 & 6.6661E-01 & 5.7486E-01\\
\hline
\multicolumn{4}{c}{ High Velocity Fit: ~ $\log_{10} v_s \gtrsim 2.8$} \\
$\Psi_{\rm keV} > 0.01$ & Standard & Moderate $B$& Strong $B$\\
$\eta_M$: & $0-0.001$ & $0.01$ & $0.10$\\
\hline
$P_0$ & -2.5390E+01 & -2.2937E+01 & -1.4213E+01\\
$P_1$ & 2.5250E+01 & 2.2586E+01 & 1.3152E+01\\
$P_2$ & -8.3685E+00 & -7.4101E+00 & -4.0532E+00\\
$P_3$ & 9.2410E-01 & 8.0990E-01 & 4.1586E-01\\
$Q_0$ & -1.0000E+01 & -1.0000E+01 & -1.0000E+01\\
$Q_1$ & 1.1093E+01 & 1.0994E+01 & 1.0343E+01\\
$Q_2$ & -4.0791E+00 & -4.0079E+00 & -3.5523E+00\\
$Q_3$ & 4.9773E-01 & 4.8496E-01 & 4.0542E-01\\
\hline
\end{tabular}
\end{center}
\label{t:psikev}
\end{table*}%

\begin{table*}[htp]
\caption{Hydrogen ionising Energy Production Parameter, $\Phi_H = I_H/(v_s\,n_H)$ (erg/H atom), $ E > 13.598$~eV }
\begin{center}
\begin{tabular}{rrrr}
\hline
\multicolumn{4}{c}{General Fit: ~ $\log_{10} v_s  \gtrsim 1.3$} \\
$\Phi_H$ & Standard & Moderate $B$ & Strong $B$\\
$\eta_M$: & $0-0.001$ & $0.01$ & $0.10$\\
\hline
\hline
$P_0$ & 1.2446E+01 & 1.3431E+01 & 1.5071E+01\\
$P_1$ & -3.3955E+01 & -3.5426E+01 & -3.4426E+01\\
$P_2$ & 2.6960E+01 & 2.7890E+01 & 2.4481E+01\\
$P_3$ & -7.5248E+00 & -7.7875E+00 & -6.3106E+00\\
$P_4$ & 4.8201E-01 & 5.0426E-01 & 3.9379E-01\\
$Q_0$ & -2.0000E+00 & -2.0000E+00 & -2.0000E+00\\
$Q_1$ & 4.0757E+00 & 4.0606E+00 & 3.6881E+00\\
$Q_2$ & -2.6150E+00 & -2.6092E+00 & -2.1720E+00\\
$Q_3$ & 5.3982E-01 & 5.4048E-01 & 4.1438E-01\\
\hline
\multicolumn{4}{c}{High Velocity Fit: ~ $\log_{10} v_s  \gtrsim 2.1$} \\
$\Phi_H$  & Standard & Moderate $B$& Strong $B$\\
$\eta_M$: & $0-0.001$ & $0.01$ & $0.10$\\
\hline
\hline
$P_0$ & -1.5307E+01 & -1.5257E+01 & -1.4851E+01\\
$P_1$ & 1.3385E+01 & 1.3608E+01 & 1.4206E+01\\
$P_2$ & -3.7046E+00 & -3.8024E+00 & -4.1461E+00\\
$P_3$ & 3.0312E-01 & 3.0968E-01 & 3.3154E-01\\
$P_4$ & 0.0000E+00 & 0.0000E+00 & 0.0000E+00 \\
$Q_0$ & 1.0000E+00 & 1.0000E+00 & 1.0000E+00\\
$Q_1$ & -7.2773E-01 & -7.4907E-01 & -8.2683E-01\\
$Q_2$ & 1.3546E-01 & 1.4230E-01 & 1.7184E-01\\
$Q_3$ & 0.0000E+00 & 0.0000E+00 & 0.0000E+00 \\
\hline
\end{tabular}
\end{center}
\label{t:phih}
\end{table*}%

\begin{table*}[htp]
\caption{Approximate power--law fits to photon production parameters at high velocity}
\label{t:psipowerlaw}
\begin{center}
\begin{tabular}{rrrr}
\hline
\multicolumn{4}{c}{$\Psi_H$ Broken Power--law Fit: ~$\log_{10} v_s \gtrsim 2.1$} \\
$\Psi_H > 1.0$ & Standard & Moderate $B$& Strong $B$\\
$\eta_M$: & $0-0.001$ & $0.01$ & $0.10$\\
\hline
\hline
$A$ & -3.1415E+00 & -3.2340E+00 & -3.6015E+00\\
$B$ & 1.5043E+00 & 1.5002E+00 & 1.5351E+00\\
$v_x$ & 4.7318E+02 & 4.8488E+02 & 6.6588E+02\\
$C$ & -1.4537E+00 & -1.5282E+00 & -1.6300E+00\\
$D$ & 8.7336E-01 & 8.6510E-01 & 8.3680E-01\\
\hline
\multicolumn{4}{c}{$\Psi_{\rm keV}$ Power--law Fit: ~ $\log_{10} v_s \gtrsim 2.8$} \\
$\Psi_{\rm keV} > 0.01$ & Standard & Moderate $B$& Strong $B$\\
$\eta_M$: & $0-0.001$ & $0.01$ & $0.10$\\
\hline
\hline
$A$ & -1.4027E+01 & -1.4081E+01 & -1.5866E+01\\
$B$ & 4.4251E+00 & 4.4035E+00 & 4.7808E+00\\
\hline
\end{tabular}
\end{center}
\end{table*}%

\begin{table*}[htp]
\caption{Approximate power--law fits to energy production parameters at high velocity}
\label{t:phipowerlaw}
\begin{center}
\begin{tabular}{rrrr}
\hline
\multicolumn{4}{c}{$\Phi_H$ Power--law Fit: ~$\log_{10} v_s  \gtrsim 2.1$} \\
$\Phi_H$ & Standard & Moderate $B$& Strong $B$\\
$\eta_M$: & $0-0.001$ & $0.01$ & $0.10$\\
\hline
\hline
$A$ & -1.4923E+01 & -1.5002E+01 & -1.5412E+01\\
$B$ & 2.1140E+00 & 2.1043E+00 & 2.1261E+00\\
\hline
\multicolumn{4}{c}{$\Phi_{\rm keV}$ Power--law Fit: ~$\log_{10} v_s  \gtrsim 2.8$} \\
$\Phi_{\rm keV}$ & Standard & Moderate $B$& Strong $B$\\
$\eta_M$: & $0-0.001$ & $0.01$ & $0.10$\\
\hline
\hline
$A$ & -2.2979E+01 & -2.3086E+01 & -2.4560E+01\\
$B$ & 4.5284E+00 & 4.5229E+00 & 4.7862E+00\\
\hline
\end{tabular}
\end{center}
\end{table*}%

\begin{table*}[htp]
\caption{X-Ray Energy Production Parameter, $\Phi_{\rm keV}$, $ E > 1.0$~keV }
\label{t:phikev}
\begin{center}
\begin{tabular}{rrrr}
\hline
\multicolumn{4}{c}{General Fit, $x \, > \, \sim 2.2$} \\
$\Phi_{\rm keV}$ & Standard & Moderate $B$& Strong $B$\\
$\eta_M$: & $0-0.001$ & $0.01$ & $0.10$\\
\hline
\hline
$P_0$ & 2.0678E+01 & 2.5092E+01 & 3.6127E+01\\
$P_1$ & -5.5036E+01 & -5.9103E+01 & -6.5857E+01\\
$P_2$ & 3.1832E+01 & 3.2936E+01 & 3.2697E+01\\
$P_3$ & -5.3056E+00 & -5.3805E+00 & -4.9340E+00\\
$P_4$ & 0.0000E+00 & 0.0000E+00 & 0.0000E+00 \\
$Q_0$ & -2.0000E+01 & -2.0000E+01 & -2.0000E+01\\
$Q_1$ & 2.4682E+01 & 2.4513E+01 & 2.3294E+01\\
$Q_2$ & -1.0026E+01 & -9.8935E+00 & -8.9511E+00\\
$Q_3$ & 1.3450E+00 & 1.3191E+00 & 1.1376E+00\\
\hline
\end{tabular}
\end{center}
\end{table*}%

\begin{table*}[htp]
\caption{H$\beta$ Photon Production Parameter, $\Psi_{\rm H\beta}$}
\begin{center}
\begin{tabular}{rrrr}
\hline
          & Standard & Moderate $B$& Strong $B$\\
$\eta_M$: & $0-0.001$ & $0.01$ & $0.10$\\
\hline
\hline
Fit (a) & \multicolumn{2}{c}{ $1.1\lesssim x  \lesssim 1.6 $}  & $1.2 \lesssim x  \lesssim 1.75 $\\
$P_0$ & -1.7288E+01 & -1.6865E+01 & -2.2404E+01\\
$P_1$ & 3.7218E+01 & 3.5602E+01 & 4.3093E+01\\
$P_2$ & -2.6477E+01 & -2.4823E+01 & -2.7462E+01\\
$P_3$ & 6.1849E+00 & 5.6810E+00 & 5.7751E+00\\
$Q_0$ & 1.0000E+00 & 1.0000E+00 & 1.0000E+00\\
$Q_1$ & -1.6623E+00 & -1.6277E+00 & -1.4547E+00\\
$Q_2$ & 6.9812E-01 & 6.6842E-01 & 5.3463E-01\\
$Q_3$ & 0.0000E+00 & 0.0000E+00 & 0.0000E+00\\
\hline
Fit (b) & \multicolumn{2}{c}{ $1.55 \lesssim x  \lesssim 1.9 $} & $1.6  \lesssim x  \lesssim 2.0 $\\
$A$ & -6.2032E+00 & -6.3601E+00 & -6.5182E+00\\
$B$ & 2.9295E+00 & 2.9677E+00 & 2.7974E+00\\
\hline
Fit (c) & \multicolumn{2}{c}{ $1.85 \lesssim x  \lesssim  3.3 $} & $2.0\lesssim x  \lesssim 3.3 $\\
$P_0$ & 3.3994E+01 & 3.5411E+01 & 2.5311E+01\\
$P_1$ & -4.5197E+00 & -2.6155E+01 & -5.0742E+01\\
$P_2$ & -1.8296E+01 & 1.0410E+00 & 2.9474E+01\\
$P_3$ & 5.7761E+00 & 1.4041E+00 & -5.1622E+00\\
$Q_0$ & 0.0000E+00 & 0.0000E+00 & 0.0000E+00\\
$Q_1$ & -1.6240E+01 & -1.0617E+01 & 1.8372E+00\\
$Q_2$ & 1.0890E+01 & 7.9867E+00 & 2.4274E-01\\
$Q_3$ & -1.0473E+00 & -1.0977E+00 & -6.5458E-01\\
\hline
Fit (c PL) &\multicolumn{2}{c}{ $x \gtrsim 2.0$}  & $x \gtrsim  2.1$\\
$A$ & -3.4514E+00 & -3.3330E+00 & -3.3376E+00\\
$B$ & 1.3349E+00 & 1.2438E+00 & 1.1453E+00\\
\hline
\end{tabular}
\end{center}
\label{t:psihbeta}
\end{table*}%

\begin{table*}[htp]
\caption{H$\alpha$ Photon Production Parameter, $\Psi_{\rm H\alpha}$}
\label{t:psihalpha}
\begin{center}
\begin{tabular}{rrrr}
\hline
          & Standard & Moderate $B$& Strong $B$\\
$\eta_M$: & $0-0.001$ & $0.01$ & $0.10$\\
\hline
\hline
Fit (a) & \multicolumn{2}{c}{ $\sim1.1 < x< \sim1.6 $}  & $\sim1.2 < x< \sim1.75 $\\
$P_0$ & -1.4746E+01 & -1.5105E+01 & -1.7540E+01\\
$P_1$ & 3.1826E+01 & 3.2108E+01 & 3.4068E+01\\
$P_2$ & -2.2788E+01 & -2.2638E+01 & -2.1964E+01\\
$P_3$ & 5.3872E+00 & 5.2695E+00 & 4.6833E+00\\
$Q_0$ & 1.0000E+00 & 1.0000E+00 & 1.0000E+00\\
$Q_1$ & -1.6575E+00 & -1.6337E+00 & -1.4752E+00\\
$Q_2$ & 6.9380E-01 & 6.7412E-01 & 5.4935E-01\\
$Q_3$ & 0.0000E+00 & 0.0000E+00 & 0.0000E+00\\
\hline
Fit (b) & \multicolumn{2}{c}{ $\sim1.55 < x< \sim1.9 $} & $\sim1.6 < x< \sim2.0 $\\
$A$ & -5.1419E+00 & -5.1763E+00 & -5.4429E+00\\
$B$ & 2.7394E+00 & 2.7051E+00 & 2.6174E+00\\
\hline
Fit (c) & \multicolumn{2}{c}{ $\sim1.85 < x< \sim3.3 $} & $\sim2.0 < x< \sim3.3 $\\
$P_0$ & 3.4422E+01 & 3.1844E+01 & 2.7794E+01\\
$P_1$ & -2.9022E+01 & -3.0091E+01 & -2.9210E+01\\
$P_2$ & 2.5642E+00 & 6.1988E+00 & 9.2034E+00\\
$P_3$ & 1.6436E+00 & 4.1653E-01 & -7.7093E-01\\
$Q_0$ & 0.0000E+00 & 0.0000E+00 & 0.0000E+00\\
$Q_1$ & -1.1172E+01 & -1.0734E+01 & -5.3971E+00\\
$Q_2$ & 8.2100E+00 & 8.3396E+00 & 4.1110E+00\\
$Q_3$ & -1.0240E+00 & -1.2855E+00 & -6.5888E-01\\
\hline
Fit (c PL) &\multicolumn{2}{c}{ $x > \sim2.0$}  & $x > \sim2.1$\\
$A$ & -2.7833E+00 & -2.9121E+00 & -2.7323E+00\\
$B$ & 1.3138E+00 & 1.3205E+00 & 1.1510E+00\\
\hline
\end{tabular}
\end{center}
\label{t:psihalpha}
\end{table*}%

\begin{table*}[htp]
\caption{Energy and Photon Efficiencies, $\varepsilon_X$ and $f_X$}
\label{t:epsilon}
\begin{center}
\begin{tabular}{rrrr}
\hline
 \multicolumn{4}{c}{$\varepsilon_X = \Phi_{\rm keV}/\Phi_{\rm H}$}\\
\hline
          & Standard & Moderate $B$& Strong $B$\\
$\eta_M$: & $0-0.001$ & $0.01$ & $0.10$\\
\hline
\hline
$P_0$ &  1.7052E+01 &  1.7988E+01 &  1.8667E+01\\
$P_1$ & -2.2330E+01 & -2.3550E+01 & -2.3939E+01\\
$P_2$ &  1.0939E+01 &  1.1539E+01 &  1.1506E+01\\
$P_3$ & -2.3771E+00 & -2.5088E+00 & -2.4576E+00\\
$P_4$ &  1.9346E-01 &  2.0436E-01 &  1.9699E-01\\
$Q_0$ &  1.0000E+00 &  1.0000E+00 &  1.0000E+00\\
$Q_1$ & -1.2440E+00 & -1.2368E+00 & -1.1812E+00\\
$Q_2$ &  5.0908E-01 &  5.0319E-01 &  4.6036E-01\\
$Q_3$ & -6.8751E-02 & -6.7558E-02 & -5.9335E-02\\
\hline
 \multicolumn{4}{c}{$f_X = \Psi_{\rm keV}/\Psi_{\rm H}$}\\
\hline
          & Standard & Moderate $B$& Strong $B$\\
$\eta_M$: & $0-0.001$ & $0.01$ & $0.10$\\
\hline
\hline
$P_0$ & 1.7122E+01 &  1.3990E+01 &  1.3982E+01\\
$P_1$ &-2.2502E+01 & -1.8006E+01 & -1.7503E+01\\
$P_2$ & 1.1115E+01 &  8.7086E+00 &  8.2407E+00\\
$P_3$ &-2.4519E+00 & -1.8817E+00 & -1.7345E+00\\
$P_4$ & 2.0452E-01 &  1.5393E-01 &  1.3825E-01\\
$Q_0$ & 1.0000E+00 &  1.0000E+00 &  1.0000E+00\\
$Q_1$ &-1.2566E+00 & -1.2394E+00 & -1.1801E+00\\
$Q_2$ & 5.1956E-01 &  5.0604E-01 &  4.5981E-01\\
$Q_3$ &-7.0893E-02 & -6.8261E-02 & -5.9272E-02\\
\hline
\end{tabular}
\end{center}
\end{table*}%

\clearpage

\section{Sample Model Values}
\label{app:structure}

\subsection{Shock Structural Parameters}
Tables \ref{t:shckprop1} and \ref{t:shckprop2} give a sample of the physical parameters of shocks from model series 8 (table \ref{t:shockgrid}): $\eta_M = 0.001$, $R = 1\times10^{6}$,a standard weak magnetic case.
The table is sampled at $0.1$dex velocity intervals -- every 5th model series velocity -- from $\log_{10}(v_s) = 1.30$ ($\sim20$km/s), to the maximum velocity;$\log_{10}(v_s) = 3.18$ ($\sim1513$km/s),
with additional 0.02\,dex interval points between 100 -- 200\,km/s, where $\Psi_H \sim 1.0$.  Full tables for all 15 grid cases at 0.02dex velocity resolution are available as extended tables in the online journal edition.

\begin{table*}[htp]
\caption{Shock Structural Parameters}
\label{t:shckprop1}
\begin{center}
\begin{tabular}{*{7}{r}}
\multicolumn{7}{l}{$ \eta_M = $ 1.00e-03, $ (2P_B/P_{\rm ram} ) $ , $B_0$ = $ 17.210~\mu$G, ${\cal M}_A = 31.62 $ }\\
\multicolumn{7}{l}{$ R = $ 1.00e+06, $ ( n_H.v_s^2 \; : \; n_H \; \mbox{\rm in cm}^{-3}, \;v_s\; \mbox{\rm in km/s}) $, $P_{\rm ram} = $ 2.357e-08 dyne/cm$^2$}\\
\hline
\multicolumn{1}{c}{Velocity}        & \multicolumn{1}{c}{Hydrogen}     & \multicolumn{2}{c}{Temperature }                                 &  \multicolumn{2}{c}{Hydrogen}                   & \multicolumn{1}{c}{Particle}\\
\multicolumn{1}{c}{$\log_{10}v_s$}  & \multicolumn{1}{c}{Density}      & \multicolumn{1}{c}{Pre--shock} & \multicolumn{1}{c}{Post--shock} &  \multicolumn{2}{c}{Ionization Fraction}        & \multicolumn{1}{c}{Mass}    \\
\multicolumn{1}{c}{(km/s)}          & \multicolumn{1}{c}{(cm$^{-3}$)}  & \multicolumn{1}{c}{(K)}        & \multicolumn{1}{c}{(K)}         &  \multicolumn{1}{c}{HI}& \multicolumn{1}{c}{HII}& \multicolumn{1}{c}{(a.m.u)} \\
\hline
\hline
  1.30 & 2.512+03 & 1.405+01 & 1.152+04 & 1.000+00 & 2.344-09 & 1.291 \\
  1.40 & 1.585+03 & 2.058+01 & 1.826+04 & 1.000+00 & 3.681-07 & 1.291 \\
  1.50 & 1.000+03 & 2.721+02 & 2.915+04 & 1.000+00 & 2.338-05 & 1.291 \\
  1.60 & 6.310+02 & 1.062+04 & 5.501+04 & 9.994-01 & 6.400-04 & 1.290 \\
  1.70 & 3.981+02 & 1.209+04 & 8.293+04 & 9.964-01 & 3.568-03 & 1.287 \\
  1.80 & 2.512+02 & 1.242+04 & 1.242+05 & 9.824-01 & 1.758-02 & 1.271 \\
  1.90 & 1.585+02 & 1.360+04 & 1.694+05 & 8.282-01 & 1.718-01 & 1.114 \\
  2.00 & 1.000+02 & 1.403+04 & 2.290+05 & 6.650-01 & 3.351-01 & 0.968 \\
  2.02 & 9.120+01 & 1.412+04 & 2.419+05 & 6.194-01 & 3.806-01 & 0.934 \\
  2.04 & 8.318+01 & 1.420+04 & 2.514+05 & 5.437-01 & 4.563-01 & 0.887 \\
  2.06 & 7.586+01 & 1.427+04 & 2.589+05 & 4.456-01 & 5.544-01 & 0.835 \\
  2.08 & 6.918+01 & 1.429+04 & 2.640+05 & 3.214-01 & 6.786-01 & 0.777 \\
  2.10 & 6.310+01 & 1.414+04 & 2.686+05 & 1.852-01 & 8.148-01 & 0.722 \\
  2.12 & 5.754+01 & 1.316+04 & 2.758+05 & 6.607-02 & 9.339-01 & 0.679 \\
  2.14 & 5.248+01 & 9.602+03 & 2.896+05 & 1.219-02 & 9.878-01 & 0.659 \\
  2.16 & 4.786+01 & 8.731+03 & 3.138+05 & 9.346-03 & 9.907-01 & 0.654 \\
  2.18 & 4.365+01 & 8.349+03 & 3.413+05 & 8.695-03 & 9.913-01 & 0.651 \\
  2.20 & 3.981+01 & 8.134+03 & 3.720+05 & 8.142-03 & 9.919-01 & 0.649 \\
  2.22 & 3.631+01 & 8.141+03 & 4.060+05 & 7.609-03 & 9.924-01 & 0.647 \\
  2.24 & 3.311+01 & 8.401+03 & 4.437+05 & 7.107-03 & 9.929-01 & 0.645 \\
  2.26 & 3.020+01 & 8.849+03 & 4.846+05 & 6.586-03 & 9.934-01 & 0.643 \\
  2.28 & 2.754+01 & 9.317+03 & 5.291+05 & 5.957-03 & 9.940-01 & 0.641 \\
  2.30 & 2.512+01 & 9.714+03 & 5.774+05 & 5.317-03 & 9.947-01 & 0.638 \\
  2.40 & 1.585+01 & 1.086+04 & 8.939+05 & 3.246-03 & 9.968-01 & 0.626 \\
  2.50 & 1.000+01 & 1.171+04 & 1.401+06 & 2.315-03 & 9.977-01 & 0.621 \\
  2.60 & 6.310+00 & 1.246+04 & 2.209+06 & 1.621-03 & 9.984-01 & 0.619 \\
  2.70 & 3.981+00 & 1.444+04 & 3.493+06 & 9.876-04 & 9.990-01 & 0.619 \\
  2.80 & 2.512+00 & 1.811+04 & 5.527+06 & 5.796-04 & 9.994-01 & 0.618 \\
  2.90 & 1.585+00 & 2.278+04 & 8.749+06 & 3.545-04 & 9.997-01 & 0.618 \\
  3.00 & 1.000+00 & 2.922+04 & 1.385+07 & 2.001-04 & 9.998-01 & 0.617 \\
  3.10 & 6.310-01 & 3.990+04 & 2.194+07 & 9.307-05 & 9.999-01 & 0.617 \\
  3.18 & 4.365-01 & 5.551+04 & 3.170+07 & 4.145-05 & 1.000+00 & 0.617 \\
\hline
\multicolumn{7}{l}{\footnotesize{Notation: $X.XXX\pm YY \equiv X.XXX\times10^{\pm YY}$.}}
\end{tabular}
\end{center}
\end{table*}%

\begin{table*}[htp]
\caption{Further Shock Structural Parameters}
\label{t:shckprop2}
\begin{center}
\begin{tabular}{*{8}{r}}
\multicolumn{8}{l}{$ \eta_M = $ 1.00e-03, $ (2P_B/P_{\rm ram} ) $ , $B_0$ = $ 17.210~\mu$G, ${\cal M}_A = 31.62 $ }\\
\multicolumn{8}{l}{$ R = $ 1.00e+06, $ ( n_H.v_s^2 \; : \; n_H \; \mbox{\rm in cm}^{-3}, \;v_s\; \mbox{\rm in km/s}) $, $P_{\rm ram} = $ 2.357e-08 dyne/cm$^2$}\\
\hline
\multicolumn{1}{c}{Velocity}       &                            & \multicolumn{2}{c}{Magnetic $\alpha$}                           & \multicolumn{2}{c}{Compression}                      & \multicolumn{2}{c}{Cooling Column Densities} \\
\multicolumn{1}{c}{$\log_{10}v_s$} & \multicolumn{1}{c}{Mach}   & \multicolumn{2}{c}{$P_B/P_{\rm gas}$}                   & \multicolumn{2}{c}{Factor}                           & \multicolumn{1}{c}{$\lambda_4$} & \multicolumn{1}{c}{$\lambda_3$}\\
\multicolumn{1}{c}{(km/s)}         & \multicolumn{1}{c}{Number} & \multicolumn{1}{c}{Pre--shock} & \multicolumn{1}{c}{Post--shock}& \multicolumn{1}{c}{Jump}& \multicolumn{1}{c}{Max.$^*$}   & \multicolumn{1}{c}{(cm$^{-2}$)}  & \multicolumn{1}{c}{(cm$^{-2}$)}   \\
\hline
\hline
  1.30 & 51.38 & 2.200+00 & 7.905-03 & 3.978 & 3.081+01 & 2.074+18 & 5.820+18 \\
  1.40 & 53.44 & 2.380+00 & 7.907-03 & 3.978 & 3.511+01 & 5.967+17 & 1.656+18 \\
  1.50 & 18.50 & 2.854-01 & 7.775-03 & 3.948 & 3.835+01 & 9.394+16 & 5.149+17 \\
  1.60 & 3.73 & 1.158-02 & 5.150-03 & 3.279 & 4.196+01 & 1.128+16 & 1.215+17 \\
  1.70 & 4.39 & 1.608-02 & 5.767-03 & 3.449 & 4.447+01 & 3.323+15 & 5.633+16 \\
  1.80 & 5.42 & 2.450-02 & 6.402-03 & 3.615 & 4.826+01 & 1.636+15 & 3.120+16 \\
  1.90 & 6.11 & 3.110-02 & 6.688-03 & 3.688 & 5.086+01 & 7.563+15 & 4.778+16 \\
  2.00 & 7.06 & 4.149-02 & 6.971-03 & 3.757 & 4.901+01 & 1.523+16 & 1.298+17 \\
  2.02 & 7.24 & 4.363-02 & 7.014-03 & 3.768 & 4.854+01 & 1.672+16 & 1.477+17 \\
  2.04 & 7.36 & 4.518-02 & 7.042-03 & 3.775 & 4.777+01 & 2.063+16 & 1.660+17 \\
  2.06 & 7.46 & 4.637-02 & 7.063-03 & 3.780 & 4.679+01 & 2.667+16 & 1.840+17 \\
  2.08 & 7.53 & 4.725-02 & 7.078-03 & 3.783 & 4.557+01 & 3.572+16 & 2.021+17 \\
  2.10 & 7.64 & 4.866-02 & 7.101-03 & 3.789 & 4.438+01 & 4.744+16 & 2.253+17 \\
  2.12 & 8.04 & 5.390-02 & 7.175-03 & 3.807 & 4.345+01 & 5.933+16 & 2.497+17 \\
  2.14 & 9.71 & 7.863-02 & 7.400-03 & 3.860 & 4.305+01 & 6.940+16 & 2.696+17 \\
  2.16 & 10.63 & 9.415-02 & 7.483-03 & 3.880 & 4.309+01 & 8.149+16 & 2.943+17 \\
  2.18 & 11.35 & 1.074-01 & 7.535-03 & 3.892 & 4.315+01 & 9.720+16 & 3.264+17 \\
  2.20 & 12.02 & 1.205-01 & 7.576-03 & 3.902 & 4.322+01 & 1.180+17 & 3.708+17 \\
  2.22 & 12.56 & 1.316-01 & 7.605-03 & 3.908 & 4.329+01 & 1.514+17 & 4.374+17 \\
  2.24 & 12.94 & 1.394-01 & 7.622-03 & 3.912 & 4.335+01 & 2.200+17 & 5.651+17 \\
  2.26 & 13.18 & 1.447-01 & 7.633-03 & 3.915 & 4.338+01 & 3.443+17 & 7.806+17 \\
  2.28 & 13.42 & 1.501-01 & 7.643-03 & 3.917 & 4.339+01 & 4.943+17 & 1.028+18 \\
  2.30 & 13.74 & 1.572-01 & 7.656-03 & 3.920 & 4.341+01 & 6.407+17 & 1.270+18 \\
  2.40 & 16.20 & 2.187-01 & 7.730-03 & 3.937 & 4.345+01 & 1.882+18 & 3.037+18 \\
  2.50 & 19.56 & 3.189-01 & 7.790-03 & 3.951 & 4.363+01 & 4.318+18 & 6.354+18 \\
  2.60 & 23.84 & 4.738-01 & 7.834-03 & 3.961 & 4.382+01 & 1.199+19 & 1.560+19 \\
  2.70 & 27.87 & 6.474-01 & 7.858-03 & 3.967 & 4.397+01 & 4.656+19 & 5.724+19 \\
  2.80 & 31.32 & 8.174-01 & 7.872-03 & 3.970 & 4.406+01 & 1.296+20 & 1.693+20 \\
  2.90 & 35.14 & 1.029+00 & 7.883-03 & 3.973 & 4.412+01 & 2.650+20 & 3.676+20 \\
  3.00 & 39.05 & 1.270+00 & 7.891-03 & 3.974 & 4.416+01 & 5.926+20 & 7.697+20 \\
  3.10 & 42.06 & 1.474+00 & 7.895-03 & 3.975 & 4.418+01 & 1.652+21 & 1.920+21 \\
  3.18 & 42.86 & 1.531+00 & 7.896-03 & 3.976 & 4.420+01 & 3.228+21 & 3.632+21 \\
\hline
\multicolumn{8}{l}{\footnotesize{Notation: $X.XXX\pm YY \equiv X.XXX\times10^{\pm YY}$.}}\\
\multicolumn{8}{l}{\footnotesize{$^*$ using density after cooling to $\log_{10} T = 3.0$}}\\
\end{tabular}
\end{center}
\end{table*}%

\subsection{Shock Radiation Parameters}

Table \ref{t:shckradiation} gives a sample of the radiation $\Psi$ and $\Phi$ parameters of shocks from model series 8 (table \ref{t:shockgrid}); $\eta_M = 0.001$, $R = 1\times10^{6}$,a standard weak magnetic case.
The table is sampled at the same velocities as \ref{t:shckprop1} and \ref{t:shckprop2}. Full tables for all 15 grid cases at 0.02dex velocity resolution are available in extended tables in the online journal edition.

\begin{table*}[htp]
\caption{Shock Radiation Parameters}
\label{t:shckradiation}
\begin{center}
\begin{tabular}{*{8}{r}}
\multicolumn{8}{l}{$ \eta_M = $ 1.00e-03, $ (2P_B/P_{\rm ram} ) $ , $B_0$ = $ 17.210~\mu$G, ${\cal M}_A = 31.62 $ }\\
\multicolumn{8}{l}{$ R = $ 1.00e+06, $ ( n_H.v_s^2 \; : \; n_H \; \mbox{\rm in cm}^{-3}, \;v_s\; \mbox{\rm in km/s}) $, $P_{\rm ram} = $ 2.357e-08 dyne/cm$^2$}\\
\hline
\multicolumn{1}{c}{Velocity}             & \multicolumn{1}{c}{Hydrogen}\\
\multicolumn{1}{c}{$\log_{10}v_s$}       & \multicolumn{1}{c}{Density}
& \multicolumn{3}{c}{Photon Parameters } & \multicolumn{3}{c}{Energy Parameters} \\
\multicolumn{1}{c}{(km/s)} & \multicolumn{1}{c}{(cm$^{-3}$)}
& \multicolumn{1}{c}{$\Psi_{\rm tot}$} & \multicolumn{1}{c}{$\Psi_H$} & \multicolumn{1}{c}{$\Psi_{\rm keV}$}
& \multicolumn{1}{c}{$\Phi_{\rm tot}$} & \multicolumn{1}{c}{$\Phi_H$} & \multicolumn{1}{c}{$\Phi_{\rm keV}$} \\
\hline
\hline
  1.30 & 2.512+03 & 9.658+04 & 1.925-09 & 0.000+00 & 1.068-12 & 4.570-20 & 0.000+00 \\
  1.40 & 1.585+03 & 5.452+04 & 3.552-07 & 0.000+00 & 1.695-12 & 1.000-17 & 0.000+00 \\
  1.50 & 1.000+03 & 3.444+04 & 2.255-05 & 0.000+00 & 2.664-12 & 6.406-16 & 0.000+00 \\
  1.60 & 6.310+02 & 1.689+04 & 5.879-04 & 0.000+00 & 5.242-12 & 1.613-14 & 0.000+00 \\
  1.70 & 3.981+02 & 1.233+04 & 3.302-03 & 0.000+00 & 8.320-12 & 8.665-14 & 0.000+00 \\
  1.80 & 2.512+02 & 1.008+04 & 1.637-02 & 0.000+00 & 1.427-11 & 4.166-13 & 0.000+00 \\
  1.90 & 1.585+02 & 1.321+04 & 1.631-01 & 0.000+00 & 2.796-11 & 4.683-12 & 0.000+00 \\
  2.00 & 1.000+02 & 2.521+04 & 3.408-01 & 0.000+00 & 3.936-11 & 1.159-11 & 0.000+00 \\
  2.02 & 9.120+01 & 2.726+04 & 3.895-01 & 0.000+00 & 4.199-11 & 1.352-11 & 0.000+00 \\
  2.04 & 8.318+01 & 2.891+04 & 4.674-01 & 0.000+00 & 4.621-11 & 1.616-11 & 0.000+00 \\
  2.06 & 7.586+01 & 2.966+04 & 5.642-01 & 0.000+00 & 5.040-11 & 1.917-11 & 0.000+00 \\
  2.08 & 6.918+01 & 3.001+04 & 6.894-01 & 0.000+00 & 5.490-11 & 2.292-11 & 0.000+00 \\
  2.10 & 6.310+01 & 3.030+04 & 8.314-01 & 0.000+00 & 5.957-11 & 2.720-11 & 0.000+00 \\
  2.12 & 5.754+01 & 3.060+04 & 9.614-01 & 0.000+00 & 6.417-11 & 3.125-11 & 0.000+00 \\
  2.14 & 5.248+01 & 3.099+04 & 1.077+00 & 0.000+00 & 6.930-11 & 3.501-11 & 0.000+00 \\
  2.16 & 4.786+01 & 3.184+04 & 1.210+00 & 0.000+00 & 7.639-11 & 3.951-11 & 0.000+00 \\
  2.18 & 4.365+01 & 3.301+04 & 1.336+00 & 0.000+00 & 8.401-11 & 4.408-11 & 0.000+00 \\
  2.20 & 3.981+01 & 3.463+04 & 1.447+00 & 0.000+00 & 9.192-11 & 4.841-11 & 0.000+00 \\
  2.22 & 3.631+01 & 3.710+04 & 1.535+00 & 0.000+00 & 1.001-10 & 5.237-11 & 0.000+00 \\
  2.24 & 3.311+01 & 4.154+04 & 1.606+00 & 0.000+00 & 1.091-10 & 5.673-11 & 0.000+00 \\
  2.26 & 3.020+01 & 4.897+04 & 1.699+00 & 0.000+00 & 1.196-10 & 6.332-11 & 0.000+00 \\
  2.28 & 2.754+01 & 5.743+04 & 1.845+00 & 0.000+00 & 1.320-10 & 7.246-11 & 0.000+00 \\
  2.30 & 2.512+01 & 6.508+04 & 2.013+00 & 0.000+00 & 1.454-10 & 8.241-11 & 0.000+00 \\
  2.40 & 1.585+01 & 1.046+05 & 2.889+00 & 2.472-09 & 2.300-10 & 1.445-10 & 4.236-18 \\
  2.50 & 1.000+01 & 1.521+05 & 3.913+00 & 1.055-06 & 3.583-10 & 2.411-10 & 1.868-15 \\
  2.60 & 6.310+00 & 2.414+05 & 5.561+00 & 7.024-05 & 5.684-10 & 3.982-10 & 1.296-13 \\
  2.70 & 3.981+00 & 5.373+05 & 7.719+00 & 2.335-03 & 8.896-10 & 6.358-10 & 4.368-12 \\
  2.80 & 2.512+00 & 1.323+06 & 9.518+00 & 2.051-02 & 1.358-09 & 9.795-10 & 3.904-11 \\
  2.90 & 1.585+00 & 2.693+06 & 1.088+01 & 6.970-02 & 2.082-09 & 1.507-09 & 1.400-10 \\
  3.00 & 1.000+00 & 4.149+06 & 1.307+01 & 1.966-01 & 3.275-09 & 2.383-09 & 4.207-10 \\
  3.10 & 6.310-01 & 5.745+06 & 1.705+01 & 4.776-01 & 5.257-09 & 3.864-09 & 1.149-09 \\
  3.18 & 4.365-01 & 7.669+06 & 2.091+01 & 8.069-01 & 7.673-09 & 5.689-09 & 2.210-09 \\
\hline
\multicolumn{8}{l}{\footnotesize{Notation: $X.XXX\pm YY \equiv X.XXX\times10^{\pm YY}$.}}\\
\end{tabular}
\end{center}
\end{table*}%

\begin{table*}[htp]
\caption{Sample Shock $4\pi$ steradian  Balmer Line Radiation}
\label{t:shckbalmer}
\begin{center}
\begin{tabular}{*{6}{r}}
\hline
\multicolumn{6}{l}{$ \eta_M =  1.00e-03$, ${\cal M}_A = 31.62 $, $ R =  1.00e+06 $}\\
\multicolumn{6}{l}{$\Phi_{\rm H\beta} = 4.0862\times 10^{-12} \Psi_{\rm H\beta}$, $\Phi_{\rm H\alpha} = 3.0268\times 10^{-12} \Psi_{\rm H\alpha}$}\\
\hline
\multicolumn{1}{c}{Velocity}        & \multicolumn{1}{c}{Hydrogen}   &                      &                  & \multicolumn{2}{c}{Ratios } \\
\multicolumn{1}{c}{$\log_{10}v_s$}  & \multicolumn{1}{c}{Density}    & \multicolumn{2}{c}{Photon Parameters }  & \multicolumn{1}{c}{Photons } & \multicolumn{1}{c}{Intensity }\\
\multicolumn{1}{c}{(km/s)}          & \multicolumn{1}{c}{(cm$^{-3}$)}& \multicolumn{1}{c}{$\Psi_{\rm H\beta}$} & \multicolumn{1}{c}{$\Psi_{\rm H\alpha}$}& \multicolumn{1}{c}{$\Psi_{\rm H\alpha}/\Psi_{\rm H\beta}$} & \multicolumn{1}{c}{$\Phi_{\rm H\alpha}/\Phi_{\rm H\beta}$} \\
\hline
\hline
    1.30 &  2.512+03 &  2.742-03 & 4.791-04 &  5.724 &  4.240 \\
    1.40 &  1.585+03 &  1.660-02 & 2.516-03 &  6.595 &  4.885 \\
    1.50 &  1.000+03 &  5.059-02 & 8.086-03 &  6.256 &  4.634 \\
    1.60 &  6.310+02 &  1.692-01 & 3.010-02 &  5.623 &  4.166 \\
    1.70 &  3.981+02 &  3.260-01 & 6.084-02 &  5.359 &  3.970 \\
    1.80 &  2.512+02 &  6.010-01 & 1.174-01 &  5.121 &  3.793 \\
    1.90 &  1.585+02 &  7.397-01 & 1.656-01 &  4.468 &  3.310 \\
    2.00 &  1.000+02 &  7.117-01 & 1.668-01 &  4.266 &  3.160 \\
    2.02 &  9.120+01 &  7.285-01 & 1.716-01 &  4.245 &  3.145 \\
    2.04 &  8.318+01 &  7.659-01 & 1.812-01 &  4.226 &  3.131 \\
    2.06 &  7.586+01 &  8.158-01 & 1.938-01 &  4.210 &  3.119 \\
    2.08 &  6.918+01 &  8.839-01 & 2.107-01 &  4.194 &  3.107 \\
    2.10 &  6.310+01 &  9.623-01 & 2.303-01 &  4.179 &  3.095 \\
    2.12 &  5.754+01 &  1.034+00 & 2.483-01 &  4.165 &  3.085 \\
    2.14 &  5.248+01 &  1.106+00 & 2.663-01 &  4.152 &  3.076 \\
    2.16 &  4.786+01 &  1.192+00 & 2.882-01 &  4.137 &  3.064 \\
    2.18 &  4.365+01 &  1.271+00 & 3.083-01 &  4.123 &  3.054 \\
    2.20 &  3.981+01 &  1.335+00 & 3.246-01 &  4.111 &  3.045 \\
    2.22 &  3.631+01 &  1.378+00 & 3.360-01 &  4.102 &  3.038 \\
    2.24 &  3.311+01 &  1.403+00 & 3.430-01 &  4.091 &  3.030 \\
    2.26 &  3.020+01 &  1.433+00 & 3.515-01 &  4.076 &  3.019 \\
    2.28 &  2.754+01 &  1.499+00 & 3.691-01 &  4.060 &  3.007 \\
    2.30 &  2.512+01 &  1.585+00 & 3.918-01 &  4.045 &  2.996 \\
    2.40 &  1.585+01 &  2.067+00 & 5.167-01 &  4.000 &  2.963 \\
    2.50 &  1.000+01 &  2.578+00 & 6.423-01 &  4.014 &  2.973 \\
    2.60 &  6.310+00 &  3.636+00 & 8.926-01 &  4.073 &  3.017 \\
    2.70 &  3.981+00 &  5.544+00 & 1.324+00 &  4.186 &  3.101 \\
    2.80 &  2.512+00 &  8.000+00 & 1.865+00 &  4.290 &  3.178 \\
    2.90 &  1.585+00 &  1.071+01 & 2.467+00 &  4.341 &  3.216 \\
    3.00 &  1.000+00 &  1.447+01 & 3.267+00 &  4.431 &  3.282 \\
    3.10 &  6.310-01 &  2.154+01 & 4.641+00 &  4.641 &  3.438 \\
    3.18 &  4.365-01 &  3.020+01 & 6.277+00 &  4.811 &  3.564 \\
\hline
\multicolumn{6}{l}{\footnotesize{Notation: $X.XXX\pm YY \equiv X.XXX\times10^{\pm YY}$.}}\\
\end{tabular}
\end{center}
\end{table*}%


\end{document}